\documentclass[11pt,a4paper]{article}
\usepackage{jcappub}
\usepackage[utf8]{inputenc}
\usepackage[table]{xcolor}
\usepackage{verbatim}
\usepackage{amsmath,amsfonts,amssymb}
\usepackage{float}
\usepackage{xcolor}
\usepackage{array, booktabs, tabularx, makecell, multirow}
\usepackage{multicol}
\usepackage{tikz}
\usetikzlibrary{shapes.geometric, arrows}

\tikzstyle{startstop} = [rectangle, rounded corners, minimum width=3cm, minimum height=1cm,text centered, draw=black, fill=red!30]
\tikzstyle{io} = [trapezium, trapezium left angle=70, trapezium right angle=110, minimum width=3cm, minimum height=1cm, text centered, draw=black, fill=blue!30]
\tikzstyle{process} = [rectangle, minimum width=3cm, minimum height=1cm, text centered, text width=3cm, draw=black, fill=orange!30]
\tikzstyle{decision} = [diamond, minimum width=3cm, minimum height=1cm, text centered, draw=black, fill=green!30]
\tikzstyle{arrow} = [thick,->,>=stealth]

\tikzstyle{lcdmparam} = [rectangle, rounded corners, minimum width=3cm, minimum height=1cm,text centered, draw=black, fill=yellow!30]
\tikzstyle{stockcode} = [rectangle, rounded corners, minimum width=3cm, minimum height=1cm,text centered, draw=black, fill=red!30]
\tikzstyle{modcode} = [rectangle, rounded corners, minimum width=3cm, minimum height=1cm,text centered, draw=black, fill=blue!30]
\tikzstyle{modparam} = [rectangle, rounded corners, minimum width=3cm, minimum height=1cm,text centered, draw=black, fill=orange!30]
\tikzstyle{goal} = [rectangle, rounded corners, minimum width=3cm, minimum height=1cm,text centered, draw=black, fill=green!50]

\usetikzlibrary{mindmap,backgrounds}
\usepackage{adjustbox}
\usepackage{tabularx}
\hypersetup{%
,urlcolor=red
,citecolor=red
,linkcolor=blue
}

\def\mnras{Monthly Notices of the Royal Astronomical Society}
\def\aap{Astronomy and Astrophysics}
\def\aj{Astrophysical Journal}
\def\jcap{Journal of Cosmology and Astroparticle Physics}
\def\apj{Astrophysical Journal}
\def\physrep{Physical Reports}
\def\prd{Physical Review D}
\def\nat{Nature}
\def\apjl{Astrophysical Journal, Letters}

\title{Cosmological gravity on all scales II: Model independent modified gravity $N$-body simulations}
\author[a]{Sankarshana Srinivasan,}
\author[b,a]{Daniel B Thomas,}%\note{On leave from XXX.}}
\author[c,a]{Francesco Pace,}
\author[a]{and Richard Battye}

\affiliation[a]{Jodrell Bank Centre for Astrophysics, School of Natural Sciences, University of Manchester, \\ Alan Turing Building, Oxford Road, Manchester, M13 9PL, United Kingdom}
\affiliation[b]{School of Physics and Astronomy, 
Queen Mary University of London \\
G O Jones Building, 327 Mile End Road, London, E1 4NS, UK}
\affiliation[c]{Dipartimento di Fisica ed Astronomia ``Augusto Righi", Alma Mater Studiorum Universit\`a di Bologna, Via Gobetti 93/2, I-40129 Bologna, Italy}

\emailAdd{sankarshana.srinivasan@postgrad.manchester.ac.uk}
\emailAdd{dan.b.thomas1@gmail.com}
\emailAdd{francesco.pace9@unibo.it}
\emailAdd{richard.battye@manchester.ac.uk}

\date{April 2020}

\abstract{Model-independent constraints on modified gravity models hitherto exist mainly on linear scales \cite{ref:IshakBin}. A recently developed formalism presented a consistent parameterisation that is valid on all scales \cite{ref:DanPF}. Using this approach, we perform model-independent modified gravity $N$-body simulations on all cosmological scales with a time-dependent $\mu$. We present convergence tests of our simulations, and we examine how well existing fitting functions reproduce the non-linear matter power spectrum of the simulations. We find that although there is a significant variation in the accuracy of all of the fitting functions over the parameter space of our simulations, the \texttt{ReACT} \cite{ref:ReactTheory} framework delivers the most consistent performance for the matter power spectrum. We comment on how this might be improved to the level required for future surveys such as \textit{Euclid} and the Vera Rubin Telescope (LSST). We also show how to compute weak-lensing observables consistently from the simulated matter power spectra in our approach, and show that \texttt{ReACT} also performs best when fitting the weak-lensing observables. This paves the way for a full model-independent test of modified gravity using all of the data from such upcoming surveys.
}


\arxivnumber{}

\begin{document}

\label{firstpage}

\maketitle

\flushbottom

\section{Introduction}

In the current cosmological paradigm, dark matter (DM) and dark energy (DE), the so-called dark sector, are the dominant components of the energy density of the Universe, but their underlying physics is still unknown. In addition, assuming that General Relativity (GR) (plus the cosmological constant $\Lambda$) is the fundamental law of gravity is an extrapolation of many orders of magnitude (from the Solar System regime where GR is well-tested). Furthermore, the cosmological constant and its theoretical foundations are yet to be completely understood \cite{ref:WeinbergGravCosmo}. Dark matter has also hitherto eluded direct detection efforts \cite{ref:SchumannWimp2019, ref:IrastorzaRedondo2019}. These problems are key parts of the theoretical motivation for testing gravity. 

The modified gravity model space is vast [see \cite{ref:CliftonReview, ref:KoyamaRev} for reviews of this topic]. As a result, in order to constrain significant volumes of this model space, one needs to formulate model-independent parameterisations that cover a large fraction of the model space.\footnote{Of course, a complete model independence is very difficult. For example, most modified gravity approaches assume a perturbed FLRW spacetime.} There are many cosmological parameterisations of modified gravity \cite{ref:BattyePearson,ref:Gubitosi,ref:Ferreira2014PPF,ref:DossetIshak2011,ref:Bertschinger2006,ref:Pogosian}, however the vast majority of these are developed in the ``linear'' regime, where perturbations (particularly density perturbations) are small. These have been constrained with multiple observational probes \cite{ref:Schmidt,ref:Perlmutter,Riess2007,ref:Planck2018,Planck2018_VIII,DES_Y1_2018,Gruen2018}, but no deviation from $\Lambda$CDM has been detected universally. However, the restriction to linear scales significantly limits what these parameterisations can achieve \footnote{Note that some recent works \cite{ref:LucasNL, ref:CliftonPoisson} are valid on non-linear scales; we will return to these and contextualise them later in this work.}.

This restriction represents a significant challenge, since upcoming cosmological surveys such as \textit{Euclid} \footnote{\href{https://www.euclid-ec.org}{https://www.euclid-ec.org}}, the Vera Rubin Observatory \footnote{\href{https://www.lsst.org}{https://www.lsst.org}}, the Nancy Roman Space Telescope \footnote{\href{https://roman.gsfc.nasa.gov}{https://roman.gsfc.nasa.gov}} and the Square Kilometre Array (SKA) \footnote{\href{https://www.skatelescope.org}{https://www.skatelescope.org}} will generate a multitude of data on non-linear cosmological scales. Until now, various studies of structure formation were carried out by performing $N$-body simulations in specific modified gravity models, such as $f(R)$ gravity \cite{ref:f(R)First} or Dvali-Gabadadze-Porrati (DGP) gravity \cite{ref:DGPFirst}; for a review, see \cite{ref:MGCodeCompare, ref:COLA}. While these studies are valuable in understanding the behaviour of the respective models in the non-linear regime, they do not allow one to rule out significant regions of the modified-gravity model space. Furthermore, these studies don't allow us to perform a robust \textit{null test} of GR, unlike model-independent approaches.

To this end, it is important to establish a model-independent approach towards testing modified gravity on all cosmological scales. Recent work \cite{ref:DanPF} used the Post-Friedmann formalism \cite{ref:Milillo} to design a rigorous approach to modified gravity that is valid on all cosmological scales. Modified gravity models are typically parameterised by two parameters in linear theory. The post-Friedmann approach extends these linear theory parameterisations in a manner that can be consistently simulated and interpreted on all scales. In this work we explore the non-linear phenomenology of this model-independent approach using $N$-body simulations. Crucially, a careful choice of the parameters ensures that we can run simulations characterised by a single parameter, and include the effect of the other in post-processing (see sections \ref{section:Review} and \ref{section:Discussion} for more details).

Understanding this behaviour, and being able to capture it using fitting functions is important if we are to use this approach to analyse the data from next generation surveys. These missions will generate a multitude of data, particularly on non-linear cosmological scales. Not only will they measure the matter power spectrum through galaxy clustering, but they will also measure cosmic shear: the apparent distortion of the shapes of galaxies due to the deflection of light by the distribution of mass along the line-of-sight. As such, being able to compute these weak-lensing observables from our simulations and any fitting functions is also of paramount importance.

The plan of this work is as follows. In section~\ref{section:Review}, we provide a brief review of the Post-Friedmann formalism and discuss some features of our approach and their impact on the simulations, followed by a description of our $N$-body pipeline in section~\ref{section:Pipeline} as well as a discussion on the performance of the different fitting-functions. We present the phenomenology of the simulations in section~\ref{section:Results}.  In section~\ref{section:Discussion}, we compute weak-lensing observables in our approach from the simulations and examine the performance of the fitting functions in this context. We present our concluding remarks in section~\ref{section:Conclusion}.

\begin{comment}
\begin{itemize}
    \item  One can, for the purpose of N-body simulations, talk about the modified gravity parameter in terms of a single parameter, $G_{\rm eff}$. 
    \item Normally there are two parameters, but since we are interested in $N$-body simulations, only one of them is relevant for us, and we can model the effect of the other using weak-lensing (signpost to section 5).
\end{itemize}
\end{comment}

\section{Modified gravity framework}\label{section:Review}

Cosmological perturbation theory is typically done in the background of the FLRW metric with small perturbations. Perturbation theory enables us to impose model-independent constraints on modified gravity models on linear scales [see for example \cite{ref:BattyePearson} and references therein]. Meanwhile, a lot of progress has been made \cite{ref:WillGrav} in constraining GR on local solar-system scales by expanding the local Minkowski metric to varying powers of $1/c$, where $c$ is the speed of light, with the aim of capturing local non-linear effects. Such an expansion is done assuming the Newtonian limit, i.e., the quasi-static approximation (where time derivatives in the metric perturbations are down-weighted). The Post-Friedmann formalism \cite{ref:Milillo} generalises the Newtonian limit to weak-field cosmology, in a manner that applies to all scales. The FLRW metric is expanded with perturbations where $1/c^2$ and $1/c^3$ terms are considered leading order. The full metric is written as 
\begin{eqnarray}
g_{00} & = & -\left[1 - 2\frac{U_{\rm N}}{c^2} + \frac{1}{c^4}(2U_{\rm N}^2 - 4U_{\rm P})\right] + \mathcal{O}\left(\frac{1}{c^6}\right) \;, \\
g_{0i} & = & -\frac{a}{c^3}B_{i}^{\rm N} - \frac{a}{c^5}B_i^{\rm P} + \mathcal{O}\left(\frac{1}{c^7}\right) \;, \\
g_{ij} & = & a^2\left[\left(1 + \frac{2V_{\rm N}}{c^2} + \frac{1}{c^4}(2V_{\rm N}^2 + 4V_{\rm P})\right)\delta_{ij} + \frac{1}{c^4}h_{ij}\right] + \mathcal{O}\left(\frac{1}{c^6}\right) \;,
\end{eqnarray}
where the spatial Cartesian coordinates are as measured in an Eulerian system of reference. We operate in the Poisson gauge, one of the few gauges that is valid on both linear and non-linear cosmological scales \cite{ref:CliftonPoisson}. The vector functions $B_i^{\rm N}$ and $B_i^{\rm P}$ are divergenceless and $h_{ij}$ is a transverse and traceless tensor (see \cite{ref:Milillo} for a detailed explanation of this expansion and the physical interpretation of each term). In this approach, the dominant constituents of the cosmological fluid are pressure-less dust and the cosmological constant. The Einstein equations and the covariant conservation of energy-momentum are used to obtain a closed system of equations.

Recently, the post-Friedmann formalism was studied in the context of validating a parameterisation of modified gravity models on both linear and non-linear scales \cite{ref:DanPF}. This work uses the similarities between the linear-perturbative and Newtonian limits, and the lack of an intermediate regime where both limits fail, to create a set of ``master'' equations that can be applied on all cosmological scales. In this work, the so-called ``re-summed potentials'' \cite{ref:Milillo}
\begin{eqnarray}
\psi_{\rm P} & = & -V_{\rm N} - \frac{2}{c^2}V_{\rm P} \, , \\
\phi_{\rm P} & = & -U_{\rm N} - \frac{2}{c^2}U_{\rm P} \, , 
\end{eqnarray}
are used to define the parameterisation for modified gravity models that we use for our simulations. As discussed in \cite{ref:DanPF}, the re-summed potentials, in combination with the fact that the weak-field approximation is valid on all cosmological scales, show that the same degrees of freedom and dynamics are operating on all cosmological scales. A parameterised Poisson equation that governs the gravitational dynamics of massive particles may be written as 
\begin{eqnarray}
\frac{1}{c^2}k^2 \tilde{\phi}_{\rm P} & = & -\frac{1}{c^2}4\pi a^2 \bar{\rho} G_{\rm N}\mu(a,k)\tilde{\Delta} \,, \label{eq:MGParam} \\
\tilde{\psi}_{\rm P} & = & \eta(a,k)\tilde{\phi}_{\rm P} \, ,
 \end{eqnarray}
where $\bar{\rho}$ is the background density, $\tilde{\Delta} = \left(\tilde{\delta} - \frac{\dot{a}}{a}\frac{3}{c^2k^2}ik_i \tilde{v}_i\right)$ is the gauge-invariant density contrast in Fourier space, FP: $G_{\rm N}$ Newton's gravitational constant and $\mu(a, k)$ is an effective dimensionless Newton's gravitational ``constant'' now promoted to a general function of space and time. In $\Lambda$CDM $\mu$ is unity, but in modified gravity models this parameter may in principle be a complicated function of space and time. 
The work in \cite{ref:DanPF} showed that these parameterised equations are valid on all cosmological scales, and encompass any modified gravity model as long as certain conditions are satisfied, such as the validity of the Newtonian limit of that theory on all non-linear scales. In  particular, $f(R)$ \cite{ref:Thomas_2015} fits within this approach. Moreover, the works (\cite{ref:LucasNL, ref:HassaniNBodyMG}) can be re-interpreted in this framework as specifying the functional forms of $\mu$ that occur for some commonly studied models with screening mechanisms.

A similar and promising extension of the linear theory parameters to a much wider range of scales was developed in \cite{ref:CliftonPRL2019}, which applies to all models that fit into a Parameterised Post Newtonian (PPN) formalism \cite{ref:WillGrav}. The equivalent $\mu$ in this approach has a functional form such that it has scale dependent features on linear scales, but is purely time-dependent on  non-linear scales, due to the restriction to theories that fit into the PPN framework. Another promising proposal to address parameterising modified gravity theories on non-linear scales was proposed in \cite{ref:LucasNL}. This theoretical setup maps specific modified gravity models to known screening mechanisms via detailed spherical collapse calculations. This formalism was recently implemented to run phenomenological $N$-body simulations of $f(R)$ gravity without having to solve for the additional dynamical scalar field \cite{ref:HassaniNBodyMG}. As discussed above, this work can be re-interpreted as a description of the functional forms of $\mu$ in our approach for known theories with screening mechanisms. We note that our approach is less restrictive in theory space than either of the two approaches discussed here, however this comes at the cost of an increased difficulty to pinpoint the type of theory responsible if a deviation is found.

With all these points in mind, in this work we take the first step toward a fully model-independent treatment of modified gravity with $N$-body simulations. Various approaches for modelling $\mu$ are considered in \cite{ref:DanPF} including, for example, using the known functional forms of $\mu$ on linear scales to transition from into the non-linear regime at $k_{\rm NL}$ (traditionally chosen to be the scale at which $k^3P(k)=1$). Such an approach would therefore have an implicit time-dependence, depending on when the transition to non-linear structure formation occurs for a given $\mu$. For our simulations, we use the ``maximally phenomenological pixels'' approach as described in \cite{ref:DanPF}, in which we run $N$-body simulations where $\mu$ is a piece-wise constant function composed of independent bins or pixels in time and space which can be varied independently of each other. This allows us to be as agnostic as possible about the functional forms of the modifications to gravity. Note that this approach has been applied successfully in linear theory for modified gravity \cite{ref:IshakBin} and also generalised constraints on dark matter \cite{ref:Kopp_2018}. Since the evolution of matter is completely determined by the Poisson equation given in eq.~(\ref{eq:MGParam}), we can run the $N$-body simulations with a single modified gravity parameter $\mu$, and only require the second parameter when computing the observables from the simulation outputs (see section \ref{section:Discussion} for a more detailed discussion of how our simulations can be used to probe both parameters). Since we restrict ourselves to sufficiently sub-horizon scales, the Newtonian approximation allows us to set $\tilde{\Delta} = \tilde{\delta}$, as is typical in Newtonian $N$-body simulations.

 We will focus on models of modified gravity with a purely time-dependent $\mu$. Whilst the full time- and scale-dependent analysis would employ similar methods and techniques outlined in this work, it requires a more involved alteration of the methods used to evolve the $N$-body simulation. We will present this in future work. More importantly, the purely time-dependent $\mu$ case is of considerable interest in its own right for several reasons. The time-dependent case is a bridge between linear and non-linear studies, because the linear growth factor remains purely time-dependent. This allows the non-linear scale dependent phenomenology due to an evolving strength of gravity to be clearly studied separately from the additional complexity introduced even on linear scales when $\mu$ is scale dependent. This was shown in \cite{ref:PaceHorndeski2019} for a general class of modified gravity models on linear scales. We will return to this point later in this work (see eq.~\eqref{eqn:GrowthFactor} and the discussion following for further details on the utility of maintaining scale independence on linear scales). We leave combined scale- and time-dependent $\mu$ simulations to future work.

Previous studies of phenomenological $N$-body simulations were carried out in \cite{ref:Cui1, ref:Cui2, ref:StabJainPoisson, ref:LazloBeanPoisson, ref:Thomas2011viability}. As discussed in \cite{ref:DanPF}, the interpretation of these simulations is not always entirely clear, in terms of how they relate to linear parameterisations, and how they relate to the equations on non-linear scales in particular modified gravity theories. By working from eq.~\eqref{eq:MGParam} and the derivation and framework of \cite{ref:DanPF}, we avoid these possible problems, although we note that some of these earlier simulations are justified \textit{post-hoc} by \cite{ref:DanPF}. The works \cite{ref:StabJainPoisson, ref:LazloBeanPoisson} restrict themselves to smaller regions of parameter space, while the studies \cite{ref:Cui1, ref:Cui2} are similar to ours in the sense that they concentrate on a purely time-dependent $\mu$, although these simulations were run with a constant value for $\mu$ throughout the simulation. For this simplified setup, the authors in \cite{ref:Cui1} derive a fitting function to calculate the power spectrum for arbitrary values of $\mu$ (see the appendix in \cite{ref:Cui1} for details). In this work we explore the validity of the fitting function derived in \cite{ref:Cui1}, and other fitting functions in the literature, which attempt to reproduce the non-linear matter power spectrum $P(k)$ for modified gravity models.

Our simulations are the first to explore the phenomenology of $\mu$ having \textit{different values in multiple redshift bins} through the simulation. This means that our method allows one to be sensitive to any time variation that $\mu$ may assume, irrespective of theoretical prejudices/biases.

As noted, an advantage of a purely time-dependent $\mu$ is that the linear growth factor $D = \delta/a$ is purely time-dependent, as in $\Lambda$CDM. We use this property in section \ref{section:Results} to separate whether phenomenology is sensitive to the era in which the modified growth occurred, and provide analytic checks of the simulations on large scales. It is also useful for comparing to different $\Lambda$CDM simulations, and when considering the ``pseudo spectrum'' in \texttt{ReACT} (see section \ref{section:Pipeline}).  We solve for the linear growth factor $D(z)$ using
\begin{equation}\label{eqn:GrowthFactor}
 D^{\prime\prime} + \left(2 + \frac{H^{\prime}}{H}\right)D^{\prime} - \frac{3}{2}\frac{\mu\Omega_{\rm m}a^{-3}}{E(z)^2}D = 0 \,,
\end{equation}
where $\prime$ denotes a derivative with respect to the logarithm of the scale factor $a$ and $E(z)$ is the dimensionless Hubble function.

\section{Methodology}\label{section:Pipeline}
\subsection{\texorpdfstring{$N$-body}{N-body} simulations}
\begin{figure}
 \centering
 \includegraphics[width = 0.7 \textwidth]{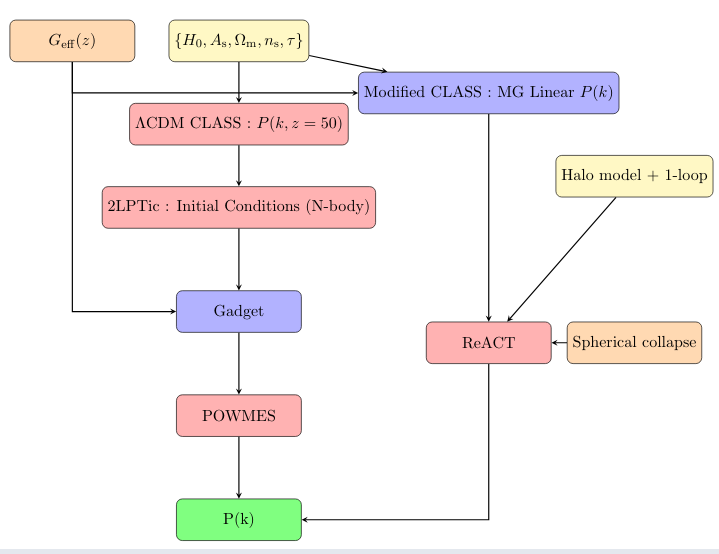}
 \caption{Flowchart of the numerical pipeline that we implemented. We establish initial conditions at $z=50$, which are then evolved up to $z=0$ using \texttt{GADGET-2}. Blue panels represent codes that we have modified, while red panels are codes that are used in their publicly available form. The modified gravity parameters are in orange, while $\Lambda$CDM parameters are in cream. The central branch represents our numerical simulation pipeline, while the right hand branch denotes our most successful pipeline for fitting the simulations (see section \ref{section:FittingFunction} for more details).}
 \label{fig:PipelineFlow}
\end{figure}

In this section we present the details of the $N$-body simulations and code modifications, as well as the tests and consistency checks we performed. We use the publicly available $N$-body code \texttt{GADGET-2} \cite{ref:GADGET-2} as the basis for our $N$-body pipeline. \texttt{GADGET-2} and modifications of it have been used extensively in the literature to run dark matter-only simulations of a wide variety \cite{ref:Millennium, ref:Millennium2, ref:MillenniumXXL}. We restrict ourselves to dark matter-only simulations in order to understand modified gravity phenomenology. We present a summary of our numerical pipeline in the flowchart presented in fig.~\ref{fig:PipelineFlow}. We will now discuss the main branch of this pipeline in detail.

\texttt{GADGET-2} employs a TreePM algorithm that splits the force computation into a long-range and short-range force. We modify both the long-range and the short-range force in the algorithm. We follow a similar procedure as in \cite{ref:ThomasContaldi2011}, where we have a modified Poisson equation with a time-dependent $\mu$ parameter, which is determined as a function of the simulation redshift. We now describe the time evolution of $\mu$ within our simulations. 

\begin{figure}
 \centering
 \includegraphics[width = 0.7\textwidth]{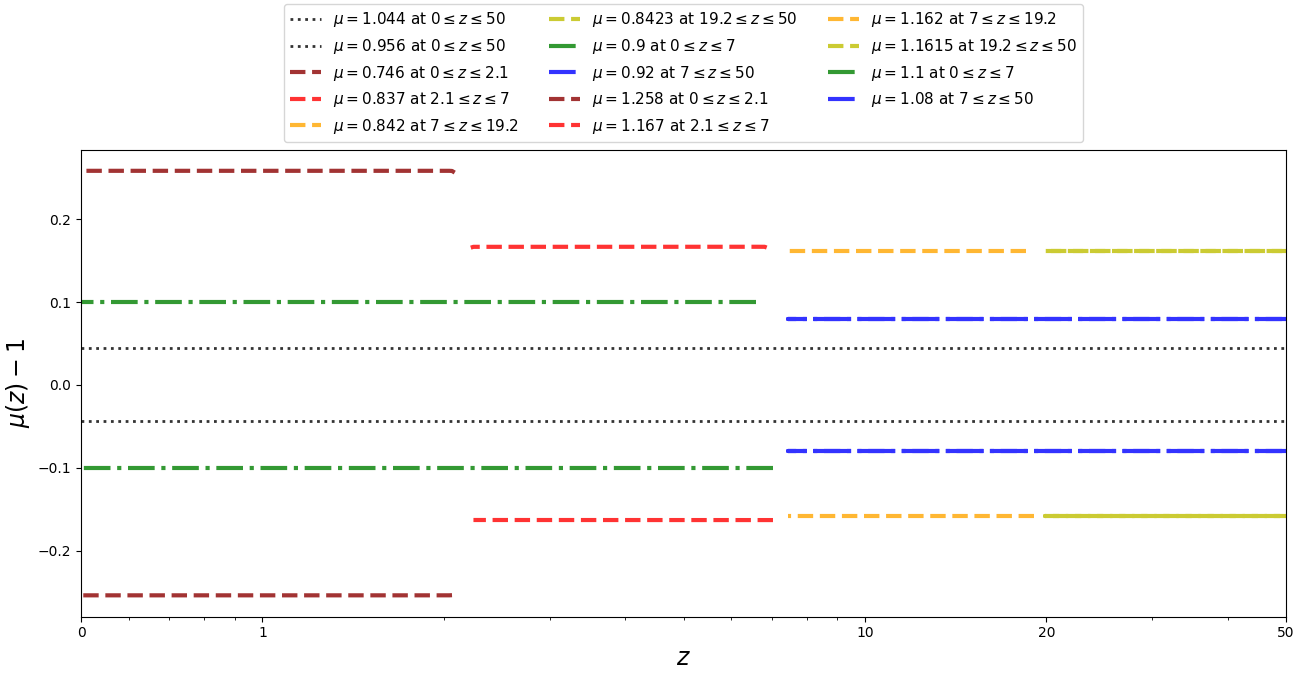}
 \caption{We illustrate the redshift evolution of $\mu$ in our simulations. Dotted lines represent the case where $\mu$ is constant through the simulation (one bin), dashed lines represent the case where there are two bins in redshift, while dot-dashed lines represent the case where there are four bins in redshift. }
 \label{fig:Geff(z)}
\end{figure}

For the phenomenological pixels that we are considering, the $\mu$ function is split into a set of piece-wise constant redshift bins, i.e., each with a constant value for $\mu$. In order to explore the phenomenology of this approach as well as make contact with earlier work \cite{ref:Cui1}, we run a suite of simulations with a simple ``1-2-4'' hierarchy of pixels. In other words, we split the redshift interval $0 \leq z \leq 50$ into 1, 2, and 4 redshift bins of equal $\Lambda$CDM growth. In order to isolate as far as possible the non-linear behaviour (of the matter power spectrum) from the linear theory behaviour, we choose the values of $\mu$ in each bin such that the linear growth factor at $z=0$ is equal when each bin is ``switched on'' exclusively, with $\mu=1$ in all the other bins (see table \ref{tab:binning} and fig.~\ref{fig:Geff(z)} for the redshift bins and the respective $\mu$ values that correspond to each of them). The widths of each bin are chosen such that, for our cosmological parameters, there is equal growth in each bin in a $\Lambda$CDM universe. See Appendix~\ref{appendix:binning} for full details of this procedure. This allows an additional consistency check for our simulations, since the matter power spectrum at $z=0$ for all our simulations should differ from the standard $\Lambda$CDM result by a factor $[D(z)]^2$.   

\begin{table}[t]
 \centering
 
 \begin{tabular}{|c|c|c|c|}
  \hline
  Number of bins & Redshift & \multicolumn{2}{c|}{$\mu$}\\
  \hline
  1 & 0-50 & 1.044 & 0.956 \\
  \hline
  \multirow{2}{*}{2} & 0-7.0 &1.100 & 0.900 \\& 7.0-50  & 1.080 & 0.920  \\
  \hline
  \multirow{4}{*}{4} & 0-2.1 & 1.256 & 0.746 \\ 
                     & 2.1-7.0 & 1.167 & 0.837 \\
                     & 7.0-19.2 & 1.162 & 0.842 \\
                     & 19.2-50.0 & 1.161 & 0.843 \\
  \hline
 \end{tabular}
 \label{tab:binning}
 \caption{The redshift bins of equal $\Lambda$CDM growth according to the procedure described in the text, where we have the same linear growth at $z=0$ when one of the bins is ``switched on'' for one, two and four bin(s). The $\mu$ values in each bin such that $P(k)$ at $z=0$ when only that bin is turned on, is identical to $P(k)$ at $z=0$ in the case when only the reference bin is turned on.}
\end{table}

We use 2nd order Lagrangian perturbation theory \cite{ref:Jenkins2LPT} to generate our initial conditions at $z=50$ for all our simulations, via the \texttt{2LPTic} code \cite{ref:2lptic}. We only consider modifications to gravity from the starting redshift of the simulations. This allows us to use the same initial conditions (and therefore realisations) for modified gravity and $\Lambda$CDM, allowing us to focus on the phenomenology due to the modified gravitational evolution. We take advantage of these identical initial conditions as this allows us to present our results in terms of ratios of matter power spectra \cite{ref:McDonaldRatio2006} (see Appendix \ref{appendix:binning} for further details). Taking ratios of matter power spectra ensures that we cancel out any realisation dependent effects that individual models, i.e., $\mu$ values and redshift bins are insensitive to, up to a wavenumber $k \sim 5\, h\,{\rm Mpc^{-1}}$ that was validated by convergence tests (see appendix \ref{appendix:Convergence} for more details). For all our simulations, we use a $\Lambda$CDM initial power spectrum created using the \texttt{CLASS} code \cite{ref:CLASS} at $z=50$, with the following parameters $\Omega_{\rm m} = 0.315, \Omega_{\Lambda} = 0.685$, $h=0.674$ and $\sigma_8 = 0.811$.  

We also run a separate set of simulations with modified initial conditions such that we obtain $\Lambda$CDM matter power spectra with the same linear growth at $z=0$ as our modified gravity simulations. These simulations have a modified value of $\sigma_8 = 0.883$ and $\sigma_8 = 0.742$, for the simulations with $\mu > 1$ and $\mu<1$, respectively. This allows us to match the linear growth at $z=0$ of all our simulations with the corresponding re-scaled $\Lambda$CDM simulation. This is important as the \texttt{ReACT} fitting function that we explore later in section \ref{section:FittingFunction} predicts exactly the ratio of matter power spectra in modified gravity w.r.t. to $\Lambda$CDM spectra with matched linear growth at the same redshift. We refer to these $\Lambda$CDM spectra as `pseudo' spectra. These simulations have the same initial seed as our standard $\Lambda$CDM simulations. 

To measure the matter power spectrum from the output of our simulations, we use the publicly available \texttt{powmes} code \cite{ref:powmescode,ref:ColombiPowmes}. For notational convenience, we define the ratio $S(k) = P(k)/P_{\rm \Lambda CDM}(k)$ to be the ratio of the matter power spectrum measured from our simulations to the $\Lambda$CDM power spectrum and $R(k) = P(k)/P_{\rm pseudo}(k)$ to be the ratio of the modified gravity matter power spectrum from the simulations to the $\Lambda$CDM power spectrum with the same linear growth factor.   

To test the convergence of our simulations, we run them increasing the particle numbers for an identical box size. We then plot the ratio of the matter power spectra from each simulation to the power spectrum from the one with highest resolution, which we choose to be our reference. Our reference simulation contains $1024^3$ particles in a comoving box of side $250\,h^{-1}\,{\rm Mpc}$. We demonstrate convergence in our $\Lambda$CDM simulations by showing that the power spectrum from the one with the second-highest resolution ($512^3$ particles) agrees with that of highest resolution up to $k=4.8 \, h\,{\rm Mpc}^{-1}$, shown by the vertical green line in fig.~\ref{fig:Convergence} (see appendix \ref{appendix:Convergence} for further details). We use a gravitational softening length equal to 2\% of the inter-particle-distance, and a comoving mesh with the same number of grid points and particles. 

We ensure that $\mu$ is a well-defined, continuous function of redshift by smoothing at the bin edges. In order to quantify the effect of such a smoothing procedure on the power spectrum, we first worked with the \texttt{CLASS} code, before implementing an identical algorithm in \texttt{GADGET-2}. Based on these results, the same smoothing procedure was implemented in \texttt{GADGET-2} (see Appendix \ref{appendix:binning} for details). These results indicate that there is very little difference in the simulation results, regardless of whether $\mu$ is a smooth, continuous function of redshift or not, for our chosen parameters. 

 Note that we have primarily discussed the central branch of the flowchart in fig.~\ref{fig:PipelineFlow} in this section; we now turn to the right hand side branch of the pipeline in section \ref{section:FittingFunction}.  
 
\subsection{Predicting the power spectrum on non-linear scales}\label{section:FittingFunction}
$N$-body simulations are too computationally expensive to be used to forecast and perform data analyses for upcoming surveys. Thus in this section we test the performance of various theoretical formulations that attempt to predict the full non-linear matter power spectrum. We do this by comparing their respective accuracy in the reproduction of the ratio of the modified gravity matter power spectra with respect to the $\Lambda$CDM matter power spectrum as measured from our simulations. This comparison of ratios allows us to remain independent of realisation-dependent effects.

The results in \cite{ref:Cui1} provide hints that knowledge of the linear power spectrum is insufficient to completely calculate the matter power spectrum on non-linear scales. This has important consequences for the fitting functions. In section \ref{section:Results} we show explicitly that in our framework simulations with identical linear matter power spectra can show significant differences in their non-linear power spectra. 

We consider the following formalisms as candidates for predicting the non-linear matter power spectrum:
\begin{itemize}
    \item case 1 - the fitting function obtained in \cite{ref:Cui1};
    \item case 2 - the standard halo model as in eqs.~\eqref{eq:1halo},\eqref{eq:2halo} with a Sheth-Tormen mass function \cite{Sheth1999,Sheth2002};
    \item case 3 - the \texttt{halofit} fitting procedure \cite{ref:halofit1};
    \item case 4 - the halo model reaction formalism \cite{ref:ReactTheory} (\texttt{ReACT}).  
\end{itemize}
We will now briefly describe each case. 

To our knowledge, the only fitting function in the literature based on an arbitrary $\mu$ is that presented in \cite{ref:Cui1}. As noted earlier, this fitting function was estimated assuming a constant $\mu$ through the simulations, which is not the case in our simulation (see appendix \ref{appendix:reaction} for more details). Nevertheless, we investigate the performance of this fitting function in the context of our simulation results.

\begin{figure}
 \centering
 \includegraphics[width = 0.45\textwidth]{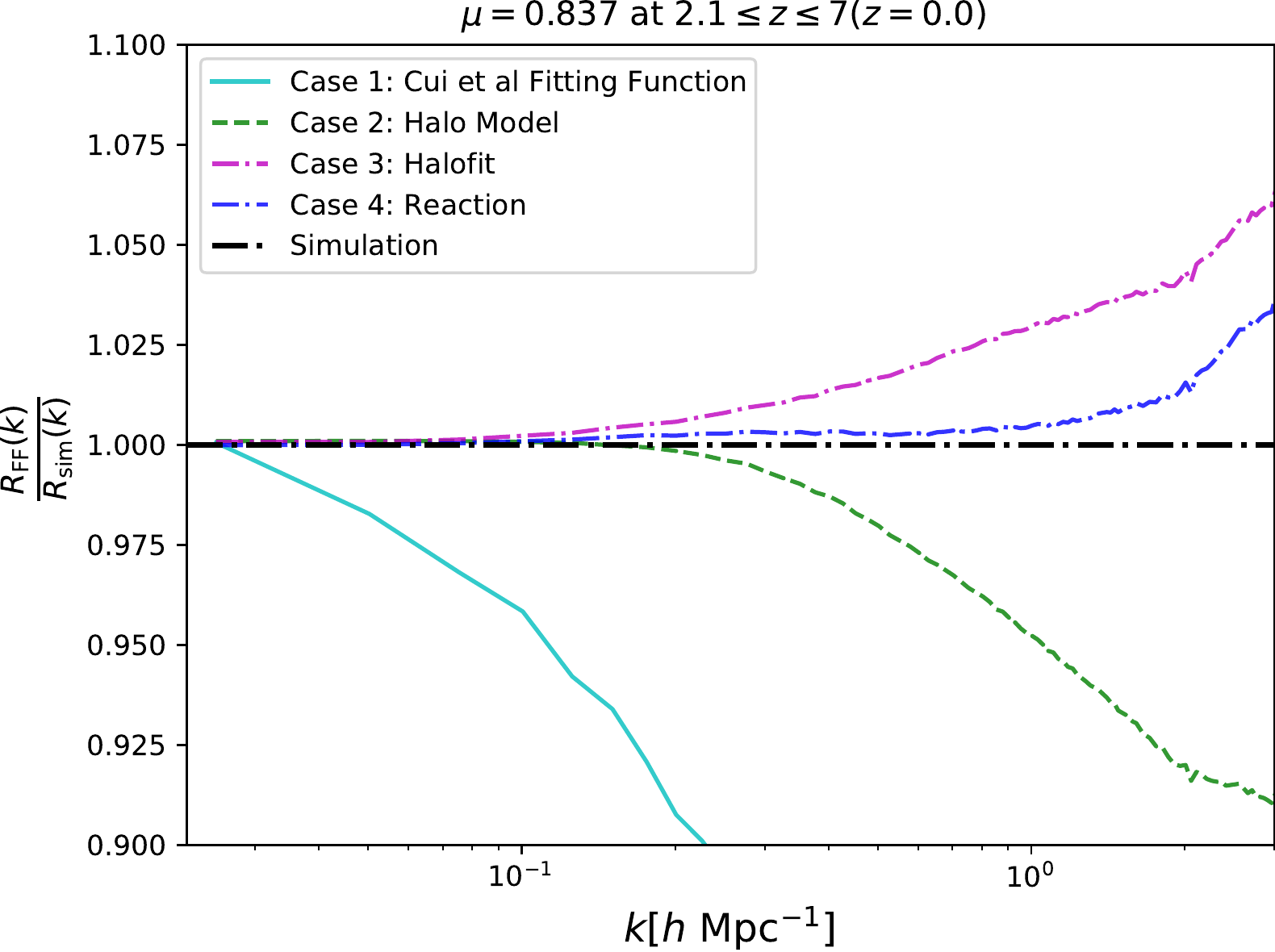}
 \includegraphics[width = 0.45\textwidth]{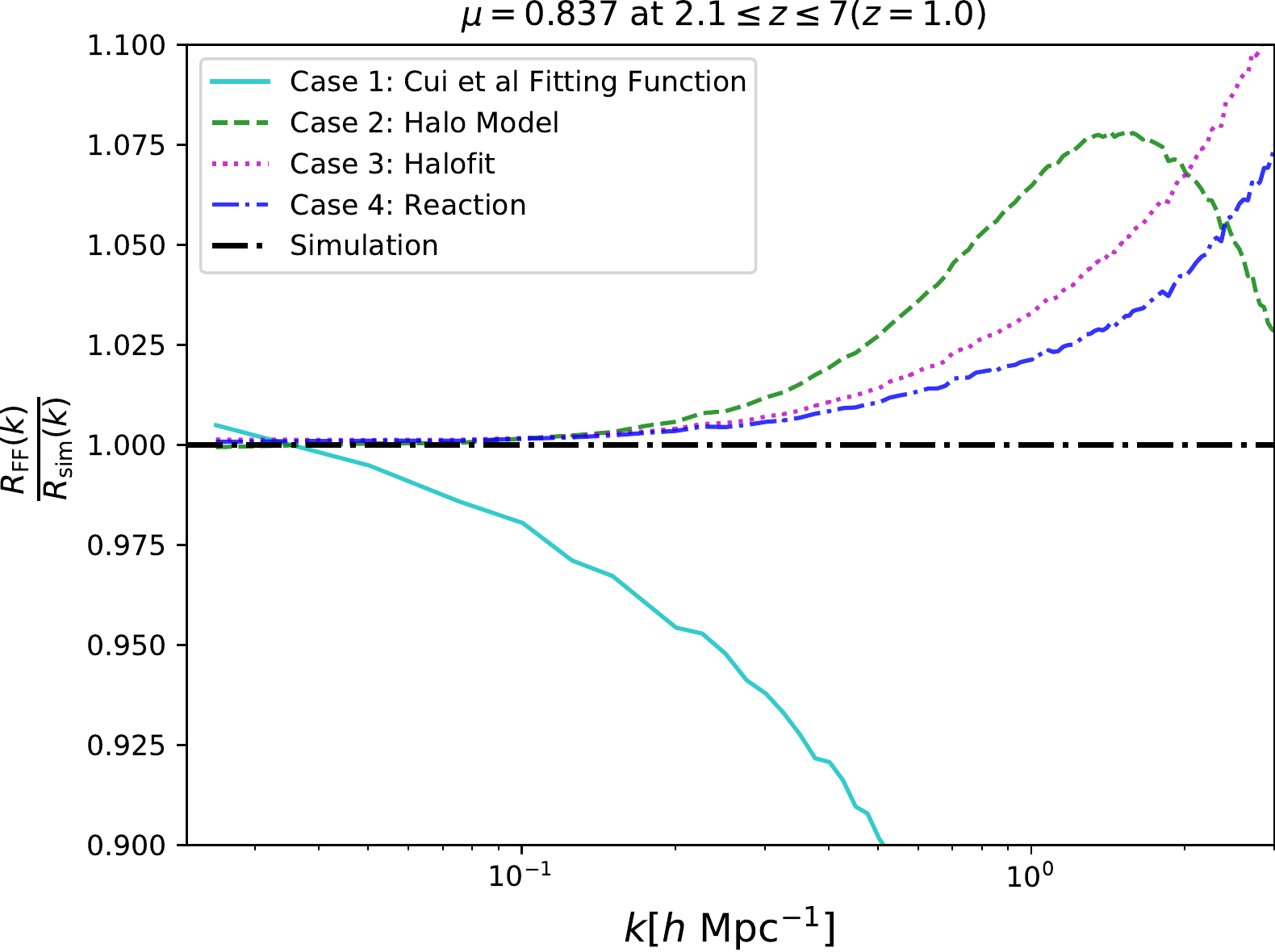}
 \includegraphics[width = 0.45\textwidth]{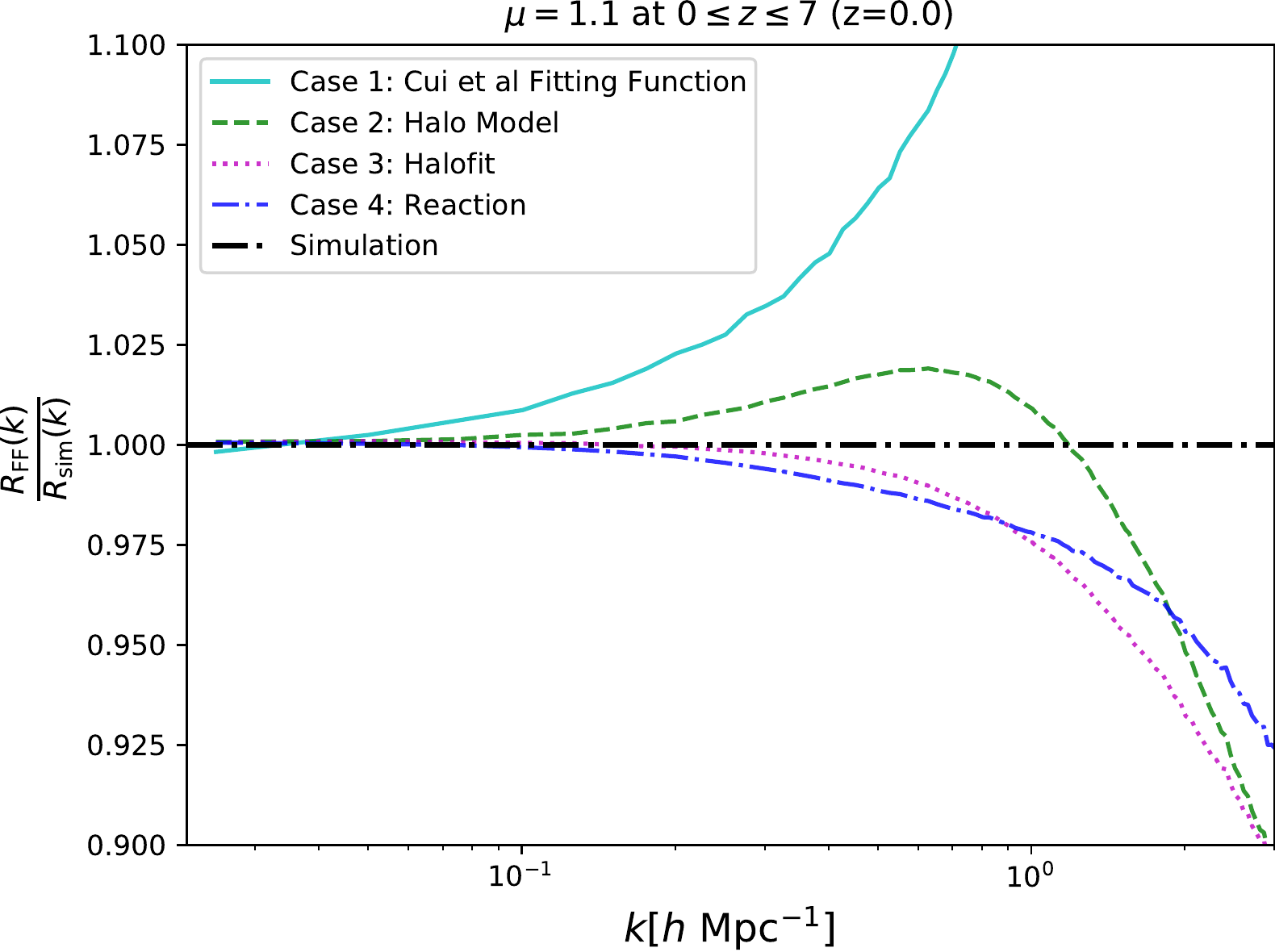}
 \includegraphics[width = 0.45\textwidth]{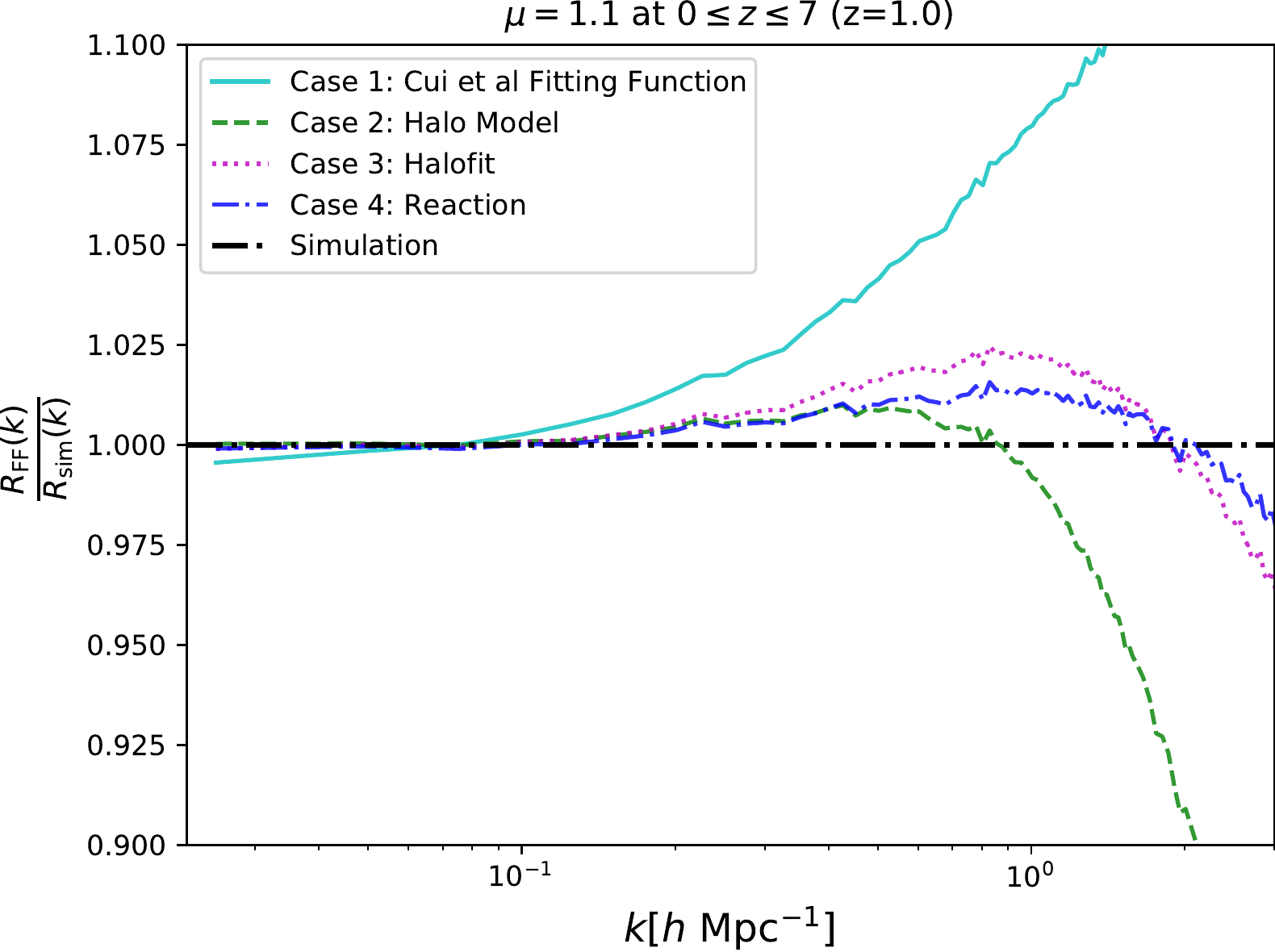}
 \caption{The ratio of the quantity $R(k)$ as predicted by the various fitting functions (as indicated by the superscript `FF') with respect to the same quantity computed from the simulations (as indicated by the superscript `sim'). It is clear that the cyan line from the fitting function of \cite{ref:Cui1} fails for all cases. The green dashed curve is the simple application of the halo model as shown in eqs.~\eqref{eq:1halo} and \eqref{eq:2halo}. It is important to note that the \texttt{halofit} prediction (magenta dotted) is actually identical for all the simulations, since they all have identical linear power spectra. In all the panels, we can clearly see the \texttt{ReACT} curve (blue dot-dashed) is consistently within 5\% of the black dot-dashed line which indicates our reference, the simulations.}
 \label{fig:react1}
\end{figure}

\begin{figure}
 \centering
 \includegraphics[width = 0.45\textwidth]{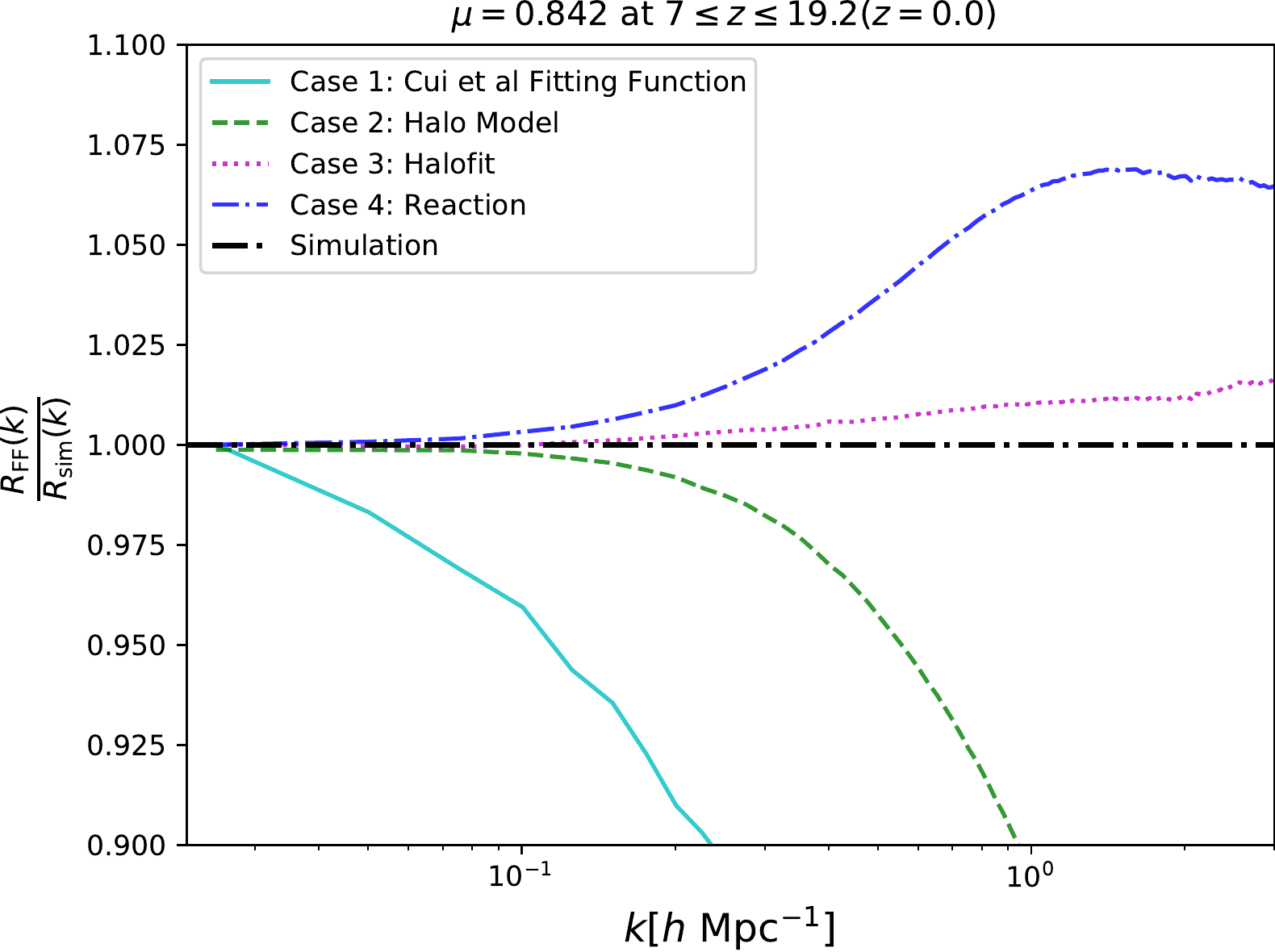}
 \includegraphics[width = 0.45\textwidth]{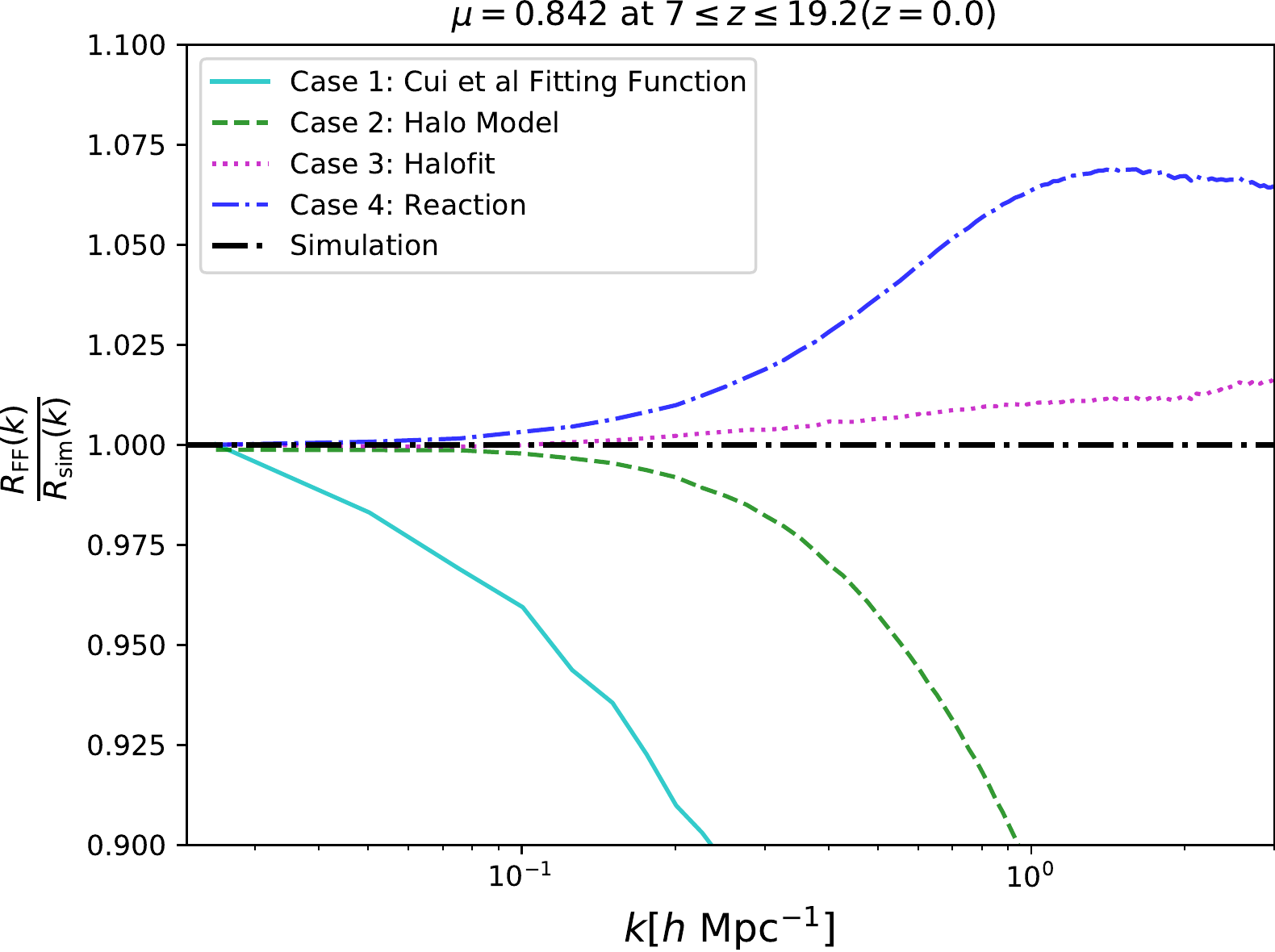}
 \includegraphics[width = 0.45\textwidth]{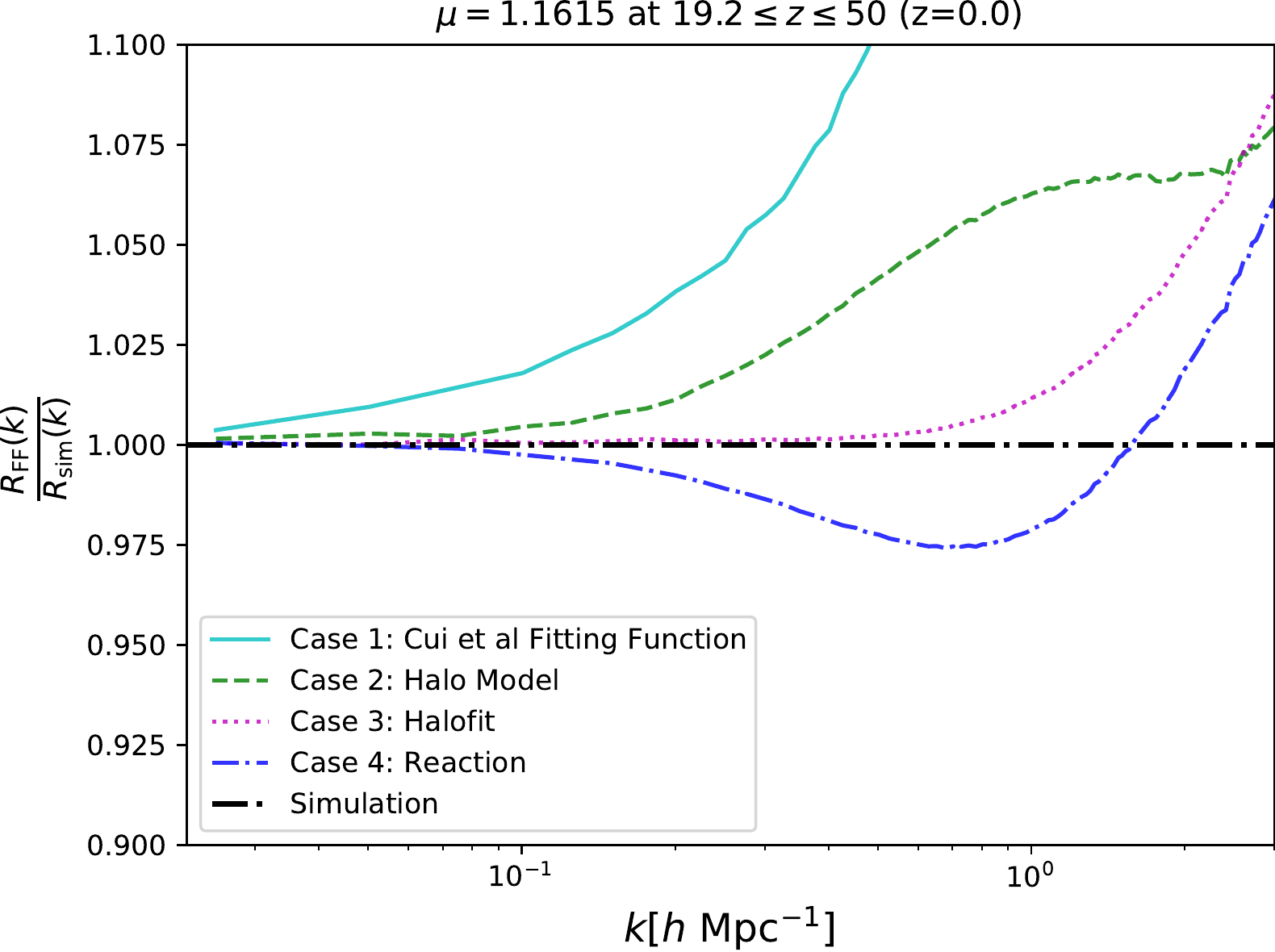}
 \includegraphics[width = 0.45\textwidth]{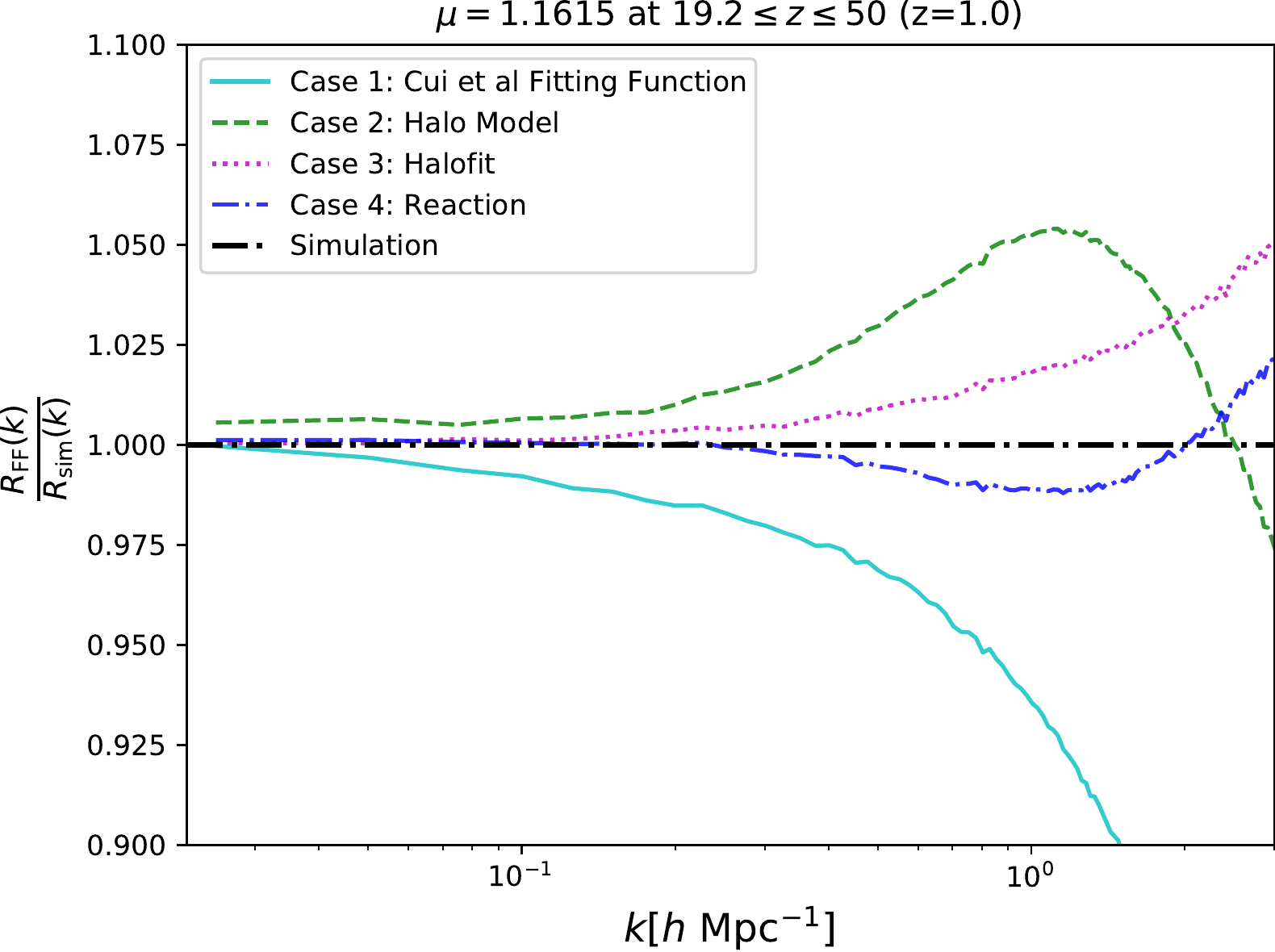}
 \caption{As in fig.~\ref{fig:react1}, but where $\mu$ has a different value from GR in one of the earlier redshift bins.}
 \label{fig:react2}
\end{figure}

As suggested in \cite{ref:Cui1}, the halo model provides a potential explanation of the results obtained in the simulations. We implement the standard halo model to predict the matter power spectrum [see appendix \ref{appendix:reaction}]. We also consider the \texttt{halofit} fitting procedure \cite{ref:halofit1, ref:halofit2, ref:halofit3} that is commonly used for a $\Lambda$CDM Universe.

However, it is well known that the basic implementation of the halo model (the simple addition of the 2-halo and 1-halo terms, or equivalently, the addition of the 1-halo term to the linear matter power spectrum) is inaccurate on quasi-linear scales due to halo correlation effects that require more detailed modelling \cite{ref:CooraySheth}. To circumvent the problems associated with the halo model on quasi-linear scales, the authors in \cite{ref:ReactTheory, ref:reactionCataneo} modified the halo model. This can be seen in the following equation 
\begin{equation}\label{eq:Reaction}
 P_{\rm mg}(z, k) = \mathcal{R}(z, k)P_{\rm pseudo}(z, k)\,, \\
\end{equation}
where $\mathcal{R}(z,k)$ is the so-called halo model reaction and $P_{\rm pseudo}$ is the $\Lambda$CDM `pseudo' power spectrum with identical linear growth to $P_{\rm mg}$. The reason for the use of the pseudo spectrum is that since it has identical linear clustering with respect to the modified gravity spectrum, the inaccuracy in the transition from linear to non-linear scales is slightly alleviated (see appendix \ref{appendix:reaction} for a more detailed explanation of the various fitting functions). 

\begin{figure}
 \centering
 \includegraphics[width = 0.45\textwidth]{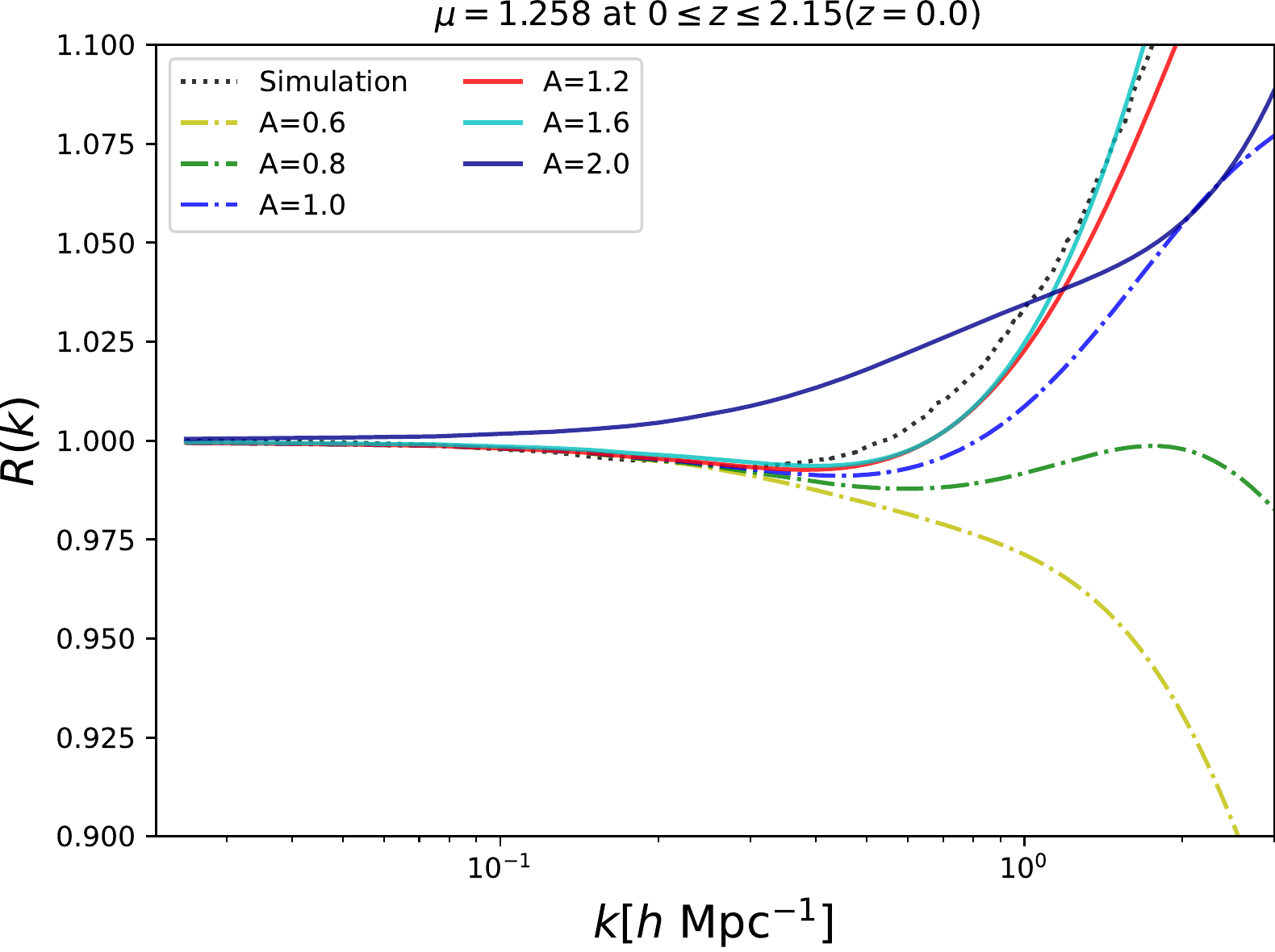}
 \includegraphics[width = 0.45\textwidth]{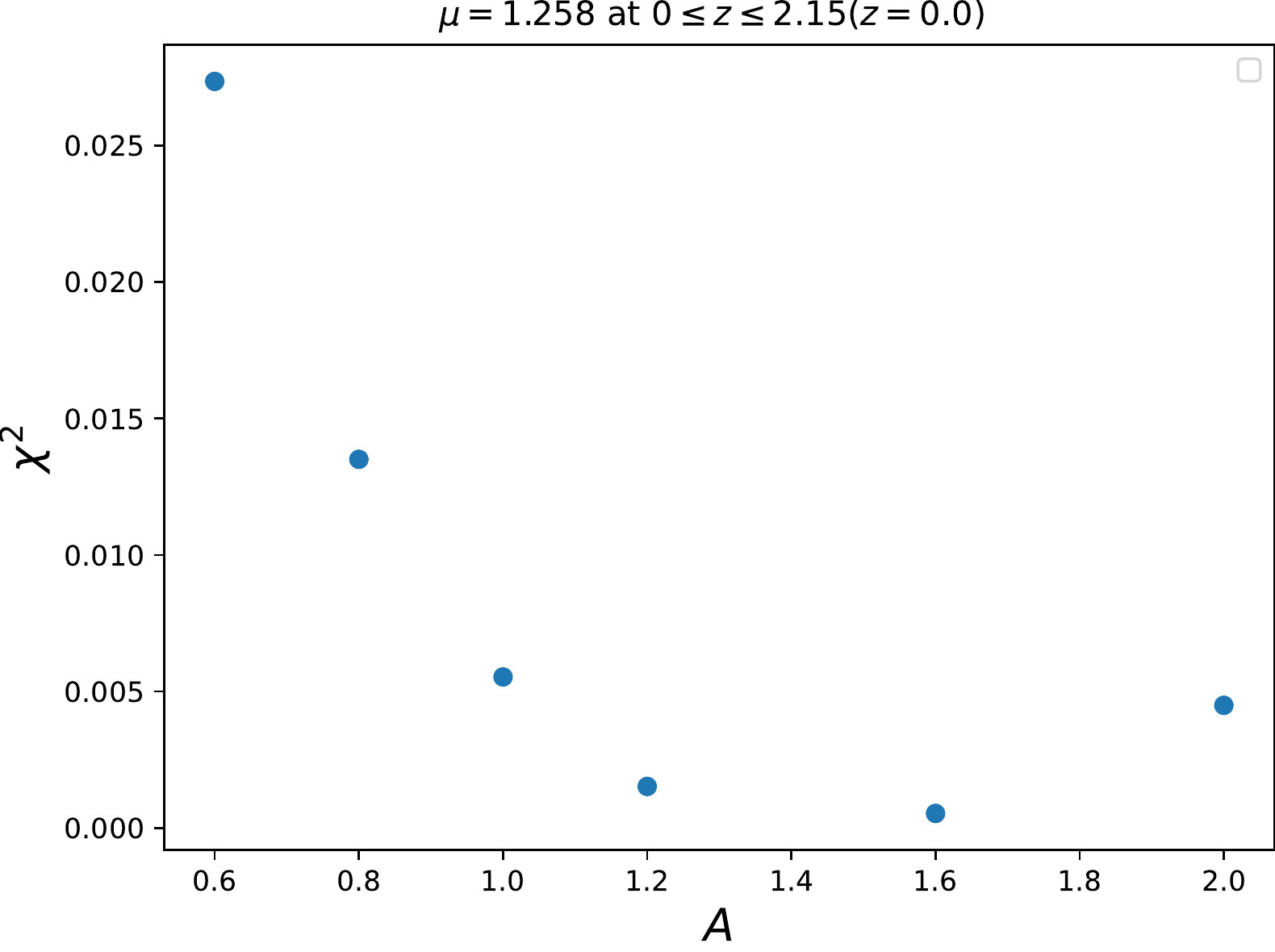}
 \caption{We display the ratio of the power spectra with respect to one of our simulations for several values of the parameter $A$ in the left panel. The solid cyan line (which corresponds to $A=1.6$) is the best fit, with the smallest $\chi^2$ parameter, as can be seen by the plot on the right panel. While this behaviour was hinted at in various works in the literature \cite{ref:Cui1}, to our knowledge this is the first time that such an analysis has been carried out for the general model-independent case time-varying $\mu$. We leave the analysis of varying $A$ as a function of $\mu$ and the resulting rigorous validation of this modification to future work.  } 
 \label{fig:concentrationMod}
\end{figure}

The authors also invoke 1-loop perturbation theory corrections to quasi-linear scales for the halo model (see Appendix \ref{appendix:reaction} for more details). We note that the authors in \cite{ref:ReactTheory} modify the $\Lambda$CDM concentration-mass relation for the dark energy models under consideration (see Appendix \ref{appendix:reaction} for more details and further discussion of the theoretical formulation of \texttt{ReACT}. See also  fig.~\ref{fig:concentrationMod} where we consider a further modification of the \texttt{ReACT} formalism).

In figs. \ref{fig:react1} and \ref{fig:react2}, we show the ratio of the quantity $R(k)$ (i.e., the ratio of the modified gravity matter power spectrum to the pseudo $\Lambda$CDM matter power spectrum) as predicted by the fitting functions with respect to the same quantity computed from the simulations. We have chosen a subset of our simulations that is representative of the performance of the different fitting functions we consider. Clearly, we see that the fitting function in \cite{ref:Cui1} fails to predict the correct non-linear behaviour when $\mu$ is not a constant value throughout the simulation. As expected, the \texttt{halofit} prediction is identical for all the simulations, since they all have identical linear growths. In other words, the simulations where the \texttt{halofit} performs better than \texttt{ReACT} are the cases where modified gravity parameters coincidentally result in a matter power spectrum that is already very close to that of $\Lambda$CDM. We show that on average, this accidental success is not typically replicated in the general case. We see that while the \texttt{ReACT} formalism doesn't always predict the non-linear trend perfectly, it appears to be consistently within 5\% of the simulations. Since we are only interested in understanding the modified gravity phenomenology, we do not vary $\Lambda$CDM parameters in this work. We leave the full validation of \texttt{ReACT} across $\Lambda$CDM and modified gravity parameter space to future work.      

\begin{figure}
 \centering
 \includegraphics[width = 0.45\textwidth]{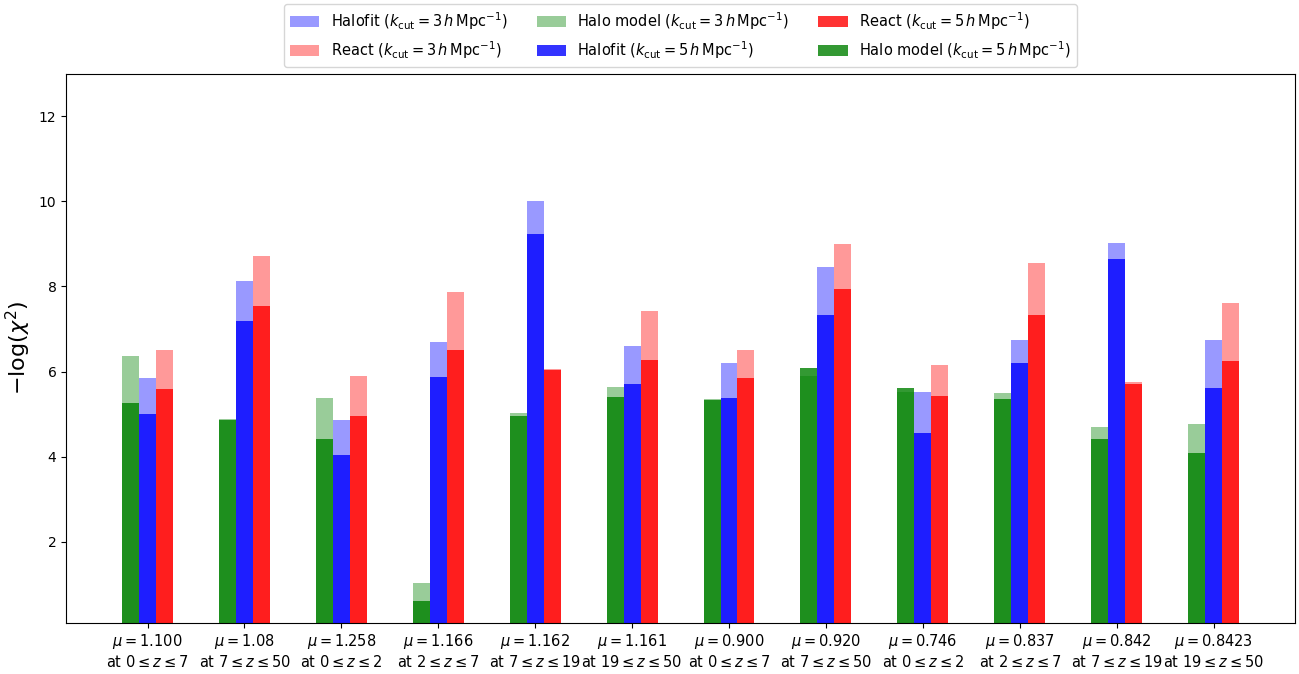}
 \includegraphics[width = 0.45\textwidth]{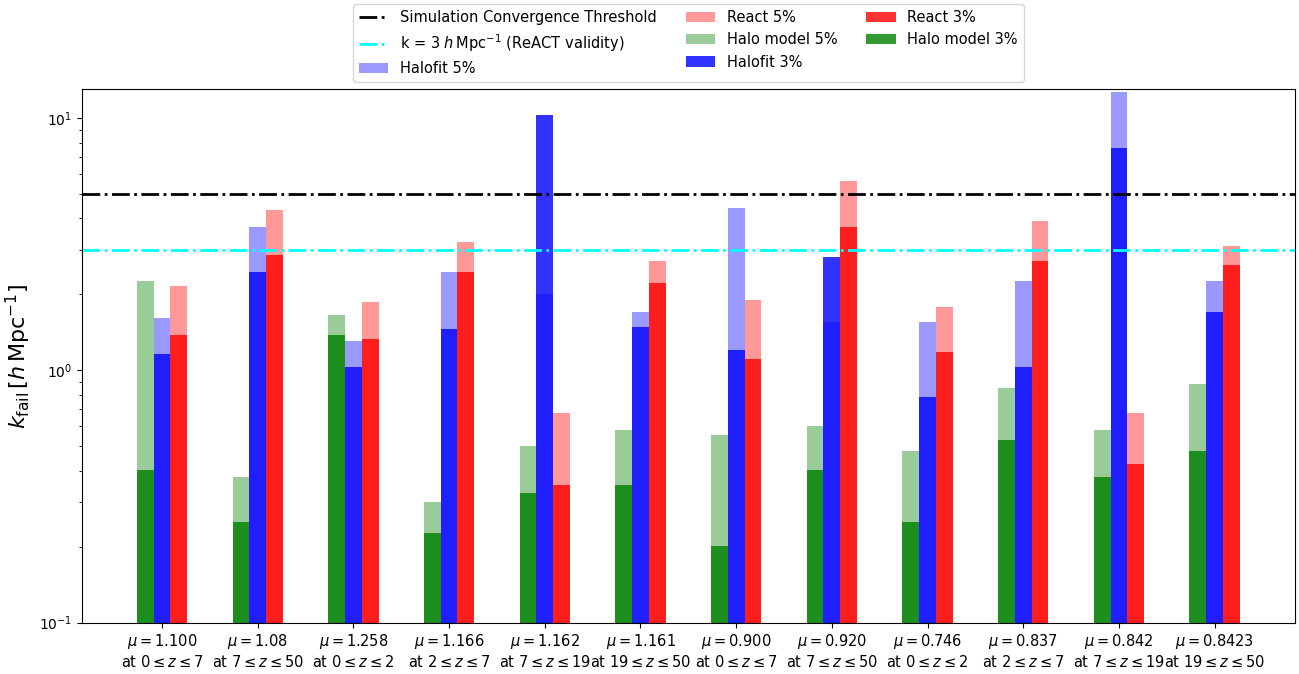}
 \caption{In the left panel, we show the values of the least square statistic for the different fitting functions as defined in eq.~(\ref{eq:chi2}). We obtained these values by employing $k_{\rm cut}=3\,h\,{\rm Mpc}^{-1}$ and $k_{\rm cut}=5\,h\,{\rm Mpc}^{-1}$, which represent the wave number up to which the \texttt{ReACT} formalism is designed to operate and the wave number up to which we have converged results from our simulations, respectively. Clearly, the red bar is larger than the other bars in the majority of the simulations which implies that the \texttt{ReACT} formalism is typically the best fit over the range of scales (note that since we plot the negative logarithm, a larger bar is a better fit). In the right panel, we plot the  wavenumber at which the quantity $R(k) - 1$ is at 3\% and 5\%, respectively. The results are consistent with the left panel, with the red bar being the largest for the majority of the simulations. This shows that the \texttt{ReACT} formalism allows one to probe deeper into the non-linear regime.}
 \label{fig:chisquaredbars}
\end{figure}

We use the following least-square statistic to quantify the agreement between the various fitting functions with our simulations
\begin{eqnarray}
 \chi^2 & = & \frac{1}{N}\sum_{k=0.02\, h\, {\rm Mpc}^{-1}}^{k=k_{\rm cut}} \left[R_{\rm sim}(k) - R_{\rm FF}(k) \right]^2 \,, \label{eq:chi2}
\end{eqnarray}
where $N = 160$, with the sampling in Fourier space being approximately logarithmic, as in \texttt{powmes} and $k_{\rm cut}$ is a wavenumber cut-off that we choose. The subscript `sim' implies the ratio as measured in our simulations and the subscript `FF' implies the ratio as predicted by the various fitting functions. We employ a cut-off at $k_{\rm cut}=3 \,h\,{\rm Mpc}^{-1}$, which is the theoretical threshold up to which the \texttt{ReACT} formalism is supposed to be accurate \cite{ref:ReactTheory}. Note that this scale is within the region of validity of our simulations [see appendix \ref{appendix:Convergence}], which is $k_{\rm cut} =  5\,h\,{\rm Mpc}^{-1}$. We show the $\chi^2$ statistic with both wavenumber cut-offs in fig.~\ref{fig:chisquaredbars}. This indicates that the \texttt{ReACT} formalism provides the best fit to our simulations, as evidenced by the fact the red line is largest in the majority of the cases presented. Note that since we plot the negative logarithm, a larger bar corresponds to a better fit. We also note that due to the failure of the Cui et~al. \cite{ref:Cui1} (see eq.~\eqref{eq:CuiFF} in appendix \ref{appendix:reaction}) even at small $k$ to capture the results of our simulations, we don't include their fitting function in this figure. 

We also show the wavenumber $k_{\rm fail}$ at which the quantity $R_{\rm FF}(k)/R_{\rm sim}(k)$ deviates from unity at the level of 3 and 5 \%, respectively in the right panel of fig.~\ref{fig:chisquaredbars}. Note that this probes the validity in $k$-space of each fitting function, while the $\chi^2$ statistic is an indicator of the accuracy of each fitting function within this region of validity. The combination of $k_{\rm fail}$ and $\chi^2$ throws light on the applicability of a fitting function to future forecasts.

In \cite{ref:Cui1}, the authors claim that in order to accurately reproduce the power spectra in simulations using the halo model, one needs to take into account that any modification to $\mu$ results in a change in the rate of structure formation, i.e., a change in the concentration parameter associated to haloes. Increasing $\mu$ causes haloes to form earlier, leading to lower concentrations in comparison to $\Lambda$CDM and vice-versa for decreasing $\mu$. 
It is also worth noting that in order to accurately reproduce power spectra in $f(R)$ and DGP models, the authors in \cite{ref:ReactTheory} modify the $\Lambda$CDM concentration-mass relationship. Therefore, we would expect that in order to achieve precision in forecasting phenomenological modified gravity, one would need to depart from the concentration-mass relationship observed in $\Lambda$CDM.   

In anticipation of this, we carry out a simple modification of the \texttt{ReACT} formalism in which we rescale the $\Lambda$CDM concentration parameters for all halo masses. This re-scaling is given by $c_{\rm vir}^{\rm MG} = c_{\rm vir}^{\Lambda \rm CDM}/A$, where $A$ is a function that would conceivably depend on $\mu$, the redshift at which it is modified and the duration in redshift over which it is modified. In fig.~\ref{fig:concentrationMod}, we demonstrate that for different values of $A$, the non-linear tail of the power spectrum is significantly affected. Indeed, we find a better fit for $A=1.6$, as can be seen in the right panel of fig.~\ref{fig:concentrationMod}. We note that this is only a preliminary examination of this effect and requires rigorous validation in future work, nevertheless we find similar qualitative behaviour to other studies.

\section{Simulation phenomenology}\label{section:Results}

\begin{figure}
  \centering
  \includegraphics[width = 0.45\textwidth]{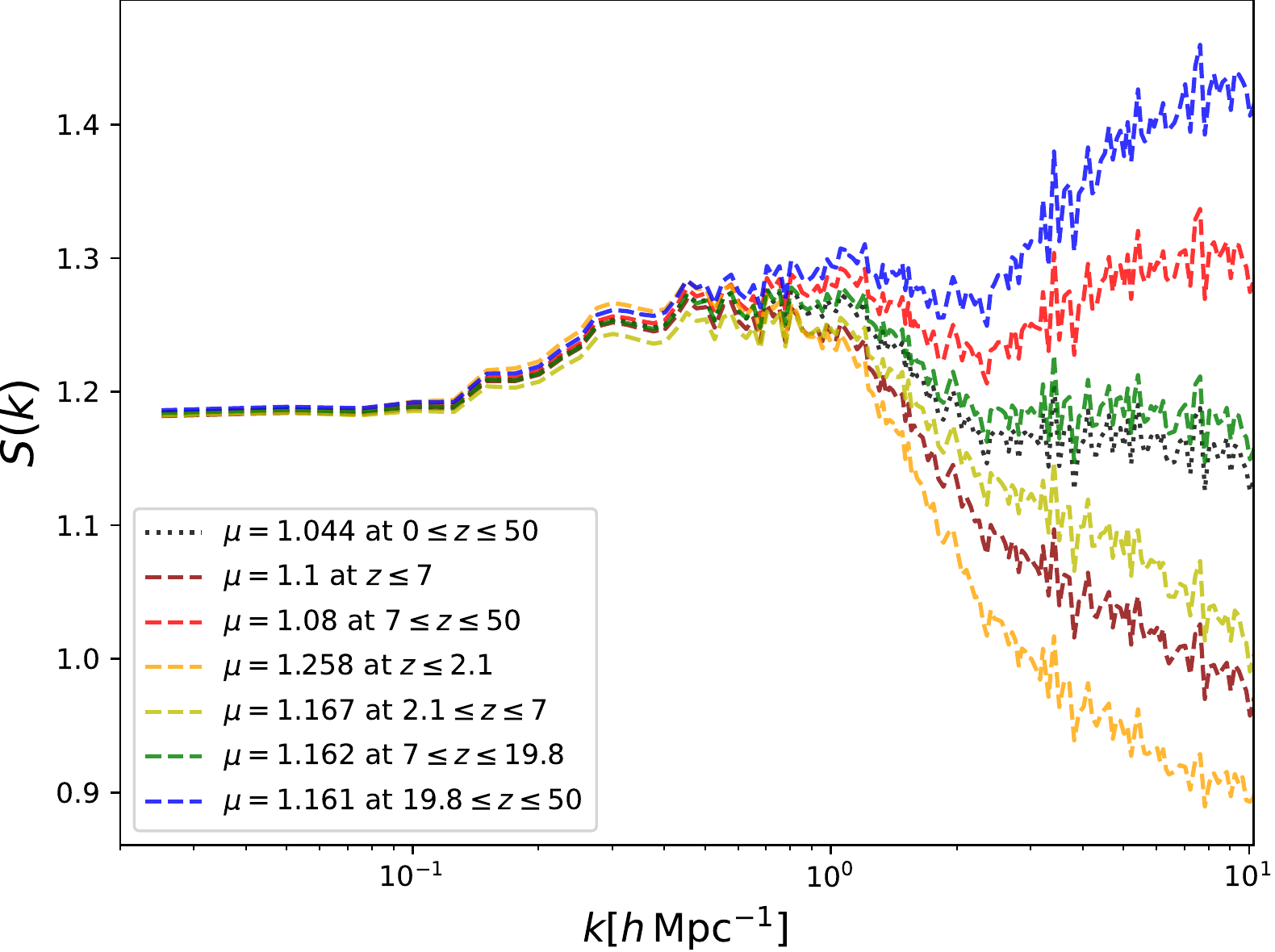}
  \includegraphics[width = 0.45\textwidth]{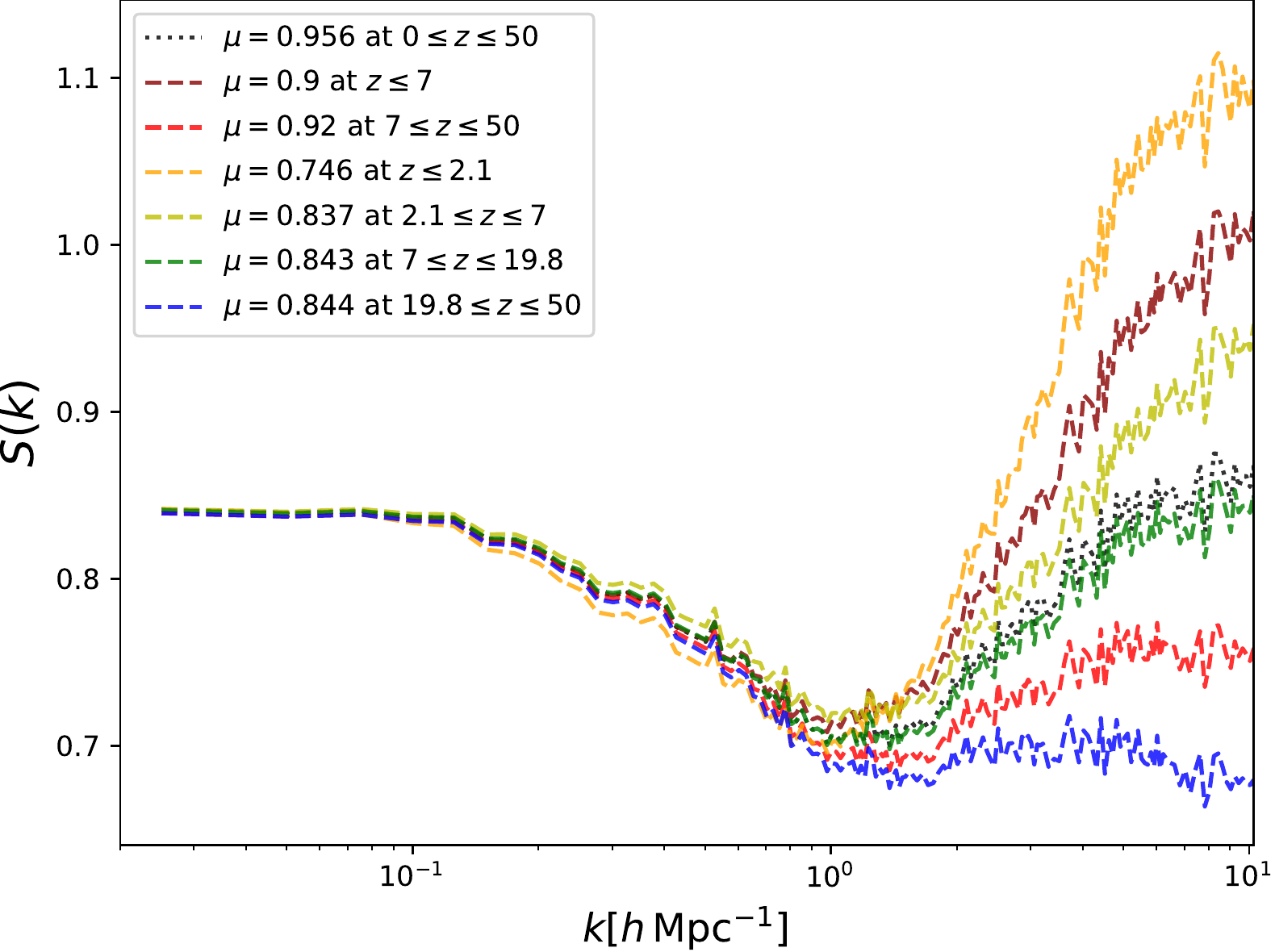}
  \includegraphics[width=0.45\textwidth]{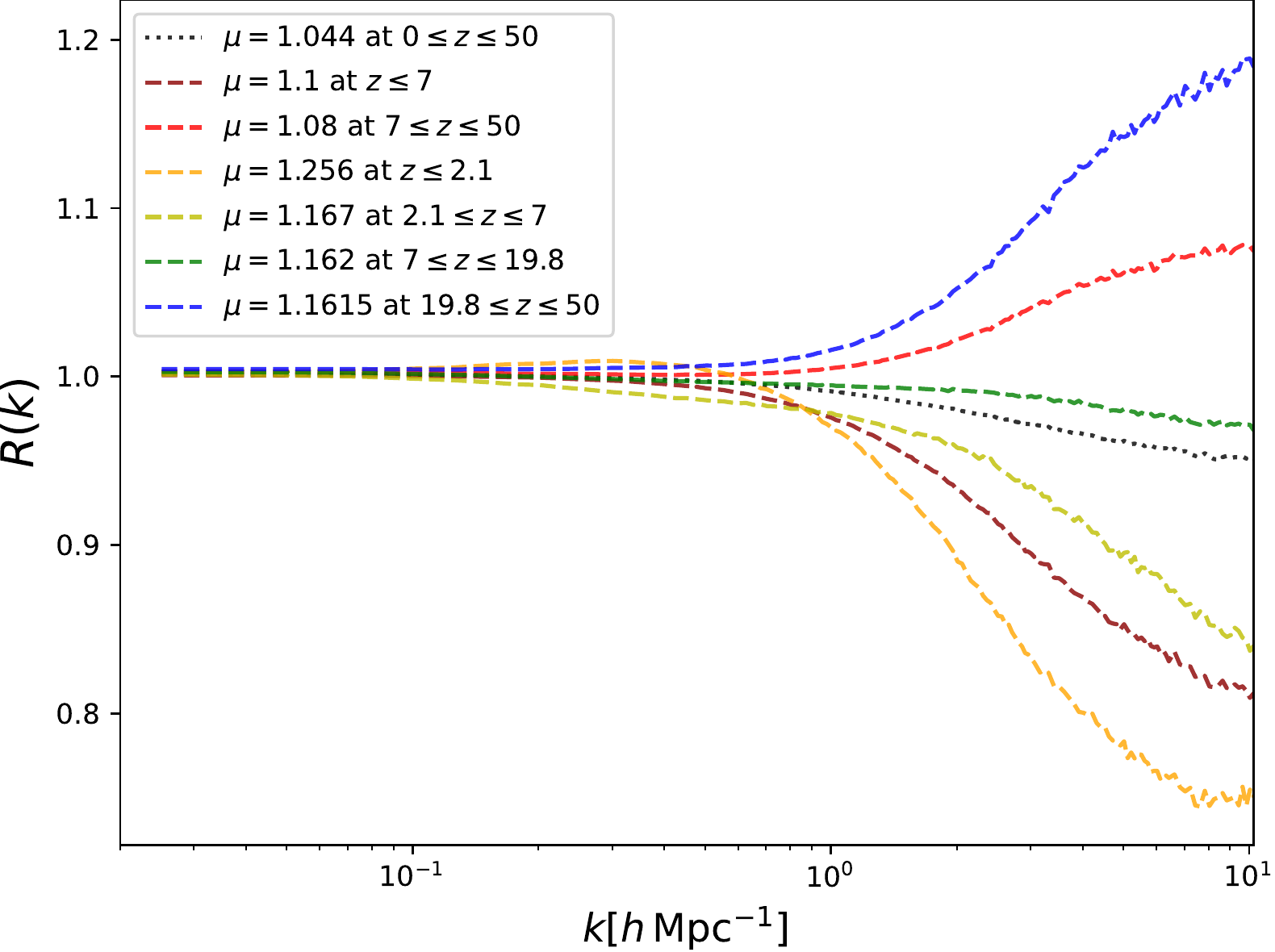}
  \includegraphics[width=0.45\textwidth]{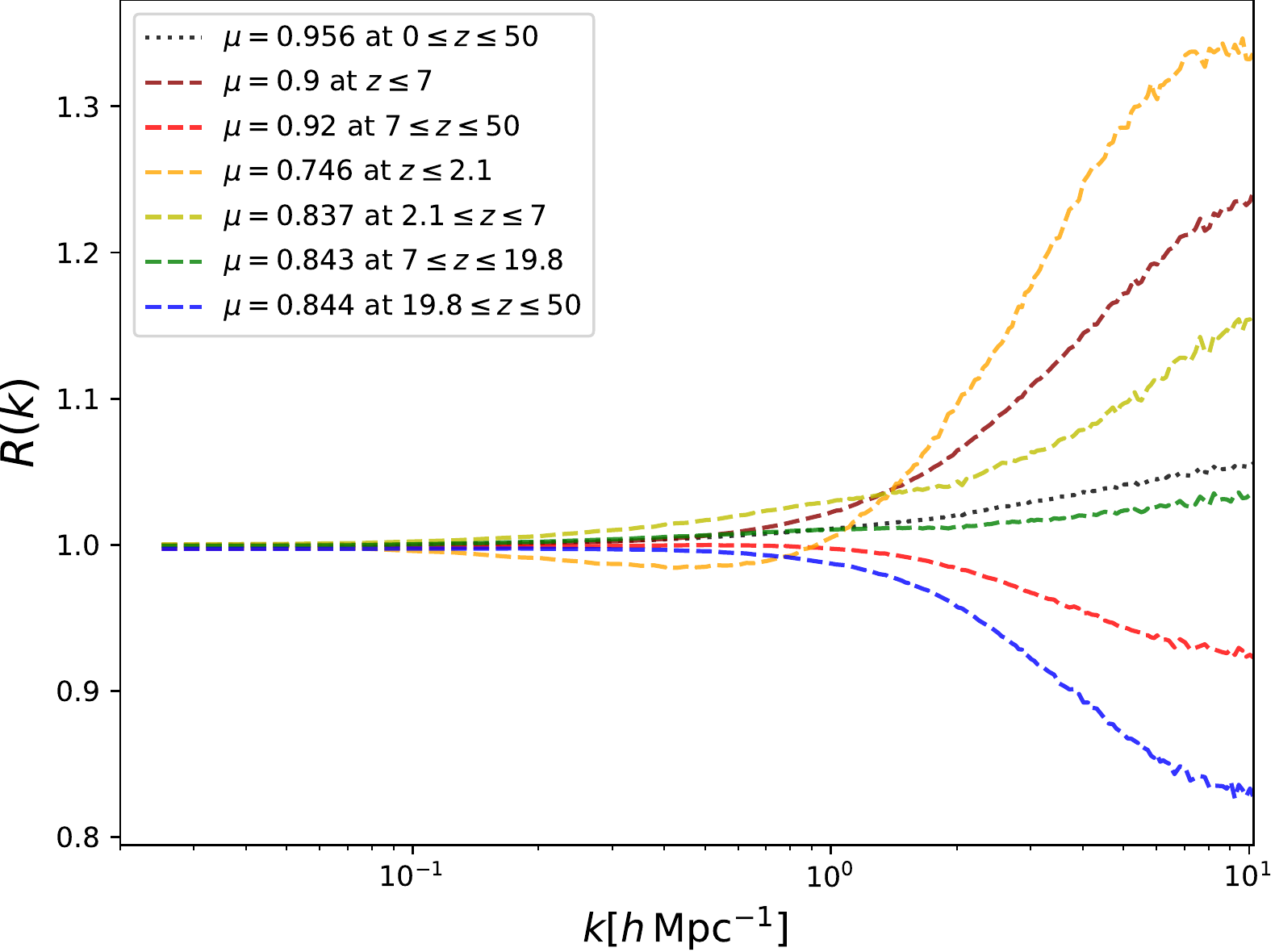}
  \includegraphics[width = 0.45\textwidth]{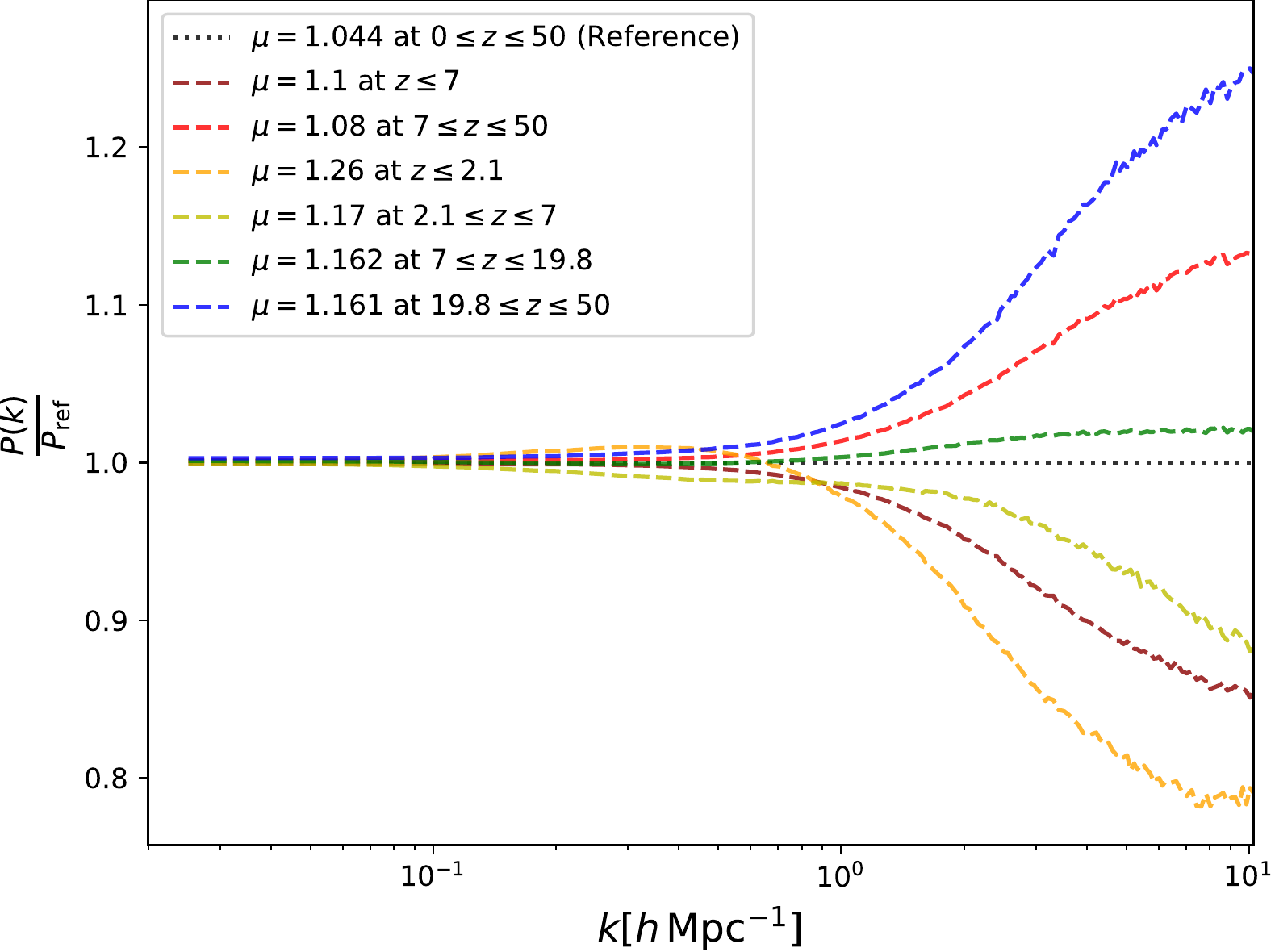}
  \includegraphics[width = 0.45\textwidth]{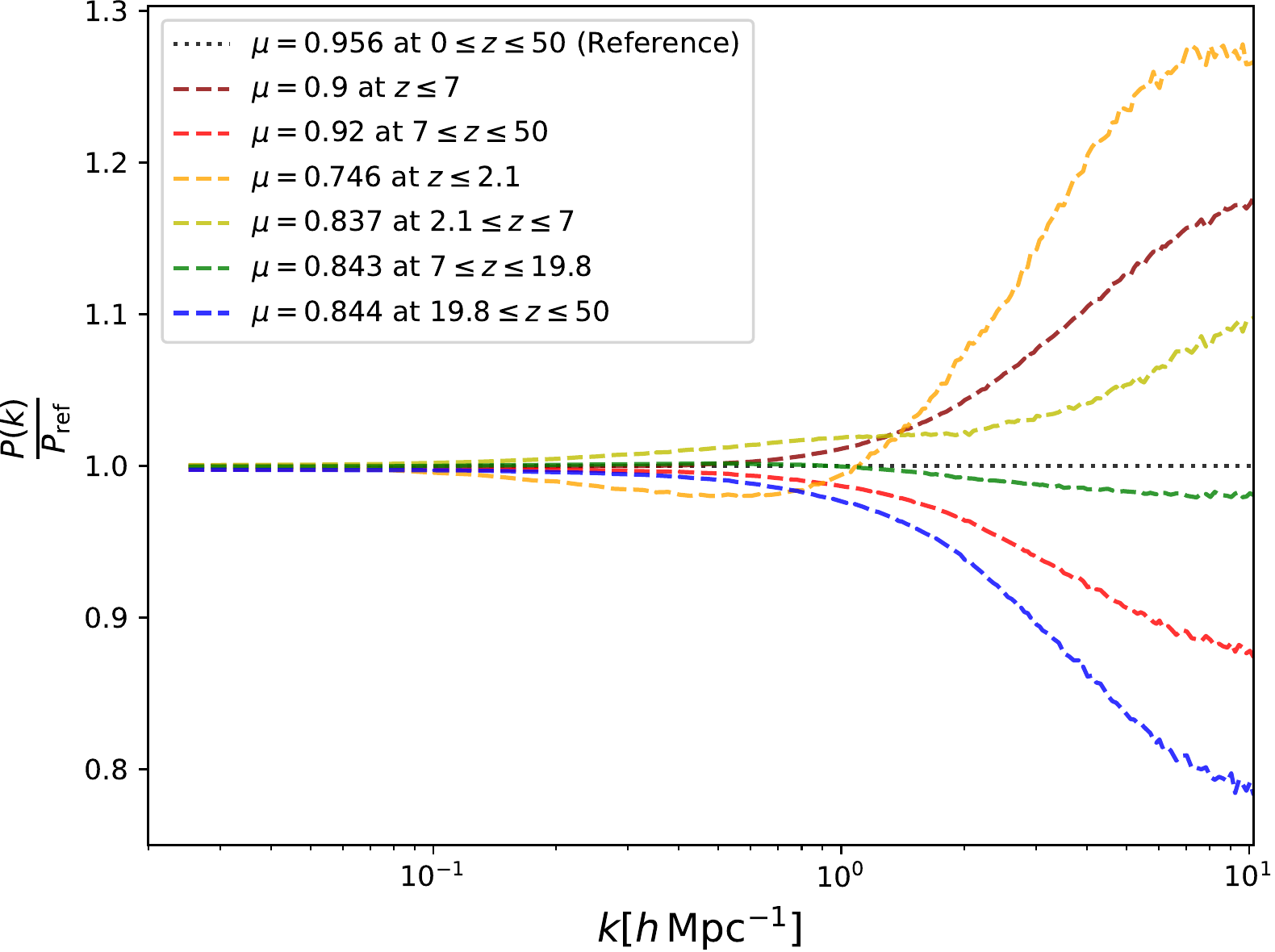}
  \caption{\textit{Top row}: The ratio of the power spectra from all the simulations to the $\Lambda$CDM power spectrum at $z=0$. Note that we see a quasi-linear bump in these due to the mismatch in linear growth, which is absent in the other rows. We see the multi-dimensionality of the parameter space in display here since we have 6 different unique non-linear power spectra for the same linear power spectrum as the case where $\mu$ is constant throughout the simulation as studied in \cite{ref:Cui1}. \textit{Middle row}: The quantity $R(k)$, the ratio of the matter power spectra from the different simulations to the pseudo $\Lambda$CDM power spectrum with the same linear growth. \textit{Lower row}: Ratio of power spectra at $z=0$ from all the simulations to the power spectrum at $z=0$ from the reference simulation with a constant $\mu$ throughout, with redshift bins according to table \ref{tab:binning}.}
  \label{fig:Sims}
\end{figure}

In this section, we discuss the phenomenology from our full suite of simulations with the fourteen different forms of $\mu(z)$. Our goal is to understand the variation in the non-linear matter power spectrum as a function of $\mu(z)$. In the top row of fig.~\ref{fig:Sims}, we present the ratio $S(k)$ (at redshift zero) of the power spectra from all the simulations to the $\Lambda$CDM spectrum with the same initial conditions. We note that while the ratio relative to the pseudo-spectrum $R(k)$ is the most appropriate ratio for the fitting functions in the previous section, here we focus on $S(k)$ because it throws light on the different structure formation histories of the simulations with respect to $\Lambda$CDM. By design, the lines in these plots are all equal on linear scales (for small $k$). We observe two distinct features in $S(k)$, present on quasi-linear scales and non-linear (large $k$) scales respectively. Firstly, we consistently see a rise (or dip, depending on whether $\mu>1$ or $\mu<1$) in power across all simulations on quasi-linear scales at the level of $\sim 20\%$ (at $0.1\, h\,{\rm Mpc}^{-1} \leq k \leq 1\, h\,{\rm Mpc}^{-1}$). This is followed by a ``split'' behaviour on small (non-linear) scales, where the power depends on the range of redshifts over which the modified gravity effects were switched on, and the simulations with $\mu \neq 1$ at large redshifts exhibit the opposite behaviour to the case where it is switched on at small redshifts. This is due to varying halo formation times and inter-halo clustering, as we will discuss later in this section. 

The quasi-linear behaviour may be attributed to the difference in growth histories between $\Lambda$CDM and the different simulations, since the effect of a different growth rate is more pronounced on quasi-linear scales compared to linear scales. We illustrate this using the middle panels of fig.~\ref{fig:Sims}, where we compute the ratio of the matter power spectrum in our simulations to the pseudo power spectrum. This quasi-linear feature is significantly diminished, since the pseudo power spectrum has the same linear growth at $z=0$, so by construction the increased effects of the changed growth are mostly removed. This is why the pseudo power spectrum is a good comparison point when examining fitting functions. The same quasi-linear behaviour is seen in the bottom panel, where we compute the ratio of the matter power spectra relative to a reference simulation where $\mu$ is held constant in redshift. 
  
We now turn to the non-linear split behaviour that we see in all of the rows. On small scales the lines diverge, depending on the redshift range during which $\mu\neq1$ in each simulation. In particular, the simulations where $\mu\neq1$ at earlier times, continues the trend expected from linear scales, that $\mu>1$ increases the power and vice versa for $\mu<1$. However, varying $\mu$ at low redshifts introduces the opposite effect to linear theory. For example, $\mu>1$ at late times actually leads to a lack of power on non-linear scales. This results in ``crossing points'' with respect to $\Lambda$CDM, i.e. scales below which the power spectrum is actually reduced, despite equal initial conditions and $\mu > 1$. This implies that the state of clustering when $\mu$ is modified, in terms of how much structure has already formed and on what scales, is important for understanding the final clustering spectrum on small scales at redshift zero. Since we obtain intrinsically different shapes for the power spectra for different values of $\mu$ with identical linear growths, the discussion on concentration factors in \cite{ref:Cui1} is insufficient to explain the physics of these simulations. These plots clearly show that any fitting procedures that predict the non-linear matter power spectrum purely from the linear spectrum will fail to capture non-linear effects accurately. 

\begin{figure}
 \centering
 \includegraphics[width = 0.4\textwidth]{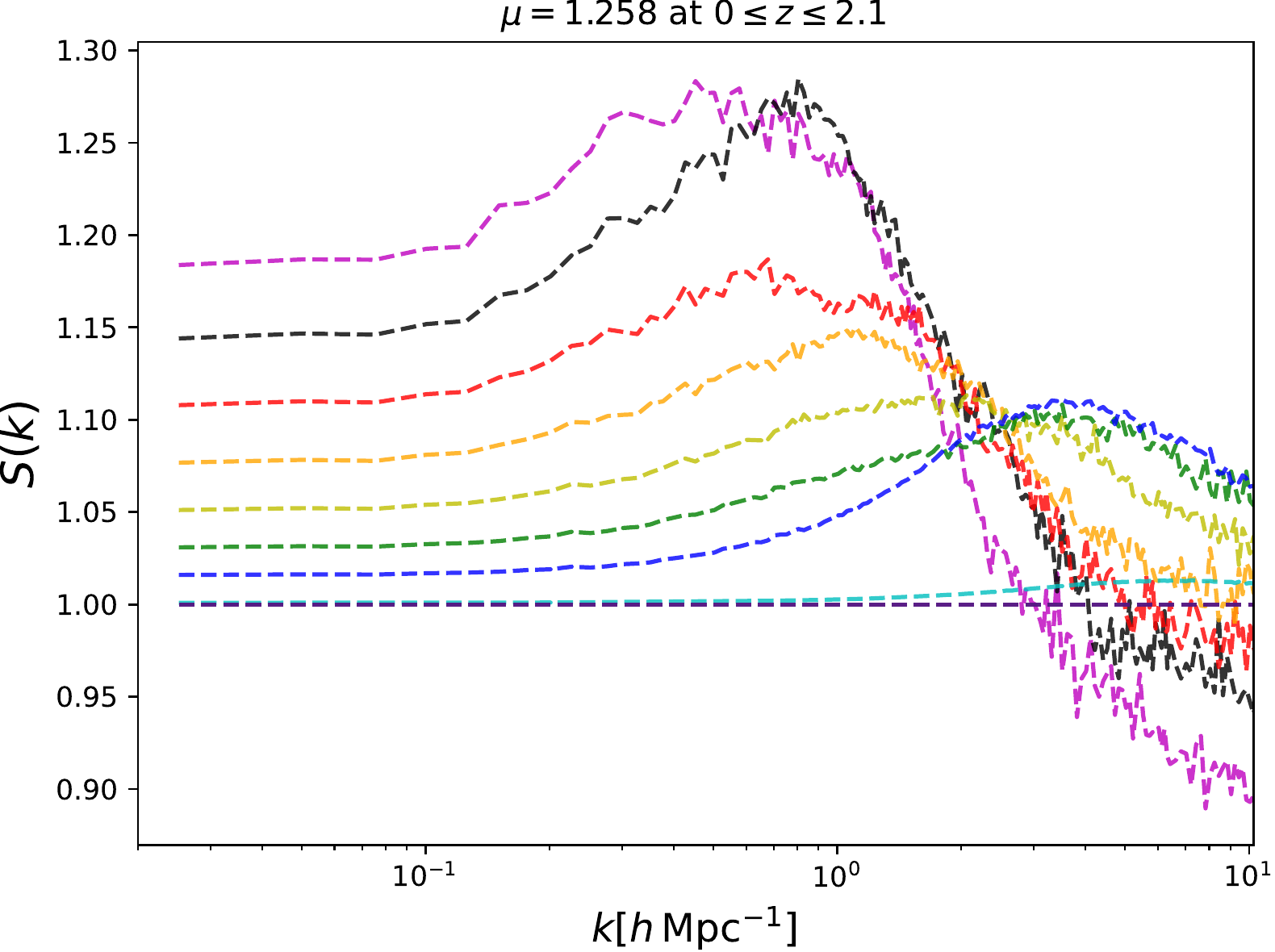}
 \includegraphics[width = 0.4\textwidth]{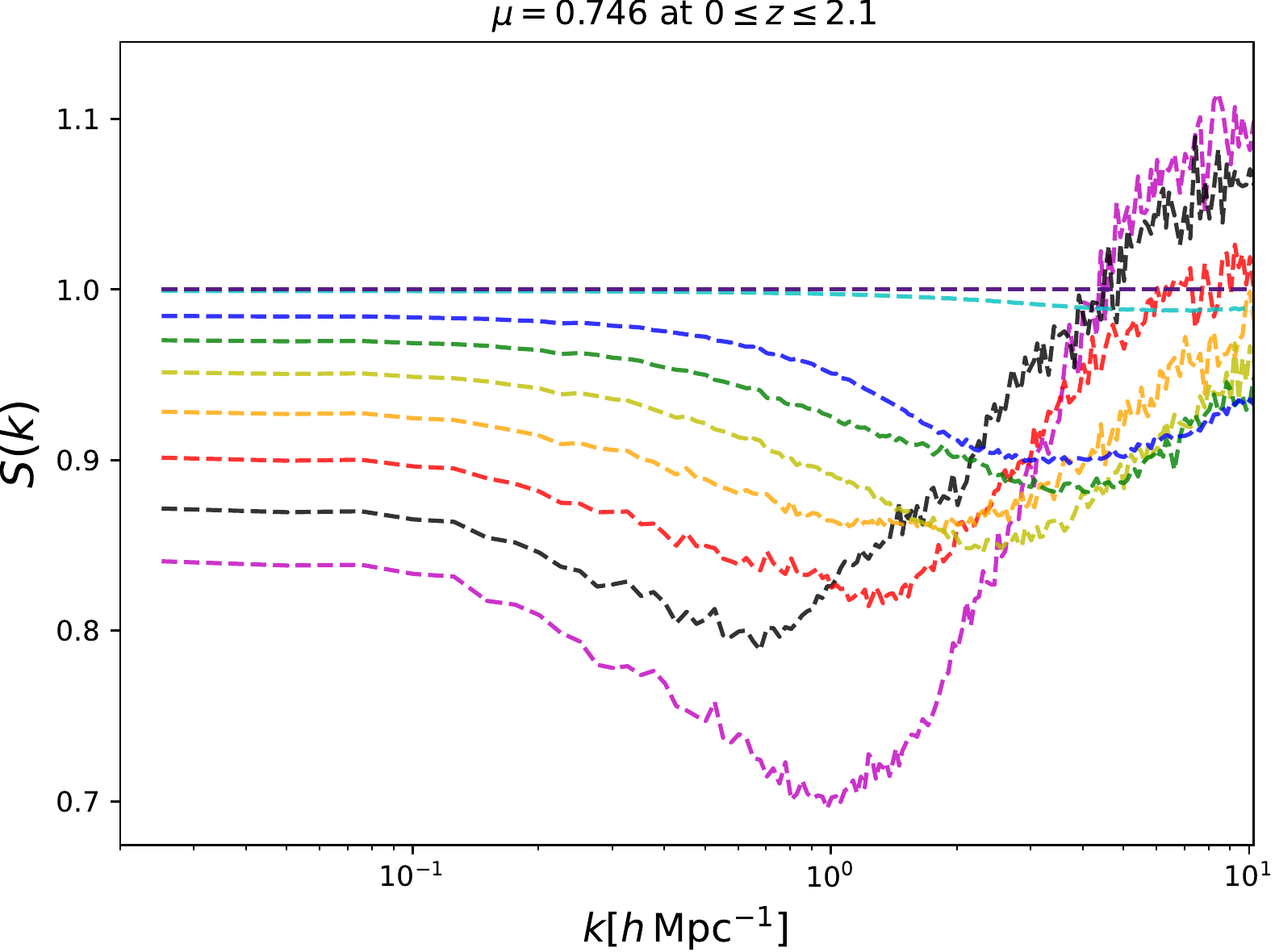}
 \includegraphics[width = 0.4\textwidth]{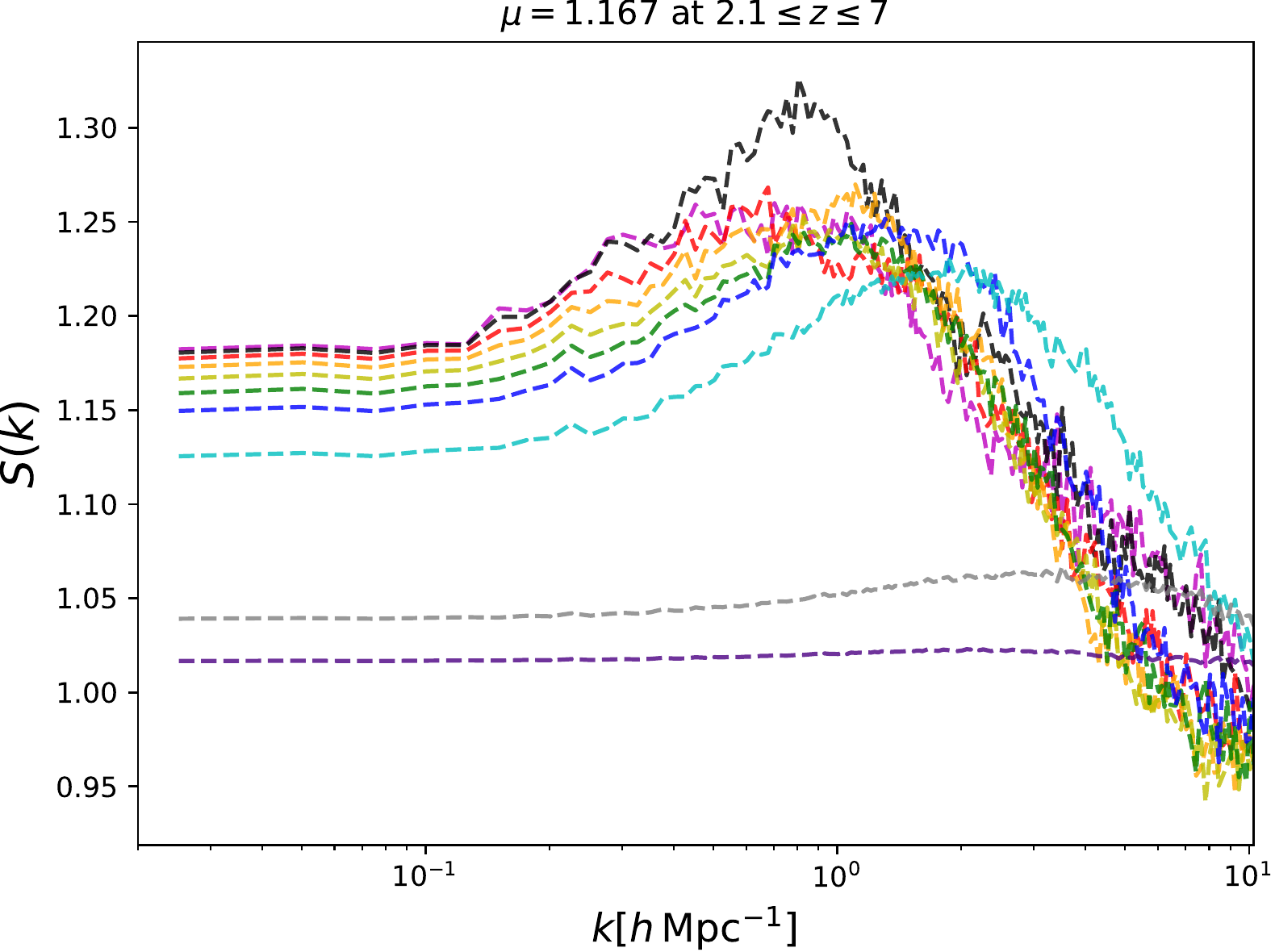}
 \includegraphics[width = 0.4\textwidth]{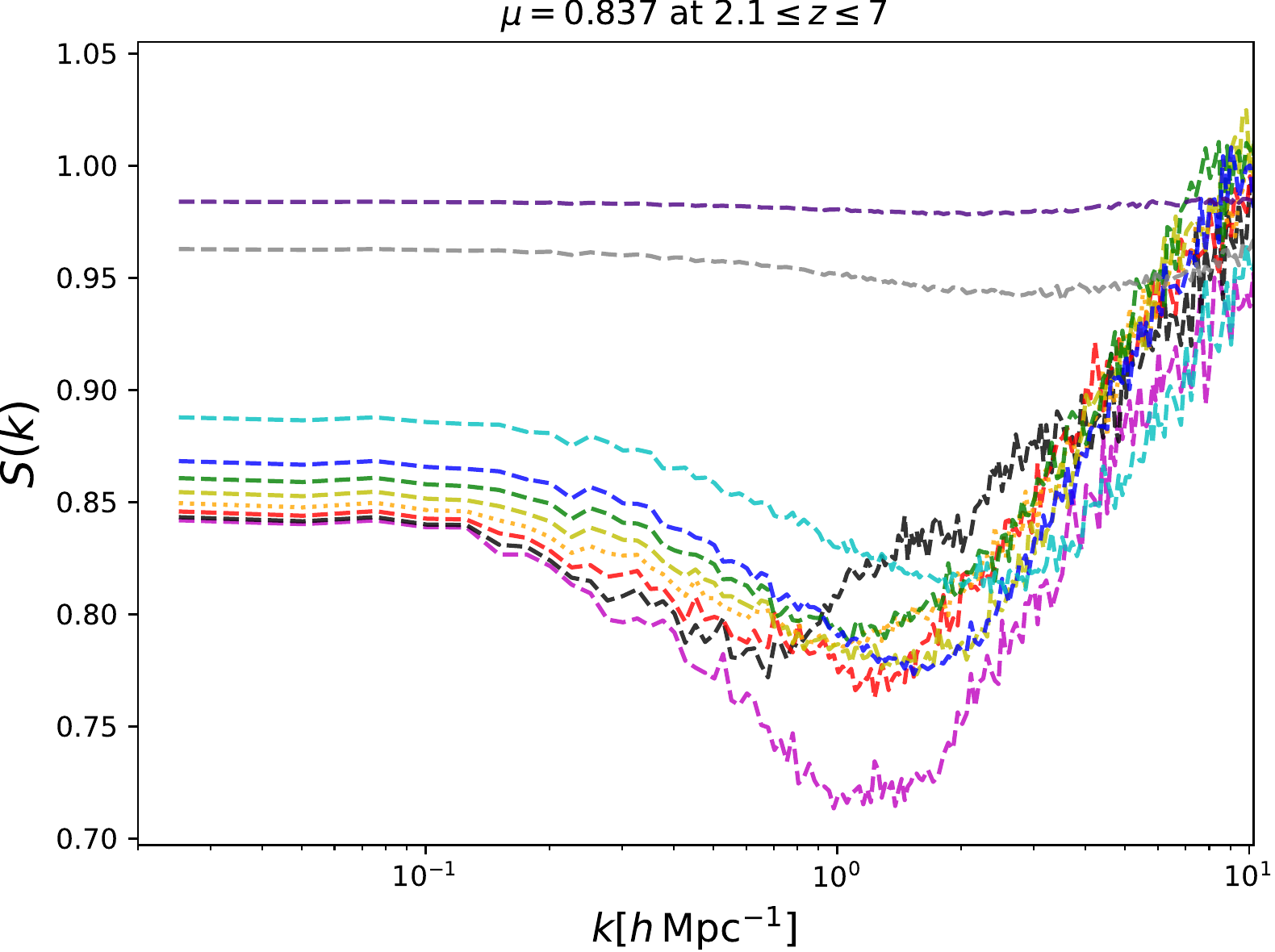}
 \includegraphics[width = 0.4\textwidth]{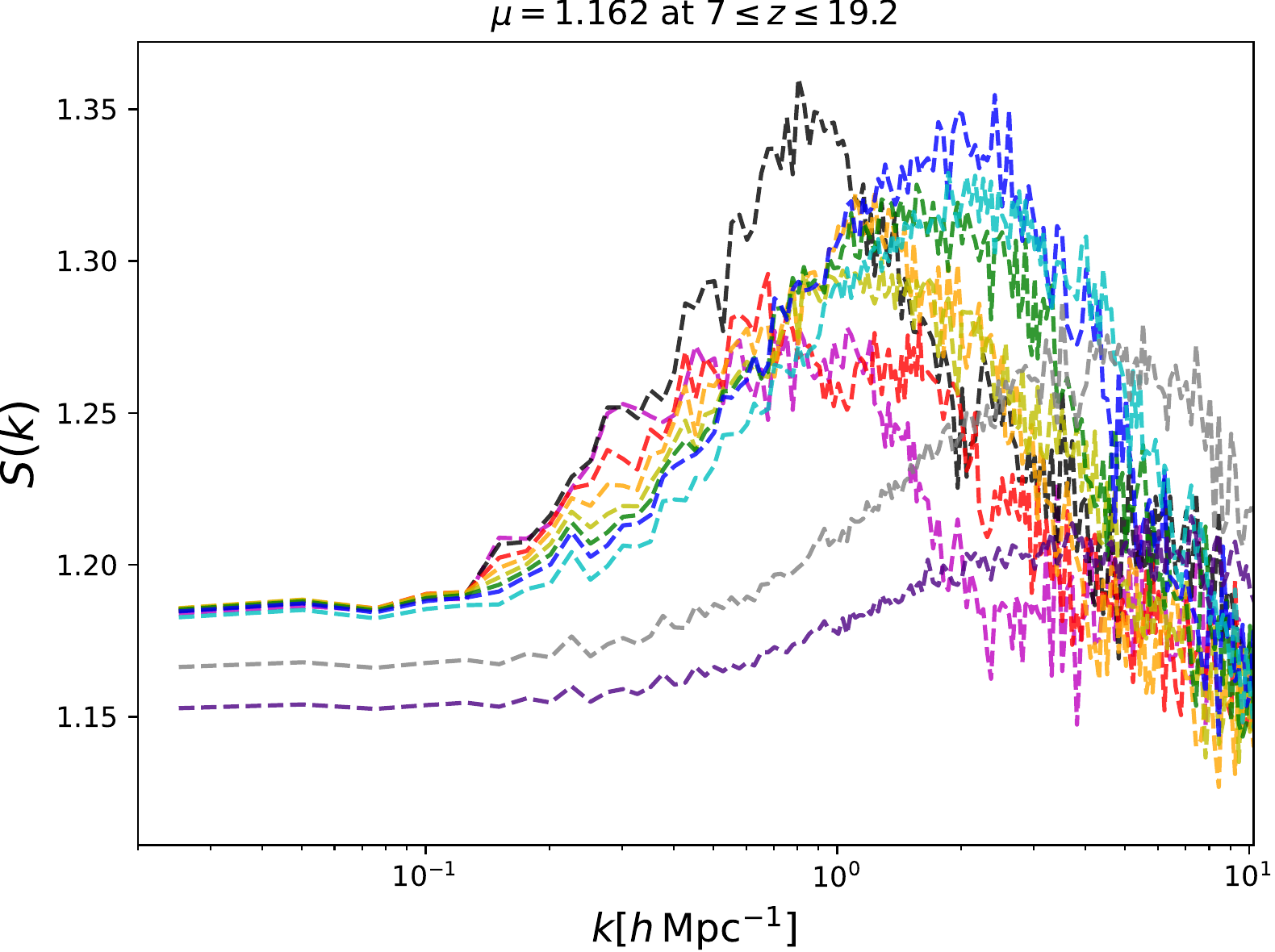}
 \includegraphics[width = 0.4\textwidth]{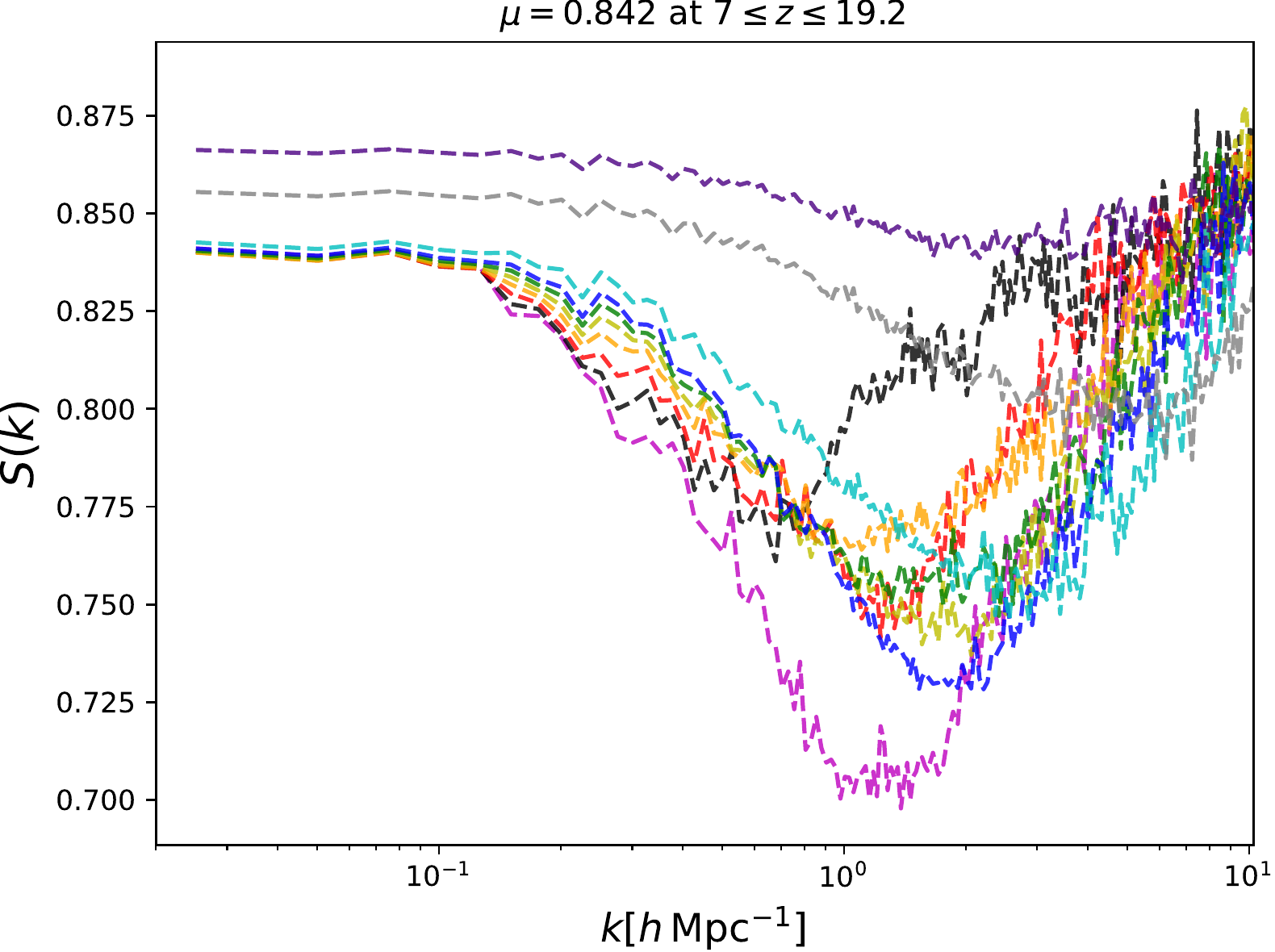}
 \includegraphics[width = 0.4\textwidth]{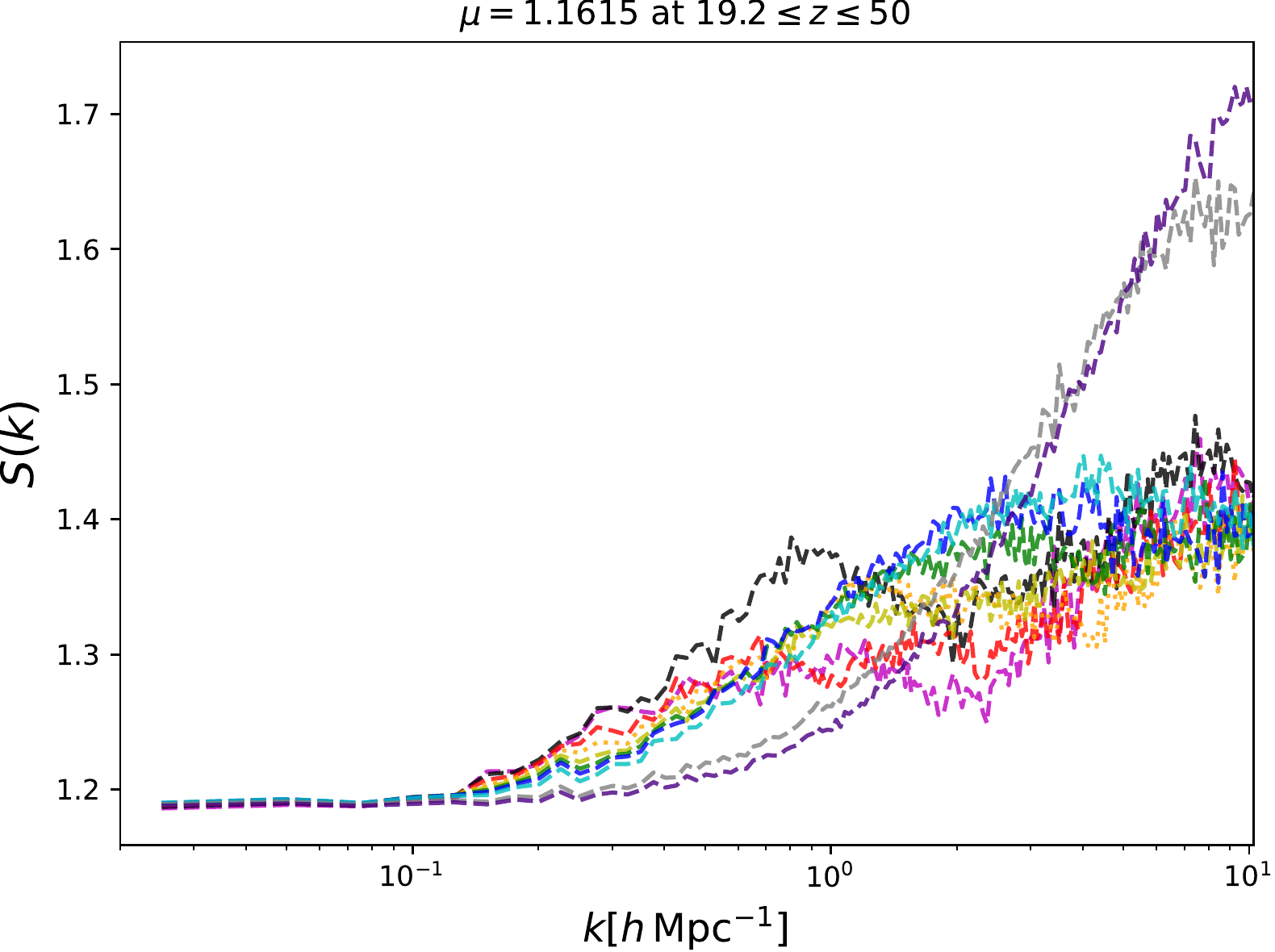}
 \includegraphics[width = 0.4\textwidth]{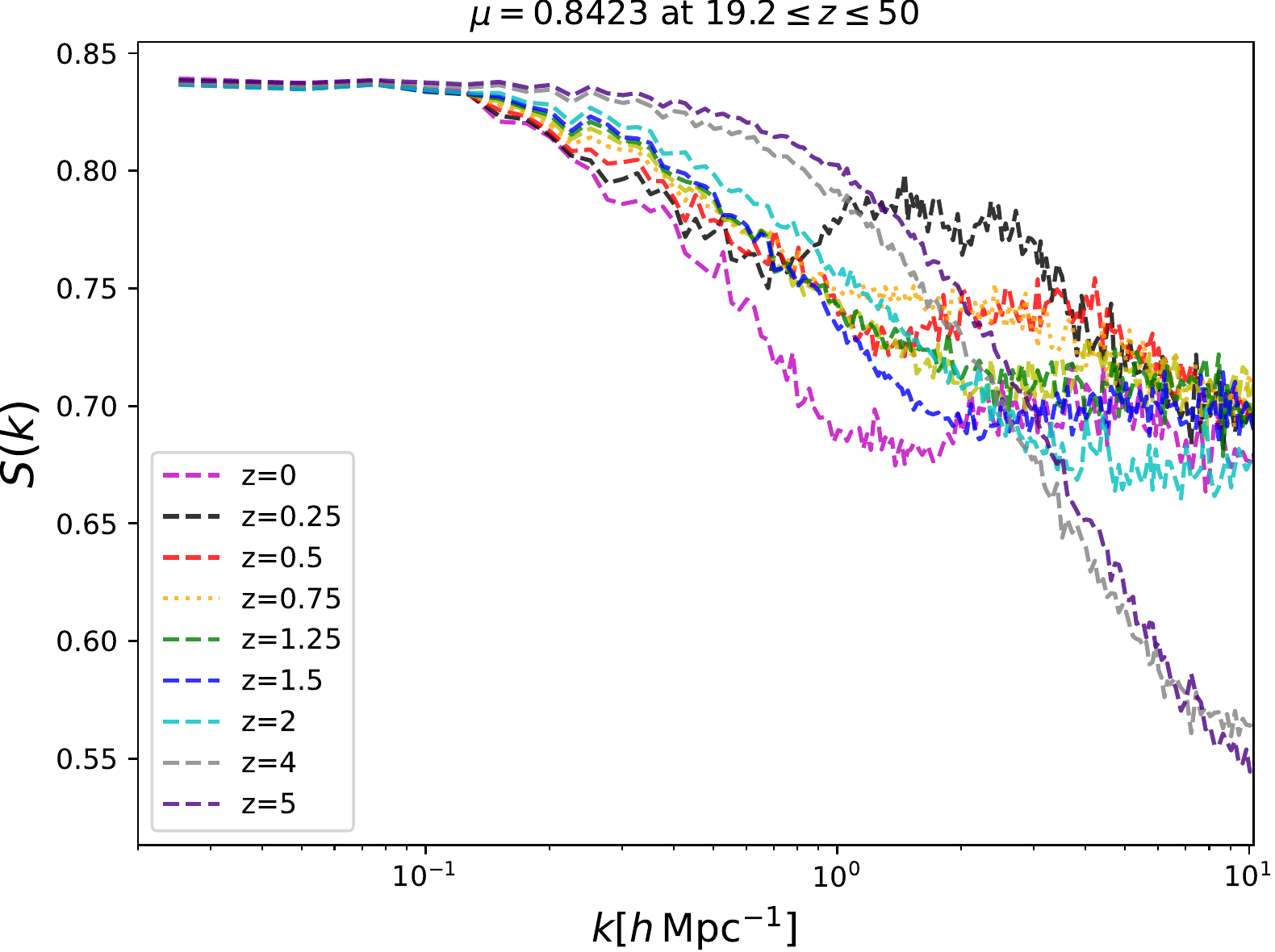}
  \caption{The ratio of the power spectra at all the redshifts output from our simulations relative to the $\Lambda$CDM power spectrum at the same respective redshifts, for the case where we have 4 bins in $\mu$ but the same linear growth at $z=0$. These figures highlight the hierarchical nature of structure formation and its varying influence on the power spectrum depending on whether $\mu$ is switched on early in the simulation or later. In the former case, we have an early modification of the non-linear power spectrum, the imprint of which can be noticed at $z=0$, while in the latter case, the late switching on of $\mu$, introduces interaction between already formed structures resulting in non-intuitive shapes for the matter power spectrum.} 
 \label{fig:Redshift_plots_4bin_11}
\end{figure}

We can examine this small-scale issue in more detail by looking at the redshift evolution of the matter power spectrum in our simulations. In fig.~\ref{fig:Redshift_plots_4bin_11}, we show the ratio of the power spectra from the simulations to the $\Lambda$CDM spectrum, at all of our output redshifts. There is a clear transfer of power over time from small scales to large scales. This is demonstrated in the top row where the peak/trough (depending on whether $\mu>1$ or $\mu<1$) in the ratio of the power spectra shifts to the left, i.e., to larger scales as one steps forward in redshift (from high redshift to low redshift). In the bottom panels, this transfer of power persists, although we see it manifested as phenomenon that `evens out' the power over large and small scales. This transfer of power indicates the formation of larger structure in the universe causing power to be deposited on larger scales. Comparing the position and height of this peak/trough provides information on the rate of structure formation and the gravitational interaction at different epochs. On comparing the behaviour of the magenta, black and red lines in the top row with the bottom row, we also see the reinforcement of our earlier result, that there is an excess of power in the non-linear regime when $\mu >1$ at high redshift, and vice versa at low redshift (this behaviour is mirrored when $\mu<1$). We now consider these two cases in more detail.

Structure formation in the Universe is hierarchical, with smaller haloes being formed before larger ones. Therefore, modifying $\mu$ at early times impacts the power spectrum on the smallest scales as demonstrated in the bottom row of fig.~\ref{fig:Redshift_plots_4bin_11} where there is a substantial peak in $S(k)$ on small scales at high redshift $z=4$ and $z=5$. This is due to the fact that only the smallest haloes were in the process of formation. Effectively, the increase in the rate of structure formation at early times causes smaller haloes to be formed earlier. The formation of larger structure only takes place when growth reverts to the $\Lambda$CDM rate. As one steps forward in redshift, one sees that this small scale peak is diminished, but not totally destroyed. As larger haloes start to form, mergers begin to take place leading to a gradual shift in power towards larger scales. However, there is still an excess of power at $z=0$ on small scales, as a signature of the enhanced structure formation at earlier times. 

Modifying $\mu$ at late times introduces additional complexity, since at late times, larger structures have started to form. Therefore, increasing/decreasing the strength of gravity massively impacts the interaction between larger haloes, i.e., halo power spectrum as seen by the aforementioned shift in the peak of the power spectrum to the left in the top row of fig.~\ref{fig:Redshift_plots_4bin_11}. Physically, this means that the interaction between haloes depletes power on smaller scales and deposits power on intermediate/large scales. Therefore, there is a transfer of power from smaller scales to larger scales (and of course the opposite happens when $\mu<1$). This further validates the behaviour of the orange, dark red and yellow lines in fig.~\ref{fig:Sims}, where the small scale behaviour is opposite to that on linear scales. We note that the middle rows in fig.~\ref{fig:Redshift_plots_4bin_11} contain a mix of the two limiting cases described in the above paragraphs.

\begin{comment}
\begin{figure}
 \centering
 \includegraphics[width = 0.45\textwidth]{dimless_k_z.png}
 \includegraphics[width = 0.45\textwidth]{dimless_k_z_zoom.png}
 \caption{The scale $k_{\rm NL}$ as a function of redshift plotted for the different runs, where $k_{\rm NL}$ is the scale associated to non-linear structure. This scale is defined to be the wavenumber at which the dimensionless matter power spectrum $\Delta^2(k) = k^3 P(k)/2\pi \approx 1$, as is typically done in the literature. \textbf{FP: What are the vertical lines? This figure is a bit weird in the formatting. With respect to the others, labels on axes are in bold and look much bigger than other figures. Also, figures go to the top of the page}}
 \label{fig:halo}
\end{figure}
\end{comment}

\section{Impact on weak-lensing observables}\label{section:Discussion}
\subsection{Computing weak-lensing observables in modified gravity}
We now discuss the impact of varying $\mu$ in different redshift bins on the observables. In this section we compute the weak-lensing convergence power spectrum from the output of our simulations, and examine the impact of having predictions that are not restricted to linear scales. Furthermore, we also discuss how one can probe our two-parameter family approach to modified gravity from the output of our simulations. 

The weak-lensing convergence power spectrum can be computed from the matter power spectrum and the modified gravity parameters from section \ref{section:Review}  using the following expression 
\begin{equation}\label{eq:convergenceSpec}
    P_{\kappa}(\ell) = \frac{9H_0^4\Omega_{\rm m}^2}{4c^4}\int_0^{\chi_{\rm max}}\mathrm{d}\chi\frac{1}{a^2(\chi)} g^2(z)\frac{\mu^2(1 + \eta)^2}{4}P_{\delta}(\ell/\chi)\, ,
\end{equation}
where $P_{\delta}$ is the matter power spectrum, $\eta$ is the second modified gravity parameter that affects the photon geodesics and $\chi$ is the comoving angular diameter distance to the source along the line of sight. This equation is derived using the standard procedure of solving for null geodesics in the perturbed FLRW metric, in order to express the convergence in terms of the metric potentials, but we then replace the potentials by substituting the modified Poisson equation in eq.~\eqref{eq:MGParam}. We stress that in order to compute the matter power spectrum, one requires no knowledge of $\eta$. Therefore, we need only model $\mu$ in our $N$-body simulations. We will return to this point shortly. In practice, the above integral is truncated at the distance $\chi_{\rm max}$ corresponding to the maximum source redshift the survey is sensitive to. In this work, we concentrate on the auto-power spectra. The additional complications involved in the analysis of the cross-power spectra requires a full statistical forecast, which is beyond the scope of this work. 

The function $g(z)$ is a filter function that depends on the redshift distribution of the background galaxies and is typically written as
\begin{equation}
 g_i(z) = \int_z^{\infty}\mathrm{d}z^{\prime}\left(1- \frac{\chi(z)}{\chi(z^{\prime})}\right)n_i(z^{\prime})\,.
\end{equation}
We use the standard expression for the source galaxy redshift distribution given by \cite{ref:Smail1994}
\begin{equation}\label{eq:smail}
 n(z) \propto \left(\frac{z}{z_0}\right)^{\alpha}\exp\left[-\left(\frac{z}{z_0}\right)^{\beta}\right]\,.
\end{equation}
We adopt a Euclid-like binning of the source number density into 10 equi-populated bins according to eq.~(\ref{eq:smail}) with $z_0 = 0.9/\sqrt{2}$, $\alpha = 2$ and $\beta = 3/2$, where we have assumed an average source density $\bar{n}_{\rm g} = 30\,{\rm arcmin}^{-2}$ \cite{ref:Hu1999, ref:Castro2005, ref:Casas2017, ref:SpurioMancini2018, ref:EuclidWL}. Due to the fact that the \texttt{ReACT} formalism is only valid at $z<2.5$, we concentrate only on the first tomographic bin, i.e., $0\leq z \leq 0.4$. The error bars are obtained by computing the following \cite{ref:EuclidWL}
\begin{equation}
 \delta C_\ell^{ij} = \sqrt{\frac{2}{f_{\rm sky}(2\ell + 1)}}\left(C_l^{ij} + \frac{\sigma_{\epsilon}^2}{\bar{n}_i}\right) \, , 
\end{equation}
where $f_{\rm sky} = 0.7$ is the fractional sky coverage, $\sigma_{\epsilon} = 0.21$ is the variance of the observed ellipticities and $\bar{n}_i$ is the surface galaxy density of each bin. Essentially, we combine the Poisson shot noise contribution with cosmic variance to obtain the total uncertainty on the power spectrum, $P_{\kappa}$. 
\begin{figure}
 \centering
 \includegraphics[width = 0.45\textwidth]{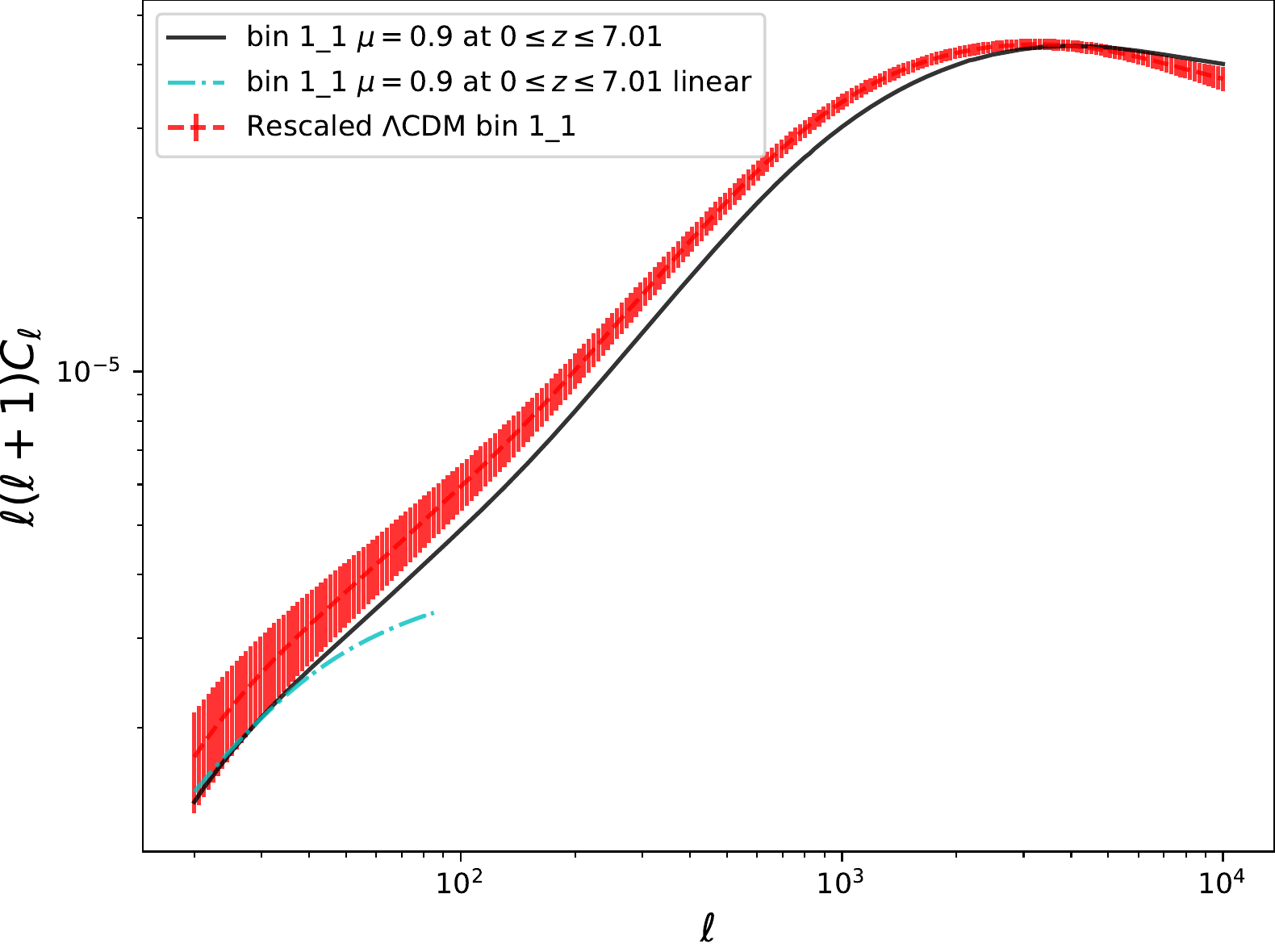}
 \includegraphics[width = 0.45\textwidth]{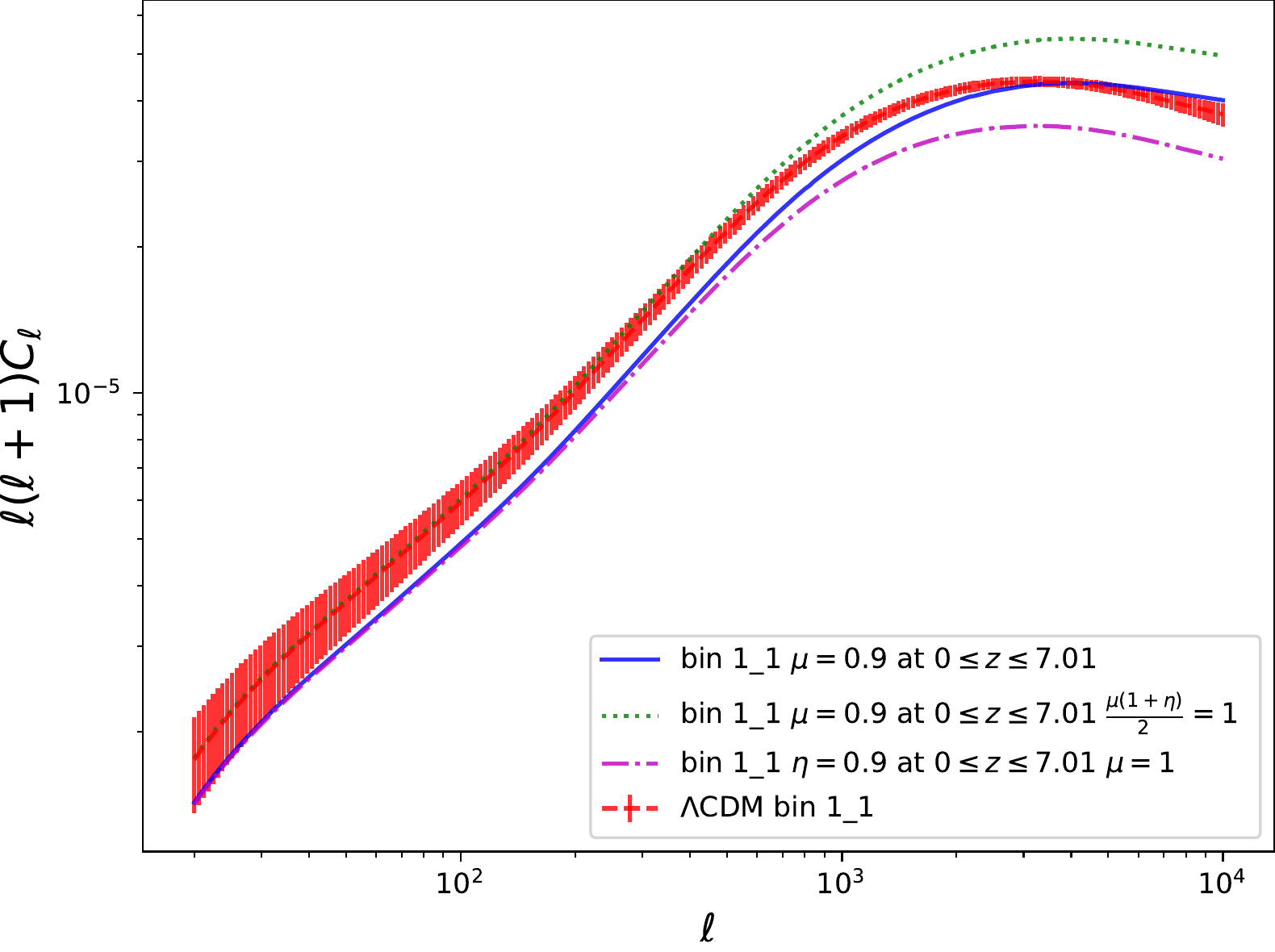}
 \caption{In the left panel, we show the weak lensing auto-power spectrum for a particular tomographic bin from $\Lambda$CDM (red-dashed) and from the simulations with $\mu = 0.9$ at $0\leq z \leq 7.01$ (blue solid). Note that we are using the $\Lambda$CDM simulation with rescaled initial conditions at $z=50$. In the right panel, we show three different curves with the same $\mu$ bin, but with different values of $\eta$. The first curve has $\eta=1$ (blue solid). The second has $\eta=0.8$ such that the prefactor $\mu(1+\eta)/2=1$ as in $\Lambda$CDM (green-dotted), such that the modified gravity effects only enter through the modified $P_\delta$. Finally, we show the case where $\mu = 1$, but $\eta = 0.8$ at $0\leq z \leq 7$, which has a different shape, i.e., different scale dependence on non-linear scales. Therefore, we show how our approach may be used by future missions to put constraints on both $\mu$ as well as $\eta$. We stress that by explicitly going to non-linear scales, we are able to break the degeneracy between $\eta$ and $\mu$ that exists in linear theory.}
 \label{fig:wlspectra}
\end{figure}

\begin{figure}
 \centering
 \includegraphics[width = 0.45\textwidth]{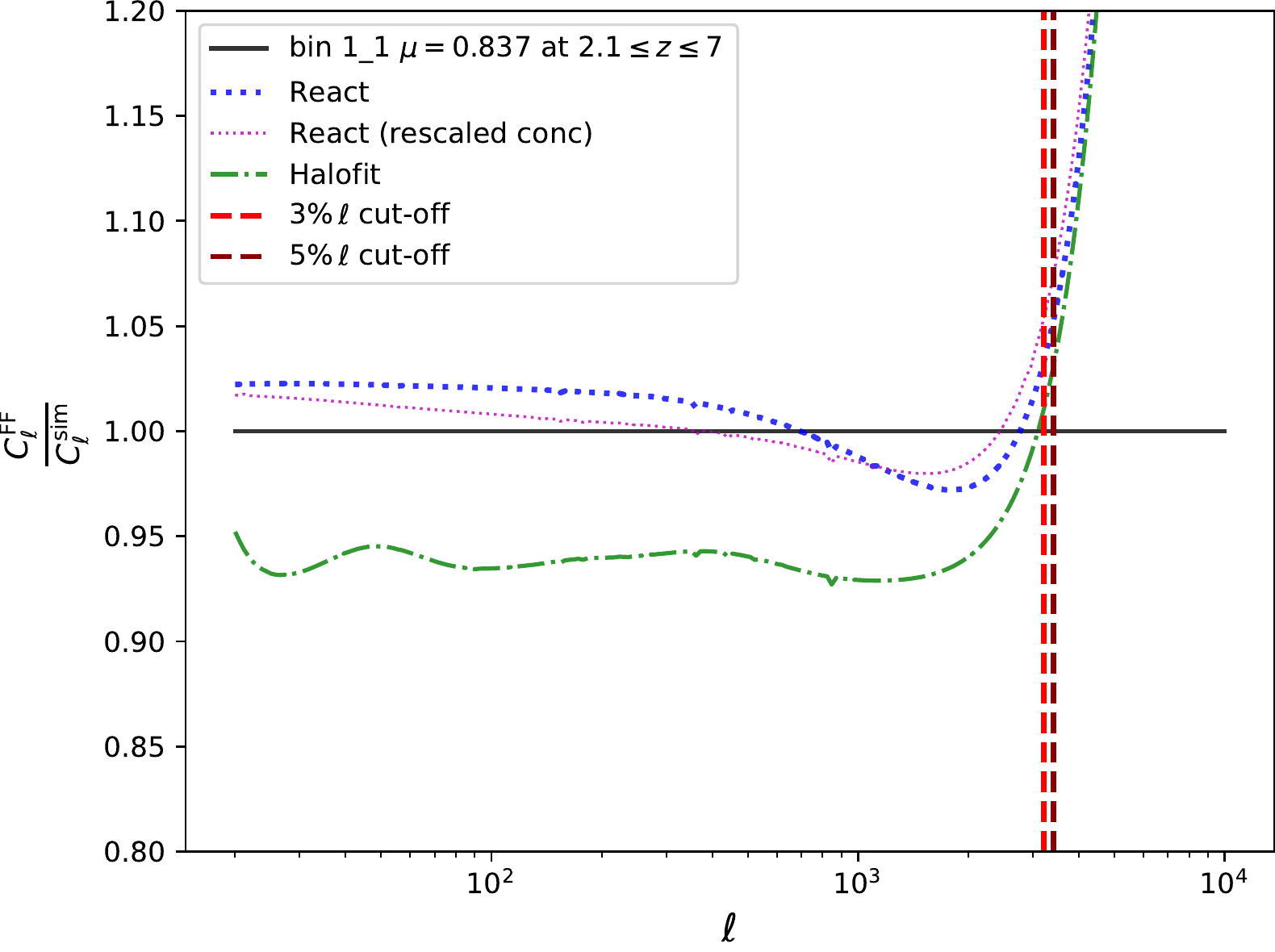}
 \includegraphics[width = 0.45\textwidth]{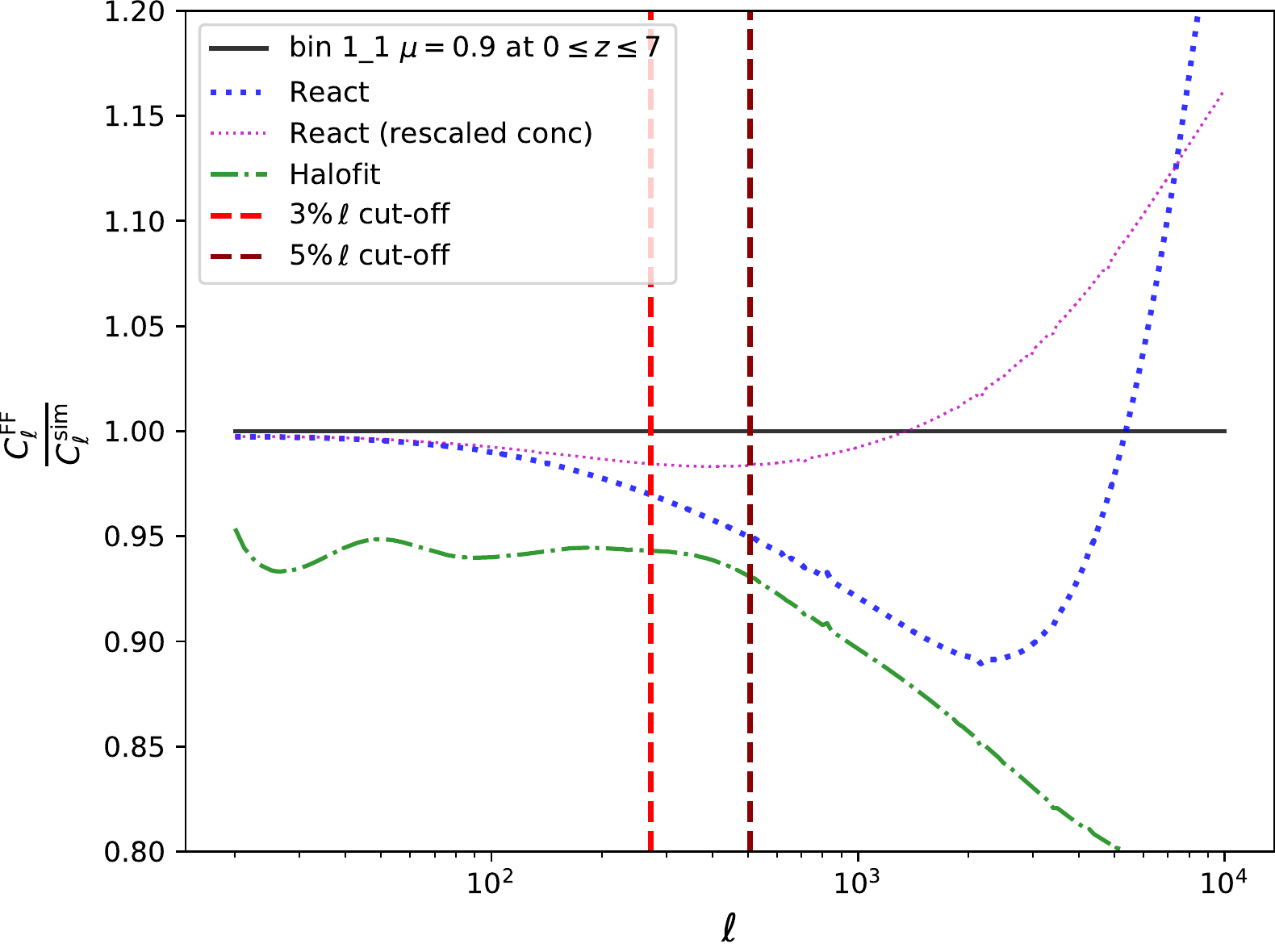}
 \caption{We show the ratio of the convergence power spectra obtained from the matter power spectra as predicted by the \texttt{ReACT} (blue-dotted) and \texttt{halofit} (green dot-dashed) formalisms with respect to those obtained from the simulation matter power spectra. The superscripts `FF' and `sim' imply `fitting function' and `simulation', respectively. We immediately see that \texttt{halofit} performs significantly worse than \texttt{ReACT} even at low $\ell$, where it fails at the level of $\sim 5\%$. The vertical lines indicate the value of $\ell$ at which the \texttt{ReACT} curve deviates from the simulation curve at the level of 3\% and 5\%, respectively. We choose two simulations that capture the variation in the $\ell$ cut-offs in our sample of simulations. In the left panel is the simulation with the largest $\ell$ cut-offs in our sample of simulations. In the right panel, we show a simulation with one of the smallest $\ell$ cut-offs. The better performance of the \texttt{ReACT} formalism over \texttt{halofit} in such a model-independent analysis of modified gravity on non-linear scales is clearly demonstrated. We also show the curve with the concentration parameter modified (see fig.~\ref{fig:concentrationMod}) which seems to further improve the \texttt{ReACT} fit, see text for details.}
 \label{fig:lensing_ratio_plots}
\end{figure}

 Note that since this involves an integration along the line of sight, one needs to compute the matter power spectrum at multiple redshifts and interpolate between them to compute the integral in eq.~(\ref{eq:convergenceSpec}). To do this, we make use of the \texttt{project\_2d} module in the \texttt{Cosmosis} numerical library \cite{ref:ZuntzCosmosis}. We compute both the linear and non-linear matter power spectrum from our simulations at $z_{\rm pk}= \{0, 0.25, 0.5, 0.75, 1.0, 1.25, 1.5, 2.0\}$ which are then used as input to perform the integral in eq.~(\ref{eq:convergenceSpec}) assuming the tomographic source redshift distribution given by eq.~(\ref{eq:smail}). Following eq.~\eqref{eq:convergenceSpec}, we multiply the input power spectra that we are interpolating between by the corresponding value of $\mu^2(1 + \eta)^2/4$ at each redshift.

We compute the convergence power spectra from both linear matter power spectra and the non-linear matter power spectra from our simulations, and compare them with the $\Lambda$CDM convergence power spectrum with Euclid-like error bars. On the left hand panel of fig.~\ref{fig:wlspectra}, we show the convergence power spectra as computed from the output of one of our simulations. Note that for this plot, we only consider a variation in $\mu$, and we set $\eta$ to its $\Lambda$CDM value ($\eta = 1$). We see that on large scales, i.e., small $\ell$, the modified gravity spectrum is within the $\Lambda$CDM error bars, while the spectrum on non-linear scales is outside the error bars. Moreover, the linear spectrum deviates from the non-linear spectrum below $\ell=100$. This shows the strong increase in constraining power that can be achieved by using non-linear scales: not only are there many more scales that can be used, but also the errors are typically smaller on those scales. These results emphasise the importance of developing simulations and fitting functions for model-independent modified gravity, as examined in section \ref{section:Pipeline}.

We note that comparing our modified gravity parameterisation (with parameters that are only time dependent) to $\Lambda$CDM for a single observable is slightly non-trivial, because one has a free choice of which redshift to make the linear matter power spectra agree at. For our weak-lensing comparison, using the lowest redshift tomographic bin, we choose to focus on the case where the two linear spectra are equal at redshift zero (i.e. the $\Lambda$CDM simulation with rescaled initial conditions at $z=50$). We make this choice for several reasons. It minimises the linear theory difference between the two curves, allowing us to focus on the additional information from non-linear behaviour/scales and, as long as the linear theory behaviour is similar at low redshift, the exact redshift at which they are equal is unimportant since lensing combines information from multiple redshifts. Our conclusions about the relative sizes of errors on difference scales, and the relative numbers of scales that contribute in each regime, are unchanged if a different choice is made. We will extend our analysis to a full parameter forecast for upcoming surveys in future work, which will allow this issue of parameter degeneracies to be examined in detail.

We now generalise our results to using our simulation output to probe the effects of both $\mu$ and $\eta$. As mentioned before, since the input matter power spectra from the simulations are unaffected by $\eta \neq 1$, we may model the effects of it in post-processing. In other words, the evolution of matter fluctuations is driven solely by $\mu$, but the trajectory of photons through this distribution is affected by both $\eta$ and $\mu$. Therefore, we can effectively ignore $\eta$ whilst running our $N$-body simulations without any loss of generality, essentially getting $P_{\delta}(k)$ as a function of $\mu$ only. We then use eq.~\eqref{eq:convergenceSpec} to compute the weak-lensing observables for both parameters. 

As shown in eq.~\eqref{eq:convergenceSpec}, the effect of varying $\eta$ is captured solely by the prefactor $\mu^2(1+\eta)^2/4$. Therefore, one would expect that a variation of $\eta$ would result in an overall shift of the power spectrum. If one were to concentrate on purely linear scales, this is exactly the same as varying $\mu$ due to the fact that a time-dependent $\mu$ also simply shifts the matter power spectrum in the vertical direction(s). As a result, there is a degeneracy between time-dependent $\eta$ and $\mu$ in linear theory. By explicitly going to non-linear scales, we break this degeneracy since we have already shown that a time-dependent $\mu$ introduces scale-dependent features in the matter power spectrum, and therefore in the weak-lensing convergence spectrum via $P_{\delta}$ in eq.~\eqref{eq:convergenceSpec}, whereas $\eta$ introduces no such scale dependence.\footnote{In principle, $\eta$ does introduce a small scale dependence due to weighting of the non-linear spectra at different times by different amounts, but this will be much smaller than the scale dependence caused by $\mu$ on non-linear scales.}

The breaking of this degeneracy is clearly shown in the right panel of eq.~\eqref{eq:convergenceSpec}, where we show the $\Lambda$CDM convergence spectrum with error bars, along with two different curves obtained from the same simulation, with identical $\mu(z)$, but different values of $\eta$. The green-dotted line is obtained by setting $\mu^2(1+\eta)^2/4 = 1$ (as in $\Lambda$CDM). The solid blue line is obtained directly from the simulation, i.e., $\eta = 1$. We see that varying $\eta$ while keeping $\mu$ constant leads to an overall different amplitude of the convergence spectra. We also explicitly show that if one reverts to the $\Lambda$CDM value of $\mu = 1$ while keeping the prefactor $\mu(1+\eta)/2$ the same as in the blue line, one obtains the same shape for the power spectrum as the $\Lambda$CDM case. This can also be seen by comparing the shapes of the blue line and the green-dotted line which have identical $\mu(z)$. Thus, it is clear that the scale dependent effects from a time-dependent $\mu$ breaks the degeneracy between the two parameters that exists on linear scales, allowing the two parameters to be distinguished from each other and from their $\Lambda$CDM values. We note that the linear theory degeneracy should persist in both the auto-power spectra and the cross-power spectra. However, we leave the full behaviour of the cross-power spectra as a function of varying $\eta$ and $\mu$ to future work.

These results not only show how weak-lensing observables can be constructed from our simulations for the full ($\mu,\eta$) parameter space, but also further highlight the importance and gain of using non-linear scales.

\subsection{Further evaluation of fitting functions}
We can extend the earlier analysis of the fitting functions, by examining how well they capture the non-linear behaviour of the weak-lensing observables. We focus on the two fitting functions that performed the best in section \ref{section:FittingFunction}, i.e., the \texttt{ReACT} formalism and the \texttt{halofit} fitting procedure.

In general we find that the better performance of \texttt{ReACT} compared to \texttt{halofit} is further improved when looking at the weak-lensing observables, compared to when directly examining the matter power spectrum. In particular, similarly to section \ref{section:FittingFunction}, we quantify the quality of the fitting functions by calculating the $\ell_\text{fail}$ value (rather than $k_\text{fail}$ as in section \ref{section:FittingFunction}) at which the fitting function first fails by $3$ (or $5$) \%. The $\ell_\text{fail}$ values at which \texttt{ReACT} fails by $3\%$ are in the range $250\leq \ell_\text{fail}\leq4000$, however \texttt{halofit} typically fails at the $3\%$ level even for very low $\ell$. For a given fitting function, we find little correlation between the $k$ and $\ell$ values at which the fitting function fails in each simulation, due to the range of $k$ and redshift values at which the matter power spectrum contributes to the weak-lensing convolution.

To illustrate this performance, we plot the results for two representative simulations: one where \texttt{ReACT} performs well to higher $\ell$, and one where it fails at much lower $\ell$, approximately corresponding to the best and worst cases we studied. These simulations are shown in fig.~\ref{fig:lensing_ratio_plots}, where we show the ratio of the convergence power spectra calculated from the matter power spectra predicted by the fitting function and the simulation respectively. These plots show that \texttt{halofit} performs significantly worse than \texttt{ReACT}, failing at the level of $\sim 5\%$ even at low $\ell$. This poor performance of \texttt{halofit} when computing weak-lensing observables is common across all simulations.

In these plots, the vertical lines show the $\ell_\text{fail}$ values, i.e., where the \texttt{ReACT} curve deviates from the simulation curve at the level of 3\% and 5\%, respectively. The left panel shows a simulation where the \texttt{ReACT} curve is accurate to $\ell \sim 5000$ while on the right, cut-offs are an order of magnitude smaller. These $\ell_\text{fail}$ values give an indication of the $\ell$ range where \texttt{ReACT} can be used to fairly reliably predict the weak-lensing observables for model-independent modified gravity studies. As described in section \ref{section:FittingFunction}, we expect this performance to be improved by adjusting the concentration parameter. To demonstrate this, we include an additional curve where the concentration parameter has been modified according to the discussion at the end of section \ref{section:FittingFunction} (see fig.~\ref{fig:concentrationMod} for more details). As expected, it appears that weak-lensing observables can be accurately computed to higher $\ell$ using this process, however we leave a detailed examination of this to future work.

These results, in combination with those in section \ref{section:FittingFunction}, show that phenomenological modified gravity analyses with current data can be carried out without restricting to linear scales (e.g. \cite{ref:BattyePearson}) or carrying out a linearistaion procedure as in \cite{ref:Abbott_2019}. For these purposes, our results show that \texttt{ReACT} is the best performing fitting function (including outperforming \texttt{halofit}), particularly when using weak-lensing observables. In future work we will extend this initial analysis to quantitatively examine a much wider range of parameter space, in preparation for analysing the data for upcoming surveys.

\section{Discussion and Conclusion}\label{section:Conclusion}

In this work, we have presented $N$-body simulations with a time-dependent strength of gravity $\mu$, based on a framework for examining modified gravity in a model-independent way across all cosmological scales \cite{ref:DanPF}. The key results of this paper are a presentation of the phenomenology of these simulations, an evaluation of existing fitting functions for capturing this phenomenology, and a demonstration of the application and importance of this framework for weak-lensing observables.

We modified the \texttt{GADGET-2} $N$-body code \cite{ref:GADGET-2}, and ran a series of simulations with piecewise-constant bins in redshift for $\mu$. See section \ref{section:Pipeline} for more details and table \ref{tab:binning} for the redshift bins and the $\mu$ values in each bin.

The only fitting function calibrated from $N$-body simulations for the matter power spectrum on non-linear scales in phenomenological modified gravity was presented in \cite{ref:Cui1}. We investigated the performance of this fitting function, as well as the $\Lambda$CDM halofit fitting procedure, the standard halo model of structure formation [see eqs.~(\ref{eq:1halo}),(\ref{eq:2halo})] and the halo model reaction \cite{ref:reactionCataneo, ref:ReactTheory}. We compared the ratio of the modified gravity matter power spectrum to the so-called `pseudo' $\Lambda$CDM power spectrum as predicted in the various formalisms to the same ratio measured from our simulations. We quantified the accuracy of each formalism by calculating the least square statistic $\chi^2$ and $k_{\rm fail}$, the wavenumber of first failure (see fig.~\ref{fig:chisquaredbars}) for each fitting function. We found that the halo model reaction formalism performed significantly better than the others, with the notable exception being the case where the modified gravity parameters are such that the resulting power spectrum is co-incidentally very similar to the $\Lambda$CDM case. We also see qualitative evidence (see fig.~\ref{fig:concentrationMod}) supporting the theoretical expectation that to achieve precision in forecasting modified gravity matter power spectra, one needs to modify the $\Lambda$CDM concentration-mass relationship. We will extend the investigation presented here to a full parameter space examination and validation of the ReACT approach and the concentration issue in future work.

We present the matter power spectra from our simulations in section \ref{section:Results}. Figs.~\ref{fig:Sims} and \ref{fig:Redshift_plots_4bin_11} show that a purely time-dependent $\mu$ induces scale-dependent features in the matter power spectrum, as reported in \cite{ref:Cui1} for a simpler case. Our results also show that the shape of the power spectrum on quasi-linear as well as non-linear scales depends not only on the value of $\mu$, but also on the redshift at which it is `switched on' (i.e., different from unity) and the duration of such a modification. Most notably, we see that introducing a modification in $\mu$ at early times produces a power spectrum that either lacks ($\mu < 1$) or has excess power on the quasi-linear scales ($\mu>1$) and vice-versa on non-linear scales, respectively. We also noticed that the peak in the power spectrum relative to $\Lambda$CDM shifts to larger scales (lower $k$) as one steps forward in time (from high redshift to low redshift). 

To understand the physics that leads to these results, we first note that modifying $\mu$ at different epochs and for different periods of time affects the redshift at which non-linear structure starts to form. Combining this with our results indicates that a \textit{transfer of power} occurs from smaller to larger scales or vice versa, depending on when and for how long $\mu$ is modified. This is due to the fact that at early times, the very first haloes are in the process of forming, which means that changing the rate of structure formation affects the smaller cosmological scales since those are the haloes forming at that time. Therefore, when one returns to $\Lambda$CDM structure formation at late times, one sees a marked difference in the power spectrum. However, if one introduces a change in the rate of structure formation at late times, when galaxies have already formed, one is then affecting the rate at which the largest haloes form, via mergers. Therefore, one sees a unique change in both the quasi-linear as well as the non-linear scales of the matter power spectrum, for a given time evolution of $\mu$.

Finally, we show the impact of these phenomenological modifications to gravity on weak-lensing observations that will be generated in future experiments like the \textit{Euclid} satellite (see fig.~\ref{fig:wlspectra} and the discussion attached). We show that the constraining power of these experiments in the context of modified gravity is strongly dependent on the data obtained from non-linear cosmological scales, both in terms of the number of scales that are accessible and the sizes of the errors on the different scales. We further show how we can use weak-lensing observables to probe the two-parameter $\mu-\eta$ family of modified gravity models using the output of our simulations [see fig.~\ref{fig:lensing_ratio_plots}]. We find that by explicitly going to non-linear scales one breaks the degeneracy between these two parameters that exists on linear scales [see right panel of fig.~\ref{fig:wlspectra}]. We also extend our analysis of the fitting functions to the weak-lensing context, again finding that the \texttt{ReACT} formalism performs the best. This shows that phenomenological modified gravity analyses with current data can be carried out without restricting to linear scales or carrying out a linearisation procedure.

The model-independent approach first elucidated in \cite{ref:DanPF}, and further developed here, is not a priori restricted to particular regions of model space or types of theories, and has the potential to be a powerful null test of the $\Lambda$CDM+GR paradigm. The simulations and results presented here are an important step towards realising this, and using all of the data in future surveys to put model-independent constraints on the laws of gravity that operate in the Universe.

\section*{Acknowledgments}
This work was performed using the DiRAC Data Intensive service at Leicester, operated by the University of Leicester IT Services, which forms part of the STFC DiRAC HPC Facility (www.dirac.ac.uk). The equipment was funded by BEIS capital funding via STFC capital grants ST/K000373/1 and ST/R002363/1 and STFC DiRAC Operations grant ST/R001014/1. DiRAC is part of the National e-Infrastructure. We also thank Lee Whittaker for valuable comments and advice. FP acknowledges the support from the grant ASI n.2018-23-HH.0.

\bibliography{References.bib}
\bibliographystyle{apsrev4-1}

\appendix
\section{Convergence Tests}\label{appendix:Convergence}
We now present the procedure we followed to test the convergence of our $N$-body simulations. In the following discussion we explicitly show convergence with respect to particle number in our simulations. Using the same methods, we also tested convergence with respect to varying box-size, softening length and also different realisations of initial conditions, respectively.   

\begin{figure}
 \includegraphics[width=0.45\textwidth]{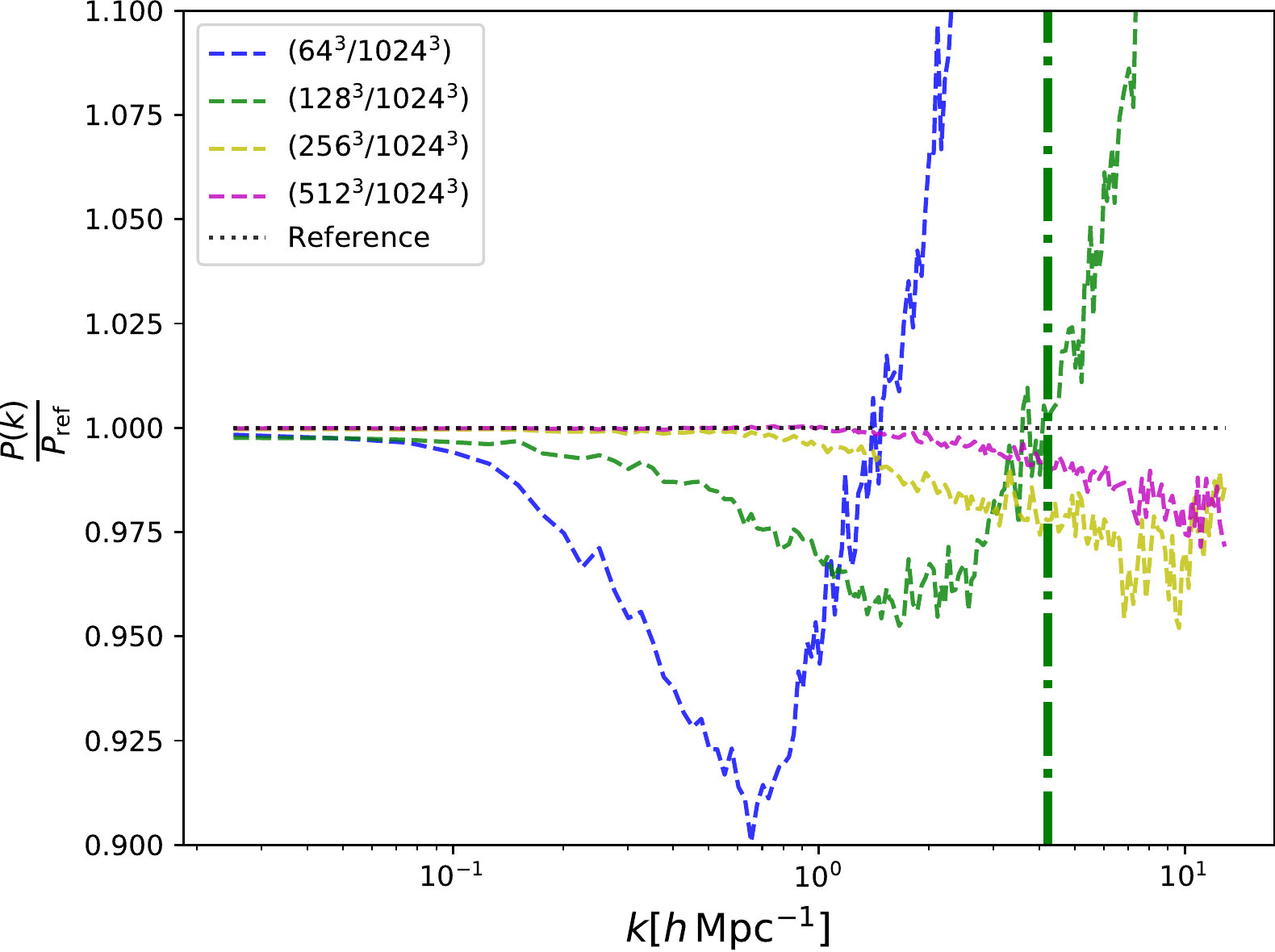}
 \includegraphics[width=0.45\textwidth]{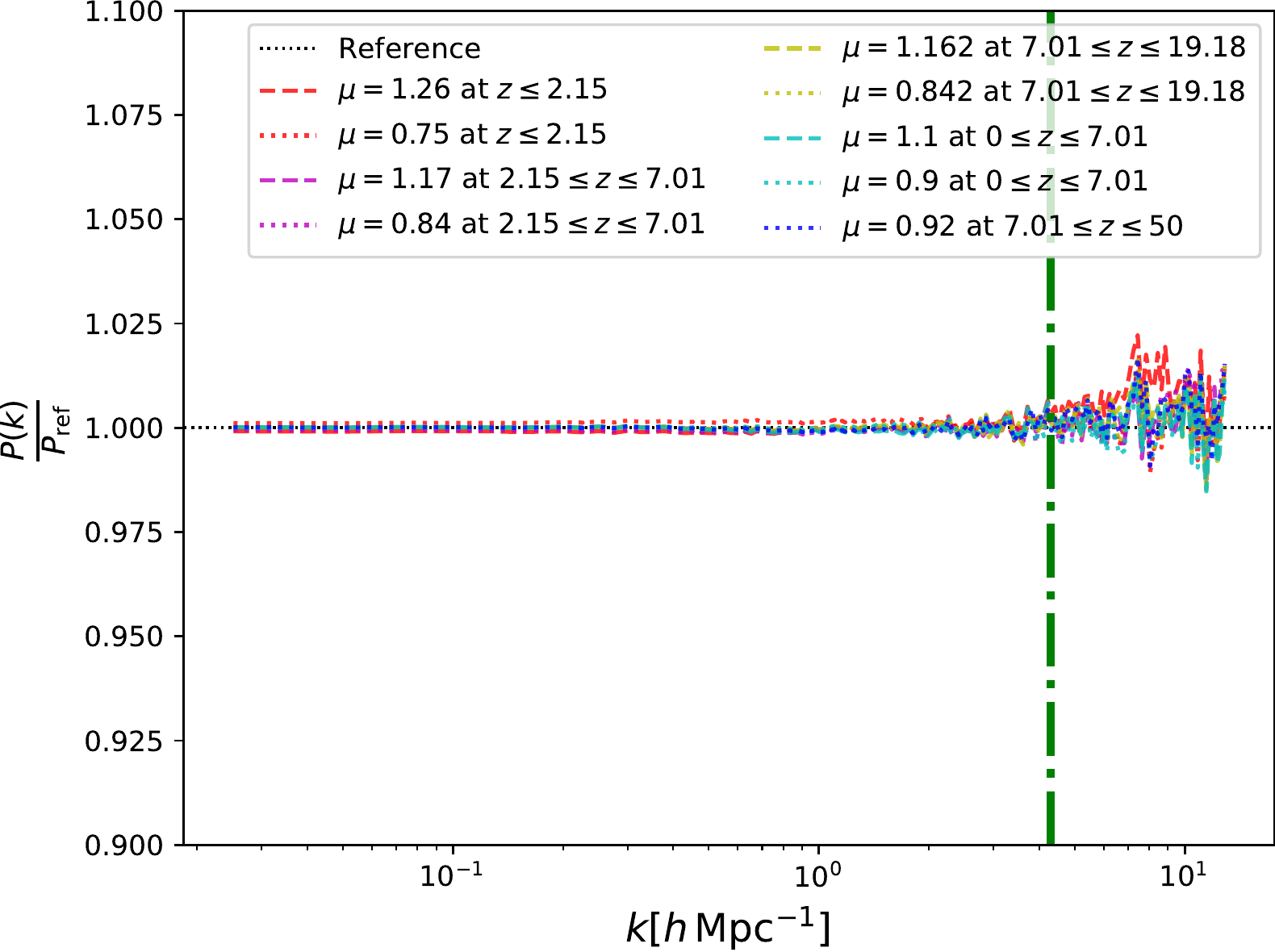}
 \caption{\textit{Left panel}: The ratio of the power spectra as a function of scale, with each line labelled by the particle number of the simulation they represent. All simulations have a box size of 250\, Mpc\,$h^{-1}$ comoving. The green vertical line represents the wavelength at which the ratio of the power spectra deviates at the level of 1\% from the simulations with the best resolution. This panel shows the level of convergence we obtain in our $\Lambda$CDM simulations. \textit{Right panel}: A similar procedure was carried out for the modified gravity simulations, but with a key difference. For each simulation, we calculate the ratio $S(k)$ of the power spectrum from the simulation to the power spectrum from the $\Lambda$CDM simulation. We then plot the ratio of $S(k)$ computed from simulations with $512^3$ particles to simulations with $1024^3$ particles. The green vertical line is plotted at the identical point as before. Therefore, we show that our modified gravity simulations have similar convergence both with respect to each other as well as with respect to our $\Lambda$CDM simulations. Note that on linear scales, all the ratios are within 0.5\% of 1, which justifies our choice of $\mu$ in each bin. For our production simulations, we use a particle number of $1024^3$ (with an equally fine mesh) and a box size of $250\,{\rm Mpc}\,h^{-1}$.}
 \label{fig:Convergence}
\end{figure}

In order to demonstrate convergence in our simulations, we present ratios of the matter power spectrum as this allows one to neglect realisation-dependent effects \cite{ref:McDonaldRatio2006}. We run all our simulations in a cosmological box of side 250 Mpc\,$h^{-1}$. In the left panel of fig.~\ref{fig:Convergence}, we compute the ratio of the matter spectrum from $\Lambda$CDM simulations with respect to our highest resolution reference simulation, increasing the particle in steps. The green line denotes the $k$ value at which the second-highest resolution simulation ($512^3$ particles) disagrees with our highest resolution ($1024^3$ particles) simulation, which is an indicator of the scale up to which our results are reliable.  

We now show that we have achieved the same level of convergence in our modified gravity simulations. In the right panel, we show the convergence of our modified gravity simulations by plotting a ratio of ratios. We evaluate the ratio of matter power spectra from the simulation with $512^3$ particles with the simulation with $1024^3$ particles in both our $\Lambda$CDM simulations as well as our modified gravity simulations. We then plot $S_{512}/S_{1024}$, and show that they agree up to the same wavenumber as the left-hand-side panel, again shown by the green line. The fact that the vertical green line is at an identical wavenumber on both our panels demonstrates that we have achieved convergence for both our $\Lambda$CDM simulations as well as our modified gravity simulations.

\section{Binning in \texorpdfstring{$\mu$}{Geff}}\label{appendix:binning}
We now describe the procedure we adopted to bin $\mu$ in redshift. We reiterate this parameter space is multi-dimensional, which simultaneously covers a large portion of the parameter space while isolating non-linear behaviour is a difficult task. We focus on the latter in this work, as our intention is to understand the phenomenology.

\begin{figure}
 \includegraphics[width = 0.45\textwidth]{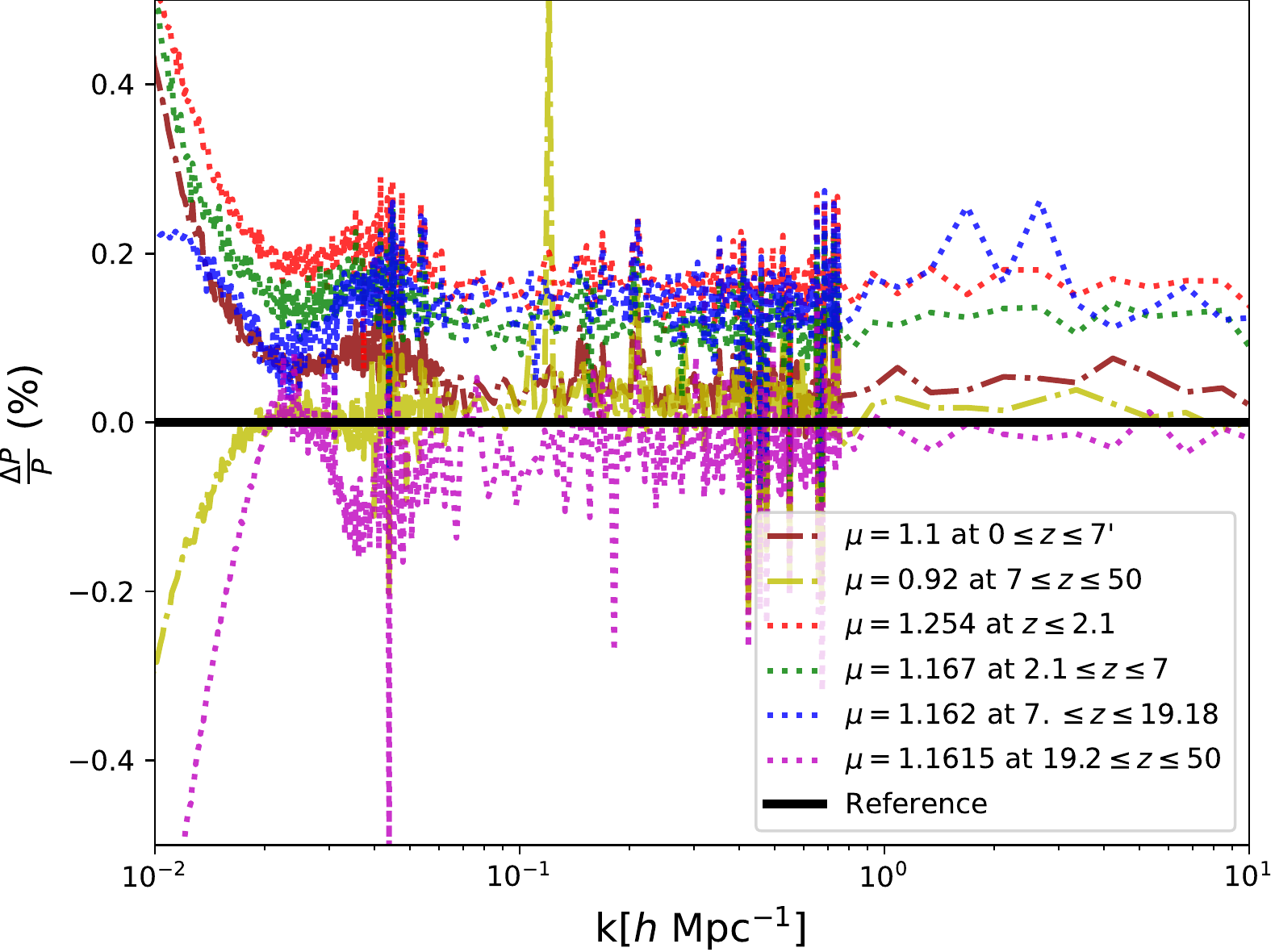}
 \includegraphics[width = 0.45\textwidth]{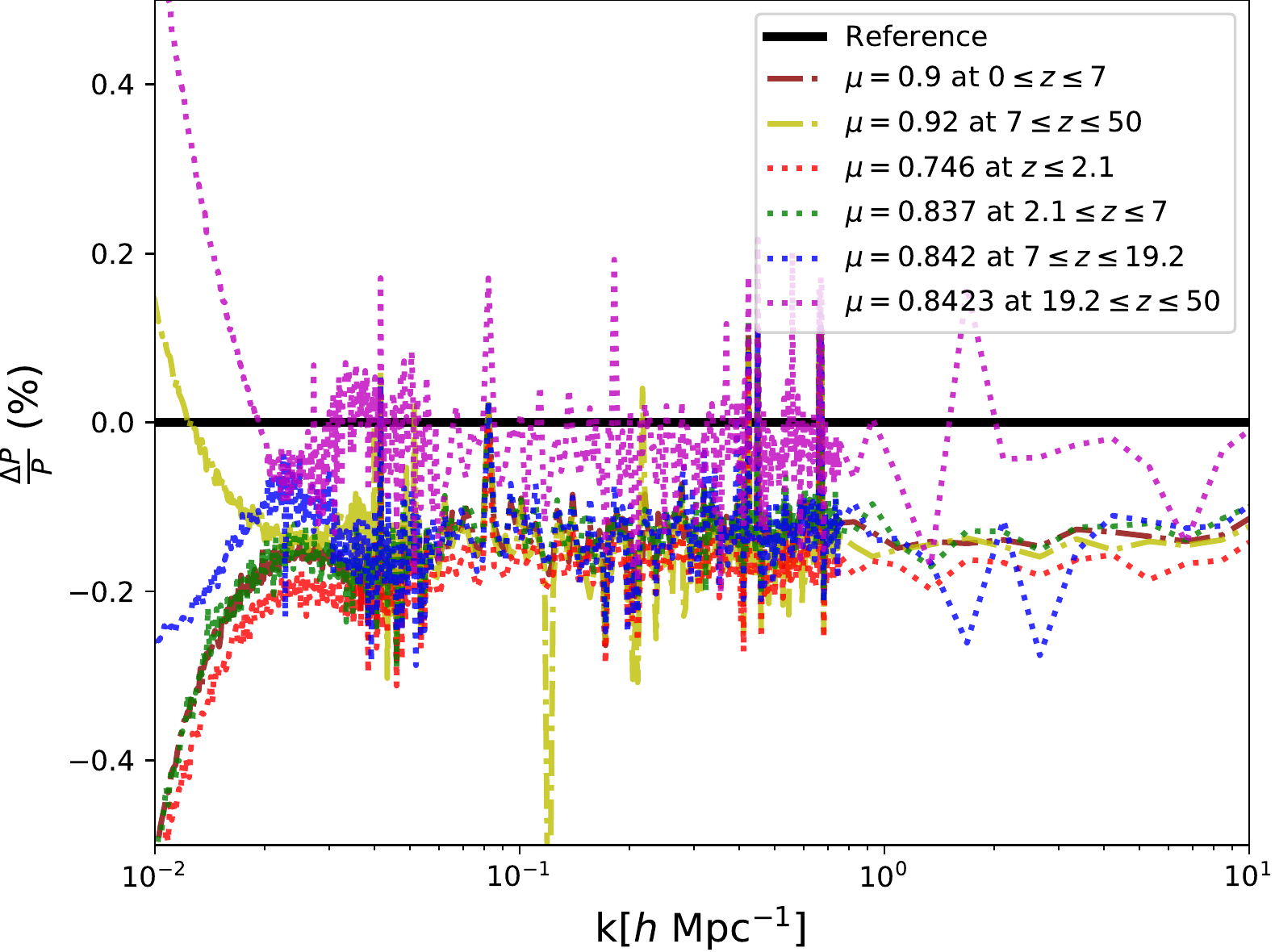}
 \caption{The relative difference in the linear power spectra at redshift zero to the reference case (chosen to be the case where $\mu$ is constant) from runs of our modified \texttt{CLASS} code. }
 \label{fig:CLASS}
\end{figure}

As mentioned in section~\ref{section:Pipeline}, we divide the redshift spanned by our simulations into 2 and 4 bins of equal $\Lambda$CDM growth. In other words, in the case where we have 2 redshift bins for $\mu$, we have that $D^{\Lambda {\rm CDM}}(z_1)/D^{\Lambda {\rm CDM}}(z_2) = D^{\Lambda {\rm CDM}}(z_2)/D^{\Lambda {\rm CDM}}(z_3)$. Setting $z_1 = 0$ and $z_3 = 50$, we obtain 
\begin{equation}
 D(z_2)^2 = D(z_1)D(z_3) \,,
\end{equation}
from which we can compute $z_2 \approx 7.01$. We follow a similar procedure in the case where we have 4 bins in redshift. We consider the bin $0 \leq z \leq 7.01$ as our reference in the context of choosing the value of $\mu$. We set $\mu = 1.1$ and $0.9$, in this bin, respectively, and compute $D(0)$. For all the other bins, we choose $\mu$ such that the growth factor at $z=0$ is identical to the value we computed for our reference bin. In this way, we obtain the $\mu$ values in table \ref{tab:binning}. In order to confirm that these values of $\mu$ indeed correspond to identical growth factors at $z=0$, we modified the equations of motion solved by \texttt{CLASS} for a time-dependent $\mu$ as outlined in \cite{ref:ZuccaMGCAMB}. In fig.~\ref{fig:CLASS} we show that the relative differences in the linear power spectra for all the cases in table \ref{tab:binning} are within 0.5\%, in accordance with the right panel in fig.~\ref{fig:Convergence}, which shows that our modified \texttt{CLASS} code performs as expected. Therefore, we run two sets of seven simulations in which $\mu\neq 1$ in one, two or four redshift bins, with the matter power spectrum measured from all the simulations being equal on linear scales at $z=0$.

We now describe the implementation of the $\mu$ smoothing in \texttt{CLASS} as well as \texttt{GADGET-2}. We assume that the input to the code includes an array of values for $\mu$ as well as an array of redshifts that denote the bins. We then use a combination of two error functions given by
\begin{eqnarray}
E_1 & = & C_1\text{erf}((z_{\rm array}[i] - z)/X) + C_2 \, , \\
E_2 & = & C_1\text{erf}(-(z_{\rm array}[i+1] - z)/X) + C_2 \, ,
\end{eqnarray}
where $C_1,\,C_2$ and $X$ are parameters that determine the amplitude, midpoint and width of the error function, respectively, to determine $\mu$ at each redshift. The iterative variable $i$ is incremented once the current redshift (associated to the timestep in the code) is lower than the midpoint of the current bin, i.e., when $z < \left(z_{\rm array}[i+1] + z_{\rm array}[i]\right)/2$. The constants $C_1$ and $C_2$ are given by
\begin{eqnarray}
C_1 & = & \mu_{\rm array}[i + 1] - \mu_{\rm array}[i]/2 \, , \\
C_2 & = & \mu_{\rm array}[i + 1] + \mu_{\rm array}[i]/2 \, , 
\end{eqnarray}
The value of $\mu$ is given by
\begin{eqnarray}
\mu & = & E_1\, \text{if} \, z>z_{\rm array}[i+1]-z[i]/2 \,, \\
\mu & = & E_2\, \text{if} \, z<z_{\rm array}[i+1]-z[i]/2 \,. 
\end{eqnarray}
We have chosen the width function $X = (1/A)\ln(\Delta z_{\rm bin})$ with $A$ being a tolerance parameter that determines the width of the transition and $\Delta z_{\rm bin}$ is the bin-width. This choice ensures that the width of the transition naturally becomes shorter at lower redshifts where the growth factor is more sensitive to redshift (where the timestep in the code is small).

\begin{figure}[H]
 \centering
 \includegraphics[width = 0.45\textwidth]{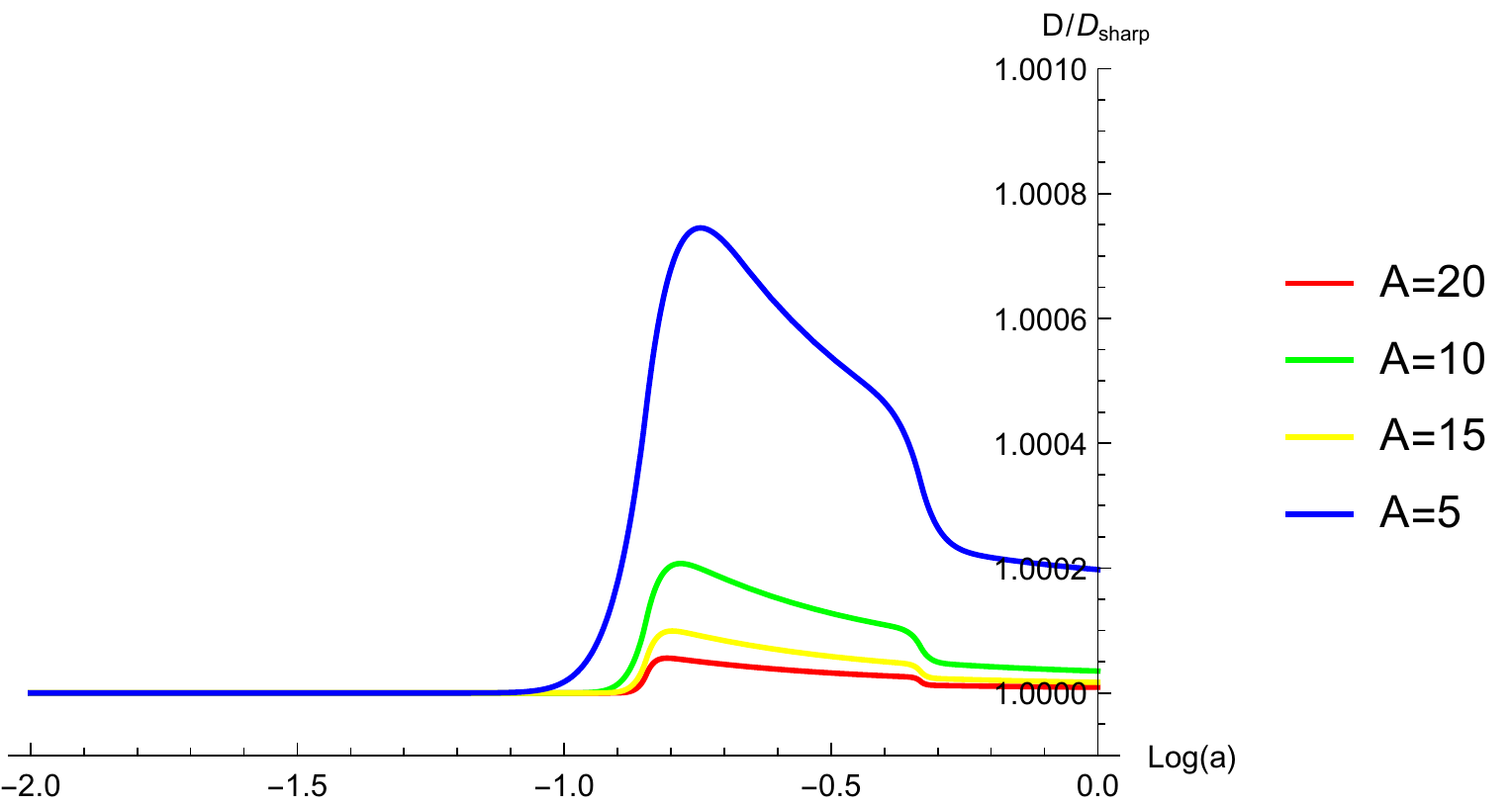}
 \includegraphics[width = 0.45\textwidth]{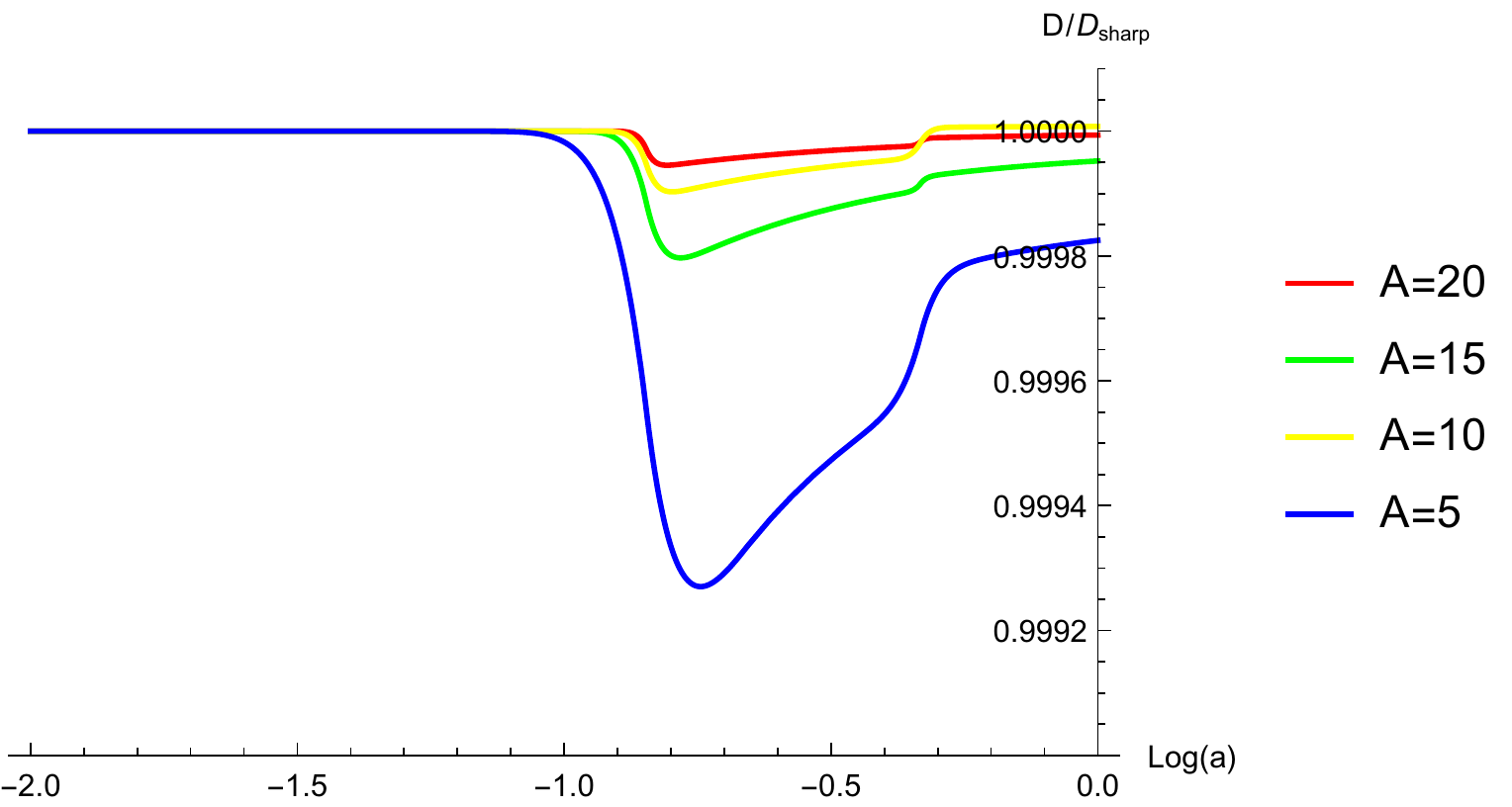}
 \caption{The plot of the growth factor as a function of redshift in the case where $\mu > 1$ (left) and $\mu < 1$ (right) at $2.15 \leq z \leq 7.01$.}
 \label{fig:smoothing}
\end{figure}

To estimate the impact of the smoothing, we have carried out two tests. The first test is the ratio of the power spectrum with the smoothed $\mu$ to the un-smooth case for different values of $A$. In this case, we have averaged the ratio over all $k$ values to obtain a ``goodness of fit" parameter, which ideally should be equal to zero. The second test was to plot the growth factor as a function of redshift, specifically over the range of redshifts where the smoothing is applied. The results from both these tests in \texttt{CLASS} are shown in fig.~\ref{fig:smoothing}.
On implementing this test in \texttt{GADGET-2}, we found that over the range of values we considered for the smoothing parameter, the differences between the smoothed and un-smoothed case were smaller than $0.1\%$.
\begin{comment}
\begin{figure}[H]
 \centering
 \includegraphics[width=0.7\textwidth]{Figures_Appendix/ratio_smoothing.pdf}
 \caption{The ratio of the smoothed power spectrum to sharp case, where we have set the bin width parameter to $A=10$ (see appendix \ref{appendix:binning} for details).}
 \label{fig:smoothingGadget}
\end{figure}

\end{comment}
\section{Fitting functions}\label{appendix:reaction}
We now present an overview of the different fitting functions we examined in section \ref{section:FittingFunction}. Note that we also examine the performance of the standard \texttt{halofit} procedure \cite{ref:halofit1,ref:halofit2,ref:halofit3} which we do not describe in detail here. Not that while this is not a detailed examination of these fitting functions, we present the procedure that we followed to actually predict $R(k)$ in order to compare with our simulations.     

\subsection{Cui et al.}

The fitting function derived in \cite{ref:Cui1} is given by
\begin{equation}\label{eq:CuiFF}
 R(x, \mu) = \exp\left\{((1 - \mu)B(\mu)x^{C(\mu)})\right\}\,,
\end{equation} 
where $R(x, \mu)$ as mentioned in the main text is the ratio of the matter power spectrum in modified gravity relative to the $\Lambda$CDM pseudo spectrum (with equal linear growth). $B(\mu) = 0.0429 + 0.133 \mu^{-4}$, $C(\mu) = 0.573$ and $x = \Delta^2(k, z, \mu=1)$ is the dimensionless power spectrum $P(k)k^3/(2\pi)$. The main feature of this function is that it predicts $R>1$ when $\mu<1$ and vice-versa when $\mu>1$. In other words, this fitting function assumes that the non-linear evolution is slower when the growth rate is increased due to $\mu$ being larger than its value in $\Lambda$CDM, and vice versa. While we have observed a similar behaviour in some of our simulations, we find that this is not true in the general case. 

\begin{figure}
 \centering
 \includegraphics[width = 0.7\textwidth]{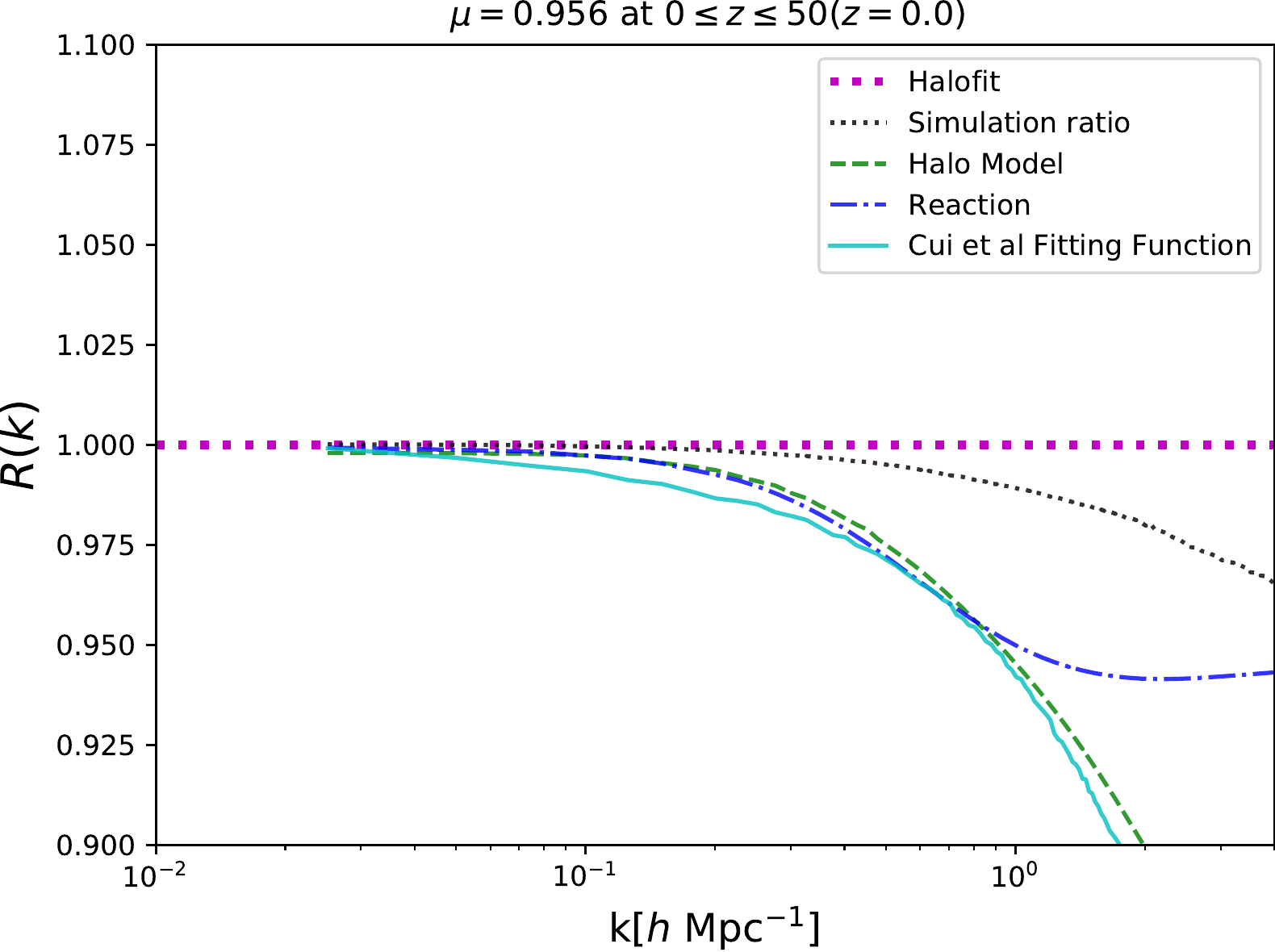}
 \caption{The ratio of the matter power spectrum in the simulation with $\mu = 0.956$ throughout the simulation period. The Cui et al. fitting function \cite{ref:Cui1} is designed to model the power spectrum in this case, where $\mu$ is constant.}
 \label{fig:CuiFF}
\end{figure}

We demonstrate in fig.~\ref{fig:CuiFF} that one can reproduce with reasonable success the matter power spectrum in the case where $\mu$ is constant throughout the simulation time period, although even in this case we see that \texttt{ReACT} outperforms the other fitting functions that we consider. 

\subsection{Halo model}
The standard halo model consists of the 1-halo term, given by
\begin{equation}\label{eq:1halo}
 P^{1\rm h}(k) = \int \mathrm{d}m n(m) \left(\frac{m}{\bar{\rho}}\right)^2|u(k|m, z)|^2\,,
\end{equation}
which is the contribution to the power spectrum from a single halo, along with the 2-halo term given by 
\begin{equation}\label{eq:2halo}
 P^{2\rm h}(k) = \int \mathrm{d}m_1 n(m_1) \left(\frac{m_1}{\bar{\rho}}\right)|u(k|m_1, z)|\int \mathrm{d}m_2  n(m_2) \left(\frac{m_2}{\bar{\rho}}\right)|u(k|m_2, z)|P_{\rm hh}(k, z|m_1, m_2)\,,
\end{equation}
which is the contribution from 2 different haloes, where $n(m)$ is the mass function, $u(k,z|m)$ is the Fourier transform of the density profile dark matter haloes of mass $m$ at redshift $z$ and $P_{\rm hh}$ is the halo power spectrum, related to the two-point-correlation function of the haloes themselves and quantifies the interaction between two haloes. Of these, the first term dominates on small non-linear scales, i.e., at large $k$, while the 2-halo term dominates on large scales. Note that the 2-halo term is usually well approximated by the linear matter powers spectrum in the literature.

In order to compute the 1-halo term, we calculate the mass function and the bias term assuming a time-dependent $\mu$ according to the Sheth-Tormen procedure \cite{Sheth1999,Sheth2002} and perform the integrals in eqs.~\eqref{eq:1halo}\eqref{eq:2halo}. We then obtain the full non-linear matter power spectrum by summing up the 1-halo term and 2-halo term. As a cross-check we ensured that our results are replicated by summing the 1-halo term and the linear matter power spectrum. 

Increasing the value of $\mu$ increases the rate of structure formation which means that haloes form earlier, resulting in smaller concentration parameters associated to the haloes. Since the Fourier transform of the density profiles is directly proportional to the concentration parameter, the contribution from the 1-halo term is smaller, resulting in a lack of power on smaller scales \cite{ref:Cui1} (with the opposite taking place for $\mu<1$).

\subsection{Halo model reaction}

The halo model reaction formalism integrates a modified halo model and the spherical collapse model together to predict the matter power spectrum on non-linear scales. As mentioned before, the halo model suffers from a lack of accuracy on quasi-linear scales as in the original recipe. The authors go one step further to address this problem via the reaction term, given by
\begin{equation}\label{eq:ReacttionCataneo}
 \mathcal{R}(a, k)  =  \frac{\left[(1 - \epsilon(z))e^{-k/k_{\ast}(z)} + \epsilon(z)\right]P_{\rm mg}^{\rm lin}(z, k) + P^{1\rm h}_{\rm mg}(z, k)}{P_{\rm pseudo}^{\rm hm}(z, k)} \,,
\end{equation}
where the subscript `pseudo' represents the pseudo spectrum and $k_{\ast}$ is computed from perturbation theory \cite{ref:Bernardeau}. Therefore, the quantity $\mathcal{R}$, the so-called reaction, is a ratio of halo-model power spectra. In order to obtain a prediction of the modified gravity power spectrum, one simply multiplies $\mathcal{R}$ by the full non-linear $\Lambda$CDM pseudo spectrum. To understand how well \texttt{ReACT} captures our simulations, we use the \texttt{ReACT} code to calculate $R(a,k)=P_{\rm mg}/P^{\rm pseudo}_{\Lambda \rm CDM} = \mathcal{R}$. \\
The quantity $\mathcal{R}$ has the following basic features:
\begin{itemize}
 \item On linear scales, $\mathcal{R}\rightarrow 1$ by construction since the pseudo-spectrum has the same linear growth as the modified gravity spectrum.
 \item On small non-linear scales the reaction is given by the ratio of the 1-halo terms, i.e., $\lim_{k\rightarrow k_{\rm nl}}\mathcal{R} \rightarrow P^{1\rm h}_{\rm mg}(z, k)/P^{1\rm h}_{\rm pseudo}(z, k)$.
\end{itemize}
The reaction term in eq.~(\ref{eq:Reaction}) may be computed from the following equations   
\begin{eqnarray}
 P_{\rm pseudo}^{\rm hm} & = & P_{\rm pseudo}^{\rm lin} + P^{1\rm h}_{\rm pseudo} \,, \\
 \epsilon & = & \lim_{k\rightarrow 0}\frac{P^{1\rm h}_{\rm mg}(z, k)}{P^{1\rm h}_{\rm pseudo}(z, k)} \,, \\
 k_{\ast}(z) & = & -\bar{k}\left( \ln\left[ \frac{A(\bar{k}, z)}{P_{\rm mg}^{\rm lin}(z, k)} - \epsilon(z) - \ln(1 - \epsilon(z))\right]\right) \,, \\
 A(k, z) & = & \frac{P_{\rm mg}^{\rm 1-loop}(z,k) + P^{1\rm h}_{\rm mg}(z, k)}{P_{\rm pseudo}^{\rm 1-loop}(z,k) + P^{1\rm h}_{\rm pseudo}(z, k)}P_{\rm pseudo}^{\rm hm}(z, k) - P^{1\rm h}_{\rm mg}(z, k) \,, \label{eq:1loop}
\end{eqnarray}
where the 1-loop terms may be computed from Fourier transforming the Poisson equation \cite{ref:reactionCataneo, ref:ReactTheory}
\begin{equation}
 -k^2\phi(a,k) = \frac{3 \Omega_{\rm m}(a) a H(a)}{2} \mu(a, k)\delta(a, k) + \tilde{S}(a, k)\,,
\end{equation}
where the source function $\tilde{S}(a, k)$ encodes non-linear corrections to the Newtonian Poisson equation that appear at higher order. Essentially, one can derive $\tilde{S}(a, k)$ for specific models and the corresponding screening mechanisms that operate on the relevant scales. The authors in \cite{ref:reactionCataneo} calculate $\tilde{S}(a, k)$ for $f(R)$ and DGP gravity in order to reproduce the full non-linear matter power spectrum in modified gravity. Note that the \texttt{ReACT} code is set up to automatically calculate $\tilde{S}(a, k)$ for these two models and $\Lambda$CDM (where it is zero). 

In contrast to \cite{ref:reactionCataneo}, our work is designed to understand the phenomenology while remaining model-agnostic and therefore, we neglect any screening mechanisms and indeed any scale dependent corrections to the Poisson equation given in eq. \eqref{eq:MGParam}. Therefore, in order to reproduce the matter power spectrum in our formalism within the \texttt{ReACT} code, we only modify the $\mu$ parameter and set $\tilde{S}(a,k)=0$ and indeed all the other modified gravity parameters within the code to their respective values in $\Lambda$CDM. This simplifies the numerical implementation of the \texttt{ReACT} while still being able to estimate the non-linear effects of modifiying $\mu$.

\label{lastpage}

\end{document}